\documentclass[fleqn,10pt]{wlscirep}
\usepackage{braket}
\usepackage{mathrsfs}
\usepackage[numbers,sort&compress]{natbib}

\title{Method of Higher-order Operators for Quantum Optomechanics}

\author[1,*]{Sina Khorasani}
\affil[1]{Vienna Center for Quantum Science and Technology, Boltzmanngasse 5, 1090 Vienna, Austria}
\affil[*]{sina.khorasani@ieee.org}


\begin{abstract}
	We demonstrate application of the method of higher-order operators to nonlinear standard optomechanics. It is shown that a symmetry breaking in frequency shifts exists, corresponding to inequivalency of red and blue side-bands. This arises from nonlinear higher-order processes leading to inequal detunings. Similarly, a higher-order resonance shift exists appearing as changes in both of the optical and mechanical resonances. We provide the first known method to explicitly estimate the population of coherent phonons. We also calculate corrections to spring effect due to higher-order interactions and coherent phonons, and show that these corrections can be quite significant in measurement of single-photon optomechanical interaction rate. It is shown that there exists non-unique and various choices for the higher-order operators to solve the optomechanical interaction with different multiplicative noise terms, among which a minimal basis offers exactly linear Langevin equations, while decoupling one Langevin equation and thus leaving the whole standard optomechanical problem exactly solvable by explicit expressions. We finally present a detailed treatment of multiplicative noise as well as nonlinear dynamic stability phases by the method of higher-order operators. Similar approach can be used outside the domain of standard optomechanics to quadratic and all other types of nonlinear interactions in quantum physics.
\end{abstract}

\begin{document}
	
	\flushbottom
	\maketitle
	
	\thispagestyle{empty}
	
	\section*{Introduction}
	
	Nonlinear quantum interactions with stochastic noise input stand among the most difficult analytical challenges to solve in the context of stochastic differential equations. While linearized interactions remain accurate for description of many experiments, a certain class of quadratic and higher-order physical phenomena cannot be normally understood under linearized approximations. While in classical problems the resulting Langevin equations are scalar functions, in quantum problems one has to deal with nonlinear operator differential equations. If expanded unto base kets, bosonic operators can assume infinite-dimensional matrix forms, rendering the solution entirely intractable.
	
	Such classes of nonlinear operator problems can be addressed by construction of Fokker-Planck or nonlinear Schr\"{o}dinger equations, among which there exists a one-to-one correspondence. The Fokker-Planck equation \cite{Fokker1,Fokker2,Fokker3,Fokker4,Fokker5} is actually equivalent to the nonlinear Schr\"{o}dinger equation with bosonic operator algebra, and its moments \cite{Moments} translate into nonlinear Langevin equations. The method of master equations \cite{Master1,Master2} also can be used in combination with the quasi-probablity Wigner functions \cite{Master3,Master4} to deal with nonlinear quantum interactions. The master equation approach is reasonably accurate as long as Born and Markov approximations are not employed \cite{Master5}. But none of these methods is probably as convenient as the method of Langevin equations \cite{Langevin1,Langevin2,Langevin3,Langevin4}, which has found popularity in the context of quantum optoemchanics \cite{Law,Macri,Kip1,Aspel1,Bowen,Opto1,Opto2,Opto3,Polariton, Multi1,Multi2,Multi3,Multi4,Multi5}. 
	
	Being an inherently nonlinear interaction among photonic and phononic baths \cite{Kerr1,Kerr2,Kerr3,Kerr4,Kerr5,Kerr6,Duffing,X2,Girvin}, the standard quantum optomechanics is normally described by linearized Langevin equations \cite{Langevin1,Langevin2,Langevin3,Langevin4}. This will suffice to address a majority of complex experimental situations such as optomechanical-induced transparency \cite{Lemonde2,OMIT1,OMIT2} and polaron anti-crossing \cite{Polaron}, but effects such as non-classical states of light \cite{Master3,Master4,Rabl,Lemonde1}, optomechanical emission of real photons from vacuum \cite{Lemonde3}, photon blockade \cite{Rabl}, nonlinear self-oscillations \cite{Self1,Self2,Self3,Self4,Self5}, and chaos \cite{Chaos1,Chaos2} are all among manifestations of nonlinear regimes in standard optomechanics, which need description using nonlinear algebra. Also, biquadratic interactions (mostly referred to as quadratic interactions) among bosonic baths remain a hurdle. In quadratic optomechanics \cite{Quad1,Quad2,Quad3,Quad4,Quad5,Quad6,Quad7,Quad8,Quad9,Quad10,Quad11,Bruschi}, which is a topic of growing interest in the recent year, having an analytical tool capable of addressing such kinds of nonlinearity is advantageous. A perturbation technique based on the expansion of time-evolution operators \cite{Bruschi} is employed to investigate quadratic interactions and it has been shown that for mechanical frequencies exceeding optical frequencies a new unexplored regime appears in which the roles of optical and mechanical partitions are interchanged.
	
	Recently, the author has reconsidered the theoretical description of optomechanics \cite{Paper1} and shown that quadratic interactions are subject to two corrections resulting from momentum conservation and relativistic effects. Such types of quadratic corrections become significant when the mechanical frequency is within the same order of or exceeds electromagnetic frequency. Furthermore, an analytical approach is proposed to tackle nonlinear quantum interactions \cite{Paper2} and a method of expansion unto higher-order operators is proposed and investigated in details.
	
	In this article, the higher-order operator approach recently proposed by the author \cite{Paper2,Paper4,Chapter,Paper5} is employed to address the standard optomechanics, and it is shown that there exists a minimal choice of higher-order operator basis which leads to exactly linear and fully separable Langevin equations with multiplicative input noise terms \cite{Noise}. We also present a full mathematical treatment of multiplicative noise terms, which turn out to play a crucial rule in higher-order quantum optomechanics. This allows one to provide an exact and explicit solution using an operator-based method to solve the optomechanical interactions in the nonlinear regime. There exists higher-order effects appearing at high optical pump rates, and can be predicted using the method discussed here. These include inequivalent red and blue detunings, higher-order resonance shift and spring effects, and also zero-point-field induced optomechanical shift of mechanical frequency. The inequivalency of red and blue detuned side-bands, which appears as a counter-intuitive difference in their respective frequency shifts, is different from the well-known anomalous Stokes-Anti-Stokes symmetry breaking \cite{Stokes1,Stokes2,Stokes3} which is connected to different scattering amplitudes. The same method of higher-order operator algebra has been recently used independently as well \cite{Liu}. 
	
	We also show for the first time that the introduced method of higher-order operators can be used to estimate the coherent population of phonons in the optomechanical cavity, here referred to as the coherent phonon number. This quantity can not only be calculated explicitly in terms of optomechanical parameters, but also, can be found by fitting the expressions of corrected spring effect to the experimental observations. Also, dynamic linear and nonlinear stability phases in red and blue-detuned drives can be well computed and estimated using the method of higher-order operators.
	
	\section*{Results}
	
	The standard optomechanical Hamiltonian reads \cite{Law,Macri,Kip1,Aspel1,Bowen}
	\begin{equation}
	\label{eq1}
	\mathbb{H}_{\text{OM}}=\hbar \Omega\hat{m}-\hbar \Delta\hat{n}-\hbar g_0\hat{n}\left(\hat{b}+{\hat{b}}^{\dagger }\right),
	\end{equation}
	\noindent
	where $\hat{n}=\hat{a}^\dagger\hat{a}$ and $\hat{m}=\hat{b}^\dagger\hat{b}$ are photon and phonon number operators with $\hat{a}$ and $\hat{b}$ respectively being the photon and phonon annihilators, $\Omega$ is the mechanical frequency, $\Delta$ is optical detuning from cavity resonance, and $g_0$ is the single-photon optomechanical interaction rate. The interaction $\mathbb{H}_{\text{OM}}$ is not quadratic, but is still cubic nonlinear. It is normally solved by a straightforward linearization \cite{Kip1,Aspel1,Bowen}, but can be also solved at the second-order accuracy using the higher-order operators described in the preceding article \cite{Paper2}. 
	
	In order to form a closed basis of operators, we may choose either the higher-order operators
	\begin{equation}
	\label{eq1a}
	\{A\}^\text{T}=\left\{\hat{a},\hat{a}\hat{b},\hat{a}{\hat{b}}^{\dagger }\right\},
	\end{equation}
	of the second-degree, which forms a $3\times 3$ system of Langevin equations, or
	\begin{equation}
	\label{eq3}
	\{A\}^\text{T}=\left\{\hat{a},\hat{b},\hat{a}\hat{b},\hat{a}{\hat{b}}^{\dagger },\hat{n},\hat{c}\right\},
	\end{equation}
	\noindent 
	which forms a $6\times 6$ system of Langevin equations. Here, we adopt the definition $\hat{c}=\frac{1}{2}\hat{a}^2$ \cite{Paper1,Paper2}. 
	
	It is easy to verify that this system is exactly closed, by calculation of all possible commutation pairs between the elements. Out of the $6!$ commutators, the non-zero ones are $\left[\hat{a},\hat{n}\right]=-\left[\hat{a}{\hat{b}}^{\dagger },\hat{b}\right]=\hat{a}$, $\left[\hat{a}\hat{b},\hat{n}\right]=\hat{a}\hat{b}$, $\left[\hat{a}{\hat{b}}^{\dagger },\hat{n}\right]=\hat{a}{\hat{b}}^{\dagger }$, and $\left[\hat{a}\hat{b},\hat{a}{\hat{b}}^{\dagger }\right]=\left[\hat{c},\hat{n}\right]=2\hat{c}$, which is obviously a closed basis. Now, one may proceed with composition of the Langevin equations. 
	
	The applicability of the basis (\ref{eq1a}) becomes readily clear by calculating the braket $[\hat{a},\mathbb{H}_\text{OM}]$ as appears in the corresponding Langevin equation. The terms involving the second-degree operators $\hat{a}\hat{b}$ and $\hat{a}\hat{b}^\dagger$ immediately show up. The key in the method of higher-order operators is to keep these operator pairs, triplets and so on together, as each combination has a clear corresponding physical process. While $\hat{a}$ and $\hat{b}$ refer to individual ladder operators, $\hat{a}\hat{b}$ and $\hat{a}\hat{b}^\dagger$ respectively construct the blue and red 1-photon/1-phonon processes. For this reason, it is probably more appropriate to call these higher-degree operator combinations as processes. 
	
	The Langevin equations for the blue $\hat{a}\hat{b}$ and red $\hat{a}\hat{b}^\dagger$ processes do not close on themselves, because of the appearance of third-order blue- and red-like processes $\hat{a}\hat{b}^2$ and $\hat{a}\hat{b}^{\dagger 2}$, describing 1-photon/2-phonon processes. Similarly, every $j$-th order blue- or red-like process such as $\hat{a}\hat{b}^j$ and $\hat{a}\hat{b}^{\dagger j}$ will lead to the $j+1$-order process. Hence, the infinite-dimensional basis $\{\hat{a}\}\cup\{\forall\hat{a}\hat{b}^j,\hat{a}\hat{b}^{\dagger j};j\in\mathscr{N}\}$ can provide an exact solution to the optomechanics. Furthermore, the convergence of solutions basis on such expansions would be questionable when $g_0<<\Omega$ is violated. In general, the $j$-th order processes correspond to the 1-photon/$j$-phonon interactions and contribute to the $j+1$-order sidebands. In this article, it has been shown that under practical conditions, it is unnecessary to take account of the processes $j\geq 2$ and the $3\times 3$ basis (\ref{eq1a}) is rather sufficient for most practical purposes. Nonetheless, $j=2$ processes contribute significantly to nonlinear stability and second-order mechanical sidebands. While the use of an infinite-dimensional basis is surprisingly unnecessary in still a higher-order formulation, using the compact minimal basis to be discussed in the following can lead to the mathematically exact solution.
    The choice of basis is not unique, and every non-degenerate linear combination of bases leads to another equivalent form. One may for instance arbitrate the three-dimensional linear basis $\{A\}^{\rm T}=\{\hat{a},\hat{b},\hat{b}^\dagger\}$ or the four-dimensional linear basis $\{A\}^{\rm T}=\{\hat{a},\hat{b},\hat{a}^\dagger,\hat{b}^\dagger\}$ as is taken in the context of linearized standard optomechanics \cite{Kip1,Aspel1,Bowen}, the five-dimensional all-Hermitian basis $\{A\}^{\rm T}=\{\hat{n},\hat{m},\hat{n}^2,\hat{n}(\hat{b}+\hat{b}^\dagger),i\hat{n}(\hat{b}-\hat{b}^\dagger)\}$ \cite{Bruschi}, and ultimately the minimal three-dimensional basis 
	\begin{equation}
	\label{eqMinimal}
	\{A\}^{\rm T}=\{\hat{n}^2,\hat{n}\hat{b},\hat{n}\hat{b}^\dagger\}=\{\hat{N},\hat{B},\hat{B}^\dagger\},
	\end{equation}
	\noindent
	assumed here, which is of the fourth-degree. We shall later observe that while (\ref{eq3}) is necessary to construct the closed Langevin equations, a second-order linearization will be needed to decouple three operators, leaving only the basis (\ref{eq1a}) in effect. Quite remarkably, however, and in a similar manner, the use of minimal basis (\ref{eqMinimal}) turns out to be fairly convenient to construct the optomechanical Langevin equations. This is not only since the Langevin equations take on exactly linear forms, but also eventually the equation for $\hat{N}$ and $\hat{B}^\dagger$ will decouple. This leaves the whole standard optomechanical interaction exactly solvable through integration of only one linear differential equation in terms of $\hat{B}$. The main difference between using various choices of higher-order operator bases \cite{Paper2} is the noise terms. It turns out that the definition and higher-order operators lead to multiplicative noise inputs, which once known, the problem will be conveniently solvable. Full mathematical treatment of multiplicative noise terms is necessary for description of some various phenomena and this will be discussed in \S{S10} of supplementary information. 
	
	\subsection*{Side-band Inequivalence}
	
\begin{figure}[ht!]
	\centering
	\includegraphics[width=5.54in]{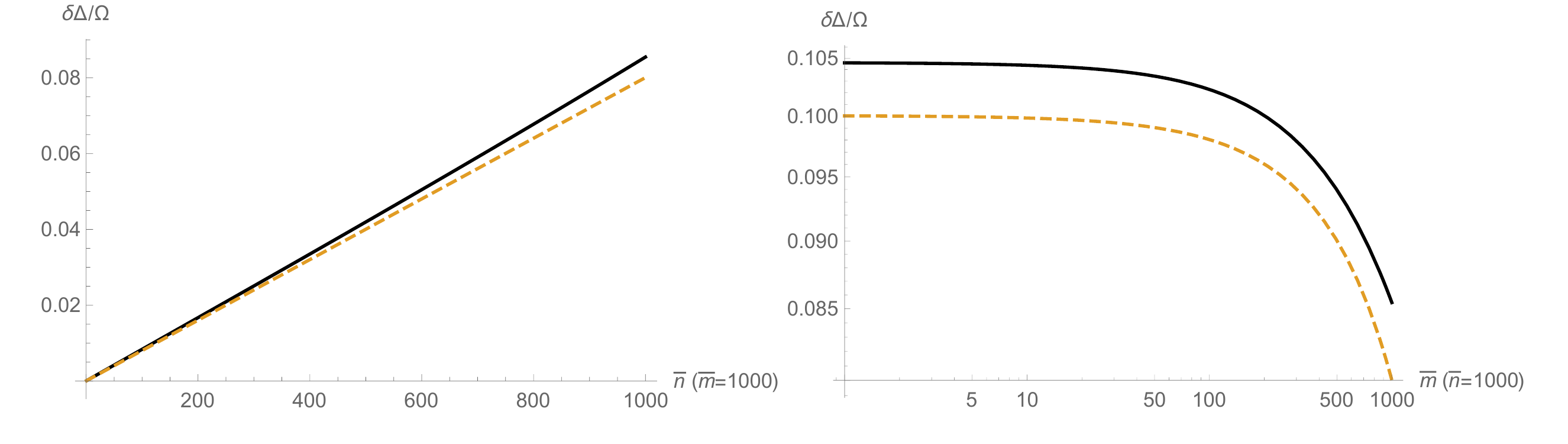}
	\caption{Normalized inequivalence $\delta\Delta/\Omega=(\Delta_\text{b}+\Delta_\text{r})/2\Omega$ of sideband frequency detunings $\Delta_\text{r}$ and $\Delta_\text{b}$ versus intracavity photon $\bar{n}$ and coherent phonon $\bar{m}$ for $g_0/\Omega=10^{-3}$. Solid lines are from exact numerical calculations and dashed lines are from the asymptotic expansion (\ref{Sideband}).\label{Fig0}}
	\end{figure}

	Defining $\Delta_b$ and $\Delta_r$ respectively as the blue and red frequency shifts of sidebands, it is possible to show that these two quantities do not necessarily agree in magnitude, such that $\Delta_b+\Delta_r\neq 0$. As shown in \S{S4} of supplementary information, an explicit relation for the side-band inequivalence $\delta\Delta=\frac{1}{2}(\Delta_\text{r}+\Delta_\text{b})$ can be found through series expansion of the eigenvalues of the coefficient matrix from (S19). With some algebra, it is possible to show that for $g_0<<\Omega$  correct to the fourth-order, we get
	\begin{equation}
	\label{Sideband}
	\frac{\delta\Delta}{\Omega}\approx \left(\frac{g_0}{\Omega}\right)^2\left(\bar{n}+\frac{1}{2}\right)-2\left(\frac{g_0}{\Omega}\right)^4\left(\bar{n}+\frac{1}{2}\right)\left(\bar{m}+\frac{1}{2}\right).
	\end{equation}
	\noindent
	Here, $\bar{n}(\Delta)$ is the intracavity photon population and $\bar{m}(\Delta)$ is the coherent phonon population given by
	\begin{equation}
	\label{m8GranCoupe} 
	\bar{m}(\Delta)\approx\frac{32 g_0^2 \Omega ^2 \left(\gamma ^2+\gamma  \Gamma +4 \Delta ^2\right)}{\left(\gamma ^2+4 \Delta ^2\right) \left(\Gamma ^2+4 \Omega ^2\right)^2}\bar{n}^2(\Delta )=g_0^2\zeta(\Delta)\bar{n}^2(\Delta),
	\end{equation}
	where $\Gamma$ is the mechanical decay rate, and $\gamma=\kappa+\Gamma$ is the total optomechanical decay rate with $\kappa$ being the optical decay rate, as proved in details in \S{S6} of supplementary information using the method of higher-order operators. Also, $\bar{n}(\Delta)$ can be found from numerical solution of a third-order algebraic equation (S9). The relationship $\bar{m}\propto\bar{n}^2$ signifies the fact that mechanical oscillations are nonlinearly driven by optical radiation pressure. A typical behavior of this phenomenon is illustrated in Fig. \ref{Fig0}.
	
	There is a related polaritonic splitting effect \cite{Lemonde2}, as a result of anti-crossing between the optomechanically interacting optical and mechanical resonances generated across either of the mechanical side-bands, amount of which happens to be exactly $2\delta\Delta$. This has nothing to do with the side-band asymmetry, which happens to occur on the two opposite sides of the main cavity resonance. It should be mentioned that observation of this phenomenon in superconducting electromechanics \cite{Mika} as well as parametrically actuated nano-string resonators \cite{Weig1,Weig2} can potentially yield the most clear results due to various experimental conditions. In fact, intracavity photon numbers as large as $10^6$ and $10^{8}$ and more are attainable respectively in superconducting electromechanics and optically-trapped nano-particle optomechanics. 
	
	A close inspection of a very high-resolution measurement on a side-band resolved microtoroidal disk\cite{Kip3} yields a side-band inequivalence of $\delta\Delta\approx 2\pi\times(142\pm 36) \text{Hz}$, which perfectly complies to (\ref{Sideband}) if $\bar{n}=(5.1\pm 1.3)\times 10^3$. Unfortunately, further such a high resolution measurements on deeply side-band resolved optomechanical cavities are not reported elsewhere to the best knowledge of authors. Nevertheless, clear signatures of side-band inequivalence can be easily verified in few other experiments \cite{Mika,Weig1,Teufel}. Remarkably, recent measurements on Stokes-Anti-Stokes scattering from multi-layered $\text{MoTe}_2$ exhibits a difference in frequency shift as large as 7\% for the five-layered sample \cite{Stokes2}, which corresponds to $0.88 \pm 0.11\text{cm}^{-1}$.

	Also, a recent landmark experiment on room-temperature quantum optomechanical correlations \cite{Vivishek} has reported measurements which coincidentally exhibit a sideband inequivalence up to $4\text{kHz}$ and roughly agree to the approximation $\delta\Delta\approx g_0^2\bar{n}/\Omega$. This issue remains, nevertheless, as an open problem in the context of experimental quantum optomechanics. 
	
	In any experimental attempt to measure this phenomenon, a side-band resolved cavity could be driven on resonance and noise spectra of the two mechanical side bands be measured with extreme precision in a heterodyne setup. Even in case of well-known thermo-optical effects and two-photon dispersion or absorption which cause drifts in the optical resonance and other optomechanical parameters \cite{Painter}, this effect should be still observable in principle. The reason is that the amount of inequivalence is actually independent of the exact pump frequency as long as intracavity photon population does not change significantly. So, it should be sufficient only if the cavity is driven on or close to the optical resonance for the side-bands to be sufficiently different in their frequency shifts.
	
	\subsection*{Higher-order Resonance Shift}
	
	The contribution of the off-diagonal terms to the mechanical frequency $\Omega$ in the coefficients matrix of optomechanical Langevin equations (S20) of supplementary information, can be ultimately held responsible for the so-called optomechanical spring effect \cite{Kip1,Aspel1,Bowen,Quad9,Spring1,Spring2,Spring3,Spring4,Spring5}. As the result of optomechanical interaction, both of the optical and mechanical resonance frequencies and damping rates undergo shifts. Even at the limit of zero input optical power $\alpha=0$ and therefore zero cavity photon number $\bar{n}=0$, it is possible to show that there is a temperature-dependent shift in the mechanical resonance frequency, markedly different from the lattice-expansion dependent effect. This effect is solely due to the optomechanical interaction with virtual cavity photons, which completely vanishes when $g_0=0$. In close relationship to the shift of resonances, we can also study the optomechanical spring effect with the corrections from higher-order interactions included.
	
	The analysis of spring effect is normally done by consideration of the effective optomechanical force acting upon the damped mechanical oscillator, thus obtaining a shift in squared mechanical frequency $\delta(\Omega^2)$, whose real and imaginary parts give expressions for $\delta\Omega$ and $\delta\Gamma$. Corrections to these two terms due to higher-order interactions are discussed in \S \ref{AppA}. Here, we demonstrate that the analysis using higher-order operator algebra can recover some important lost information regarding the optical and mechanical resonances when the analysis is done on the linearized basis $\{A\}^\text{T}=\{\hat{a},\hat{a}^\dagger,\hat{b},\hat{b}^\dagger\}$. 
	
	To proceed, we consider finding eigenvalues of the matrix $\textbf{M}$ as defined in (S19) of supplementary information. Ignoring all higher-order nonlinear effects beyond the basis $\{A\}^\text{T}=\{\hat{a},\hat{a}\hat{b},\hat{a}\hat{b}^\dagger\}$, we set $s=0$. This enables us to search for the eigenvalues of the coefficients matrix $\textbf{M}$ as
	\begin{equation}
	\label{eqSpring1Copy}
	\text{eig}[\textbf{M}]=\text{eig}\left[ 
	\begin{array}{ccc}
	i\Delta-\frac{\kappa }{2} & ig_0 & ig_0 \\ 
	i(G+f^+) &  -i(\Omega-\Delta)-\frac{\gamma }{2} & 0 \\ 
	-i(G-f^-) & 0 & i(\Omega+\Delta)-\frac{\gamma }{2}  
	\end{array}
	\right]=i\left\{
	\begin{array}{c}
	\Delta+\lambda_1+i\gamma_1 \\
	\Delta+\lambda_2+i\gamma_2 \\
	\Delta+\lambda_3+i\gamma_3 
	\end{array}
	\right\}=i\left\{
	\begin{array}{c}
	\Delta+\eta_1(\Delta,T) \\
	\Delta+\eta_2(\Delta,T) \\
	\Delta+\eta_3(\Delta,T) 
	\end{array}
	\right\},
	\end{equation}
	\noindent
	in which $G=g_0\bar{n}$, $f^\pm=g_0(\bar{m}+\frac{1}{2})\pm \frac{1}{2}g_0$, $\lambda_j=\Re[\eta_j]$ and $\gamma_j=\Im[\eta_j]$ with $j=1,2,3$ are real valued functions of $\Delta$ and bath temperature $T$. The temperature $T$ determines $\bar{m}$ while $\bar{n}$ is a function of $\Delta$ as well as input photon rate $\alpha$. 	In general, the three eigenvalues $\eta_j=\lambda_j(\Delta,T)+i\gamma_j(\Delta,T), j=1,2,3$ are expected to be deviate from the three free-running values $\psi_1=i\frac{1}{2}\kappa$, $\psi_2=-\Omega+i\frac{1}{2}\gamma$, and $\psi_3=\Omega+i\frac{1}{2}\gamma$, as $\eta_j\approx \psi_j-\Delta$ because of non-zero $g_0$. Solving the three equations therefore gives the values of shifted optical and mechanical frequencies and their damping rates compared to the bare values in absence of optomechanical interactions with $g_0=0$, given by $\delta\Omega=-\frac{1}{2}\Re[\eta_2-\eta_3]-\Omega$, $\delta\omega=-\frac{1}{2}\Re[\eta_2+\eta_3]$, $\delta\Gamma=\Im[-2\eta_1+\eta_2+\eta_3]-\Gamma$, and $\delta\kappa=2\Im[\eta_1]-\kappa$. This  method to calculate the alteration of resonances, does not regard the strength of the optomechanical interaction or any of the damping rates. In contrast, the known methods to analyze this phenomenon normally require $g<<\kappa$ and $\Gamma+\delta\Gamma<<\kappa$ \cite{Aspel1}.
	
 	\subsection*{Corrections to Spring Effect}\label{AppA}
	
	As shown in \S{S7} of supplementary information, the full expression for corrected spring effect is given as
	Put together combined, we get
	\begin{eqnarray}
	\label{eqA10Copy}
	\delta\Omega(w,\Delta)&=& \frac{g_0^2\bar{n}\Omega}{w}\left[\frac{\Delta+w}{(\Delta+w)^2+\frac{1}{4}\kappa^2}+\frac{\Delta-w}{(\Delta-w)^2+\frac{1}{4}\kappa^2}\right]\\ \nonumber
	&+&\frac{g_0^2\Re[\mu(w)]\Omega}{w}\left[\frac{\Delta+w}{(\Delta+w)^2+\frac{1}{4}\kappa^2}+\frac{\Delta-w}{(\Delta-w)^2+\frac{1}{4}\kappa^2}\right]+\frac{g_0^2\Im[\mu(w)]\Omega}{w}\left[\frac{\kappa}{(\Delta+w)^2+\frac{1}{4}\kappa^2}-\frac{\kappa}{(\Delta-w)^2+\frac{1}{4}\kappa^2}\right],\\ 
	\delta\Gamma(w,\Delta)&=&\frac{g_0^2\bar{n}\Omega}{w}\left[\frac{\kappa}{(\Delta+w)^2+\frac{1}{4}\kappa^2}-\frac{\kappa}{(\Delta-w)^2+\frac{1}{4}\kappa^2}\right]\\ \nonumber
	&+&\frac{g_0^2\Re[\mu(w)]\Omega}{w}\left[\frac{\kappa}{(\Delta+w)^2+\frac{1}{4}\kappa^2}-\frac{\kappa}{(\Delta-w)^2+\frac{1}{4}\kappa^2}\right]-\frac{g_0^2\Im[\mu(w)]\Omega}{w}\left[\frac{\Delta+w}{(\Delta+w)^2+\frac{1}{4}\kappa^2}+\frac{\Delta-w}{(\Delta-w)^2+\frac{1}{4}\kappa^2}\right].
	\end{eqnarray}
	
	Here, the second and third terms on the rights hand sides of both equations are corrections to the spring effect due to the higher-order interactions, resulting from the temperature-dependent expressions 
	\begin{eqnarray}
	\label{eqA11Copy}
	\Re[\mu(w)]&=&\frac{w}{\Omega}\left(\bar{m}+\frac{1}{2}\right)+\frac{1}{2}, \\ \nonumber
	\Im[\mu(w)]&=&\frac{\Gamma}{2\Omega}\left(\bar{m}+\frac{1}{2}\right).
	\end{eqnarray}
	The temperature-dependence of (\ref{eqA11Copy}) causes dependence of the spring effect on temperature as well. The influence of additional terms in (\ref{eqA10Copy}) due to higher-order interactions can strongly influence any measurement of $g_0$ through spring effect, as most easily can be observable in the weak coupling limit for Doppler cavities. 

	\subsection*{Weak Coupling Limit}
	
	In the weakly coupled operation mode and far Doppler regime where $g_0<<\Omega$ and $\kappa>>\Omega>>\Gamma$ hold \cite{Transduction,Verhagen}, using (S9) of supplementary information with $\bar{n}\approx(\Delta^2+\frac{1}{4}\kappa^2)^{-1}|\alpha|^2$, the spring equations are obtained from (\ref{eqA10Copy}) by setting $w=\Omega$ as
	\begin{equation}
	\label{eqA14}
	\delta\Omega(\Omega,\Delta)\approx 2\Delta g_0^2\frac{\bar{n}(\Delta)+\bar{m}(\Delta)+1}{\Delta^2+\frac{1}{4}\kappa^2}
	\approx g_0^2\left[\frac{2\Delta|\alpha|^2}{\left(\Delta^2+\frac{1}{4}\kappa^2\right)^2}\right]+ g_0^4\left[\frac{2\Delta\zeta(\Delta)|\alpha|^4}{\left(\Delta^2+\frac{1}{4}\kappa^2\right)^3}\right].
	\end{equation}
	\noindent
	Here, $|\alpha|$ is photon input rate to the cavity with $\alpha$ being complex drive amplitude, and $\Re[\mu(\Omega)]=\bar{m}+1$ and $\Im[\mu(\Omega)]\approx 0$ from (\ref{eqA11Copy}). The importance of this equation is that the optical spring effect is actually proportional to $\delta\Omega\propto g_0^2(\bar{n}+\bar{m})\propto g_0^2\bar{n}(1+g_0^2\zeta\bar{n})$ where $\zeta(\Delta)$ is already defined in (S30). This shows that if $g_0$ is to be determined from experimental measurement of the optical spring effect, then the experiment should be done at the lowest optical power possible, otherwise the term $\bar{m}\propto g_0^2\bar{n}^2$ becomes large and would result in an apparent change in $g_0$. This fact also can explain why the measured $g_0$ through optical spring effect using uncorrected standard expressions (S42) is always different from the design value, which could be attributed to the absence of the second term proportional to $g_0^4$ in the corrected optical spring effect using the higher-order algebra.
	
	The above equation together with the fact that on the far red detuning $\Delta\rightarrow+\infty$ we have $\bar{n}(\Delta)\rightarrow 0$, $\bar{m}(\Delta)\rightarrow 0$, and $\delta\Omega(\Omega,\Delta)\rightarrow 0$, provides an alternate approximation for the resonant coherent phonon number $\bar{m}(0)$ at zero-detuning as
	\begin{eqnarray}
	\label{eqA15}
	\bar{m}(0)&\approx&\frac{\kappa^2}{8g_0^2}\left[\frac{\partial\delta\Omega(\Omega,\Delta)}{\partial\Delta}\right]_{\Delta=0}-4\frac{|\alpha|^2}{\kappa^2}-1 \approx \frac{32 g_0^2 Q^2_\text{m}}{\Gamma^2}\bar{n}^2(0) \approx \frac{512 g_0^2 Q^2_\text{m}}{\Gamma^2\kappa^4}|\alpha|^4,
	\end{eqnarray}
	\noindent
	where $Q_\text{m}=\Omega/\Gamma$ is the mechanical quality factor, and the expression within the brackets can be measured experimentally, and represents the slope of frequency displacement due to the spring effect  versus detuning. The second expression proportional to $\bar{n}^2(0)$ follows (S30) from \S{S6} of supplementary information where an explicit and accurate formula for $\bar{m}(\Delta)$ is found. 
	
	Noting $|\alpha|^2\propto P_\text{op}$ reveals that while the intracavity photon number is propotional to the optical power as $\bar{n}(0)\propto P_\text{op}$, the coherent phonon population is proportional to the square of the optical power as $\bar{m}(0)\propto P_\text{op}^2$. This implies that the effects of coherent mechanical field gets important only at sufficiently high optical powers, and also marks the fact that in the low optical power limit where linear optomechanics is expected to work well, effects of coherent phonons do not appear. This also explains why this quantity has not been so far noticed in the context of quantum optomechanics. Because it does not show up anywhere in the corresponding fully linearized Langevin equations. 
	
	\section*{Discussion}
	
	
	As shown in \S{S10} of supplementary information, a fairly convenient but approximate solution to the symmetrized spectral density of output optical field due to multiplicative noise is given as
	\begin{equation}
	\label{Noise18Copy}
	S(\omega)=|Y_{11}(\omega)|^2 S_{AA}(\omega)+\frac{1}{\gamma^2}\left|\left[Y_{12}(\omega)+Y_{13}(\omega)\right]\ast\bar{a}(\omega)\right|^2 S_{BB}(\omega) +\frac{1}{\theta^2}\left|\left[Y_{14}(\omega)\ast\overline{ab}(\omega)+Y_{15}(\omega)\ast\overline{ab^\ast}(\omega)\right]\right|^2 S_{BB}(\omega),
	\end{equation}
	where spectral power densities $S_{AA}$ and $S_{BB}$ are already introduced in (S11) of supplementary information and convolutions $\ast$ take place over the entire frequency axis. In practice it is far easier to use numerical integration, however, this can cause numerical instabilities when $|\omega-\Delta|>\frac{1}{2}\Omega$. Owing to the fractional polynomial expressions for the elements of scattering matrix elements as well as the multiplicative terms, it is possible to evaluate the integrals exactly using complex residue techniques. Here, we proceed using numerical integration of the convolution integrals. The third term involving the functions $\overline{ab}(\omega)$ and $\overline{ab^\ast}(\omega)$ are unnecessary for the $3\times 3$ second-order formalism, and arise only in the $5\times 5$ third-order formalism.
	
	In the above equation, every term adds up the contribution from linear, second-order, and third-order optomechanics. These respectively are due to the processes of photon creation-annihilation $\{\hat{a},\hat{a}^\dagger\}$, the 1-photon/1-phonon blue $\{\hat{a}\hat{b},\hat{a}^\dagger\hat{b}^\dagger\}$ and red $\{\hat{a}\hat{b}^\dagger,\hat{a}^\dagger\hat{b}\}$ processes, and the 1-photon/2-phonon second-order blue-like $\{\hat{a}\hat{b}^2,\hat{a}^\dagger\hat{b}^{\dagger 2}\}$ and red-like $\{\hat{a}\hat{b}^{\dagger 2},\hat{a}^\dagger\hat{b}^2\}$ sideband processes. Apparently the Hermitian conjugate operators do not exist in the original $5\times 5$ higher-order formalism (S70) of supplementary information, since they are completely uncoupled from their Hermitian counterparts. However, calculation of the noise spectral densities necessitates their presence, so that a real-valued and positive definite spectral density has actually already taken care of these conjugate processes. Obviously, the first term contributes to the $w=\Delta$ resonance, while the second term contributes to the first-order mechanical side-bands at $w=\Delta=\pm\Omega$. Similarly, the third term constitutes the second-order mechanical sidebands at $w=\Delta=\pm2\Omega$.
	
	It has to be mentioned that the spectral density (\ref{Noise18Copy}) is not mathematically exact, since the multiplicative operators appearing behind Weiner noise terms, are approximated by their time-averaged frequency-dependent terms (S68) of supplementary information.
	
\begin{figure}[ht!]
		\centering
		\includegraphics[width=6.0in]{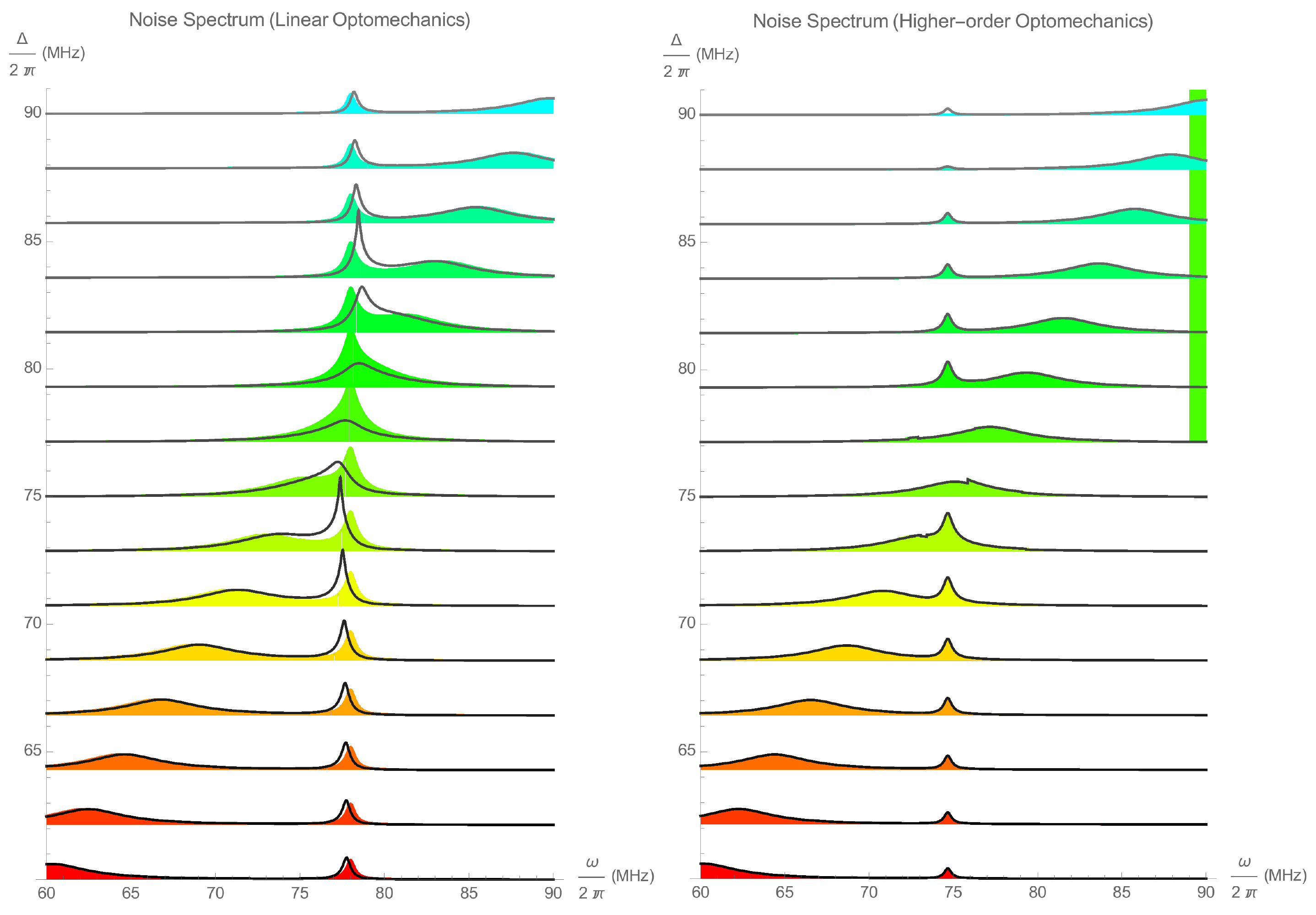}
		\caption{Noise spectrum across the red mechanical sideband and optical resonance for various detuning values ranging from $\Delta=2\pi\times 60\text{MHz}$ to $\Delta=2\pi\times 90\text{MHz}$. Left panel corresponds to the simulations using $3\times 3$ (color fills) and $4\times 4$ linearized optomechanics (black lines). Right panel corresponds to the simulations using $3\times 3$ (color fills) and $5\times 5$ higher-order optomechanics (black lines). Optomechanical system parameters were taken from a remarkable experimental article \cite{Polaron}. \label{FigPolaron}}
	\end{figure}

	As an application example, we simulate the noise spectrum across the red mechanical sideband and optical resonance generated in an optomechanical experiment on the whispering galley mode of an optical micro-toroid, reported in a very remarkable experiment \cite{Polaron}. The pump is set around the red mechanical side band for various detuning values ranging from $\Delta=2\pi\times 60\text{MHz}$ to $\Delta=2\pi\times 90\text{MHz}$, and noise spectra are observed. In Fig. \ref{FigPolaron}, the simulation results using linearized and higher-order optomechanics are illustrated. Here, the left panel shows the simulations using $3\times 3$ linear optomechanics (color fills) with the basis $\{\hat{a},\hat{b},\hat{b}^\dagger\} $ and $4\times 4$ linearized optomechanics (black lines) using the basis $\{\hat{a},\hat{a}^\dagger,\hat{b},\hat{b}^\dagger\}$. While the linear $4\times 4$ formalism is expected to be more accurate than the linear $3\times 3$ formalism, there exists a notable difference between the two approaches. Here, the optomechanical parameters were taken from the same article with some adjustment to resemble the actual experiment \cite{Polaron} as  $T=65\text{mK}$, $P_\text{op}=1.4\text{mW}$, $\Omega=2\pi\times 78\text{MHz}$, $g_0=2\pi\times 3.4\text{kHz}$, $\Gamma=2\pi\times 407\text{kHz}$, $\kappa=2\pi\times 3.54\text{MHz}$ $\eta=0.5$, and $\lambda=775\text{nm}$. 
	On the right panel of Fig. \ref{FigPolaron} the simulations using $3\times 3$ higher-order optomechanics (color fills) with the basis $\{\hat{a},\hat{a}\hat{b},\hat{a}\hat{b}^\dagger\}$ and $5\times 5$ higher-order optomechanics (black lines) with the basis $\{\hat{a},\hat{a}\hat{b},\hat{a}\hat{b}^\dagger,\hat{a}\hat{b}^2,\hat{a}\hat{b}^{\dagger 2}\}$ are shown. Agreement between the second-order $3\times 3$ and third-order $5\times 5$ formalisms is remarkably good. 
	
	It is possible to employ the method of higher-order operators to investigate the dynamic stability of optomechanical systems in the side-band resolved operation limit. A stable optomechanical system can be still perturbed by thermal effects and they appear to be dominant in driving the cavity into instability for Doppler samples. However, for side-band resolved samples, thermal effects are much less pronounced and the major contribution to the instability comes from inherent nonlinear dynamics of the optomechanical interactions. That implies that optomechanical interactions are linearly stable, but they can become nonlinearly unstable at a certain interaction order to be discussed below.
	
	The dynamic stability can be done by inspecting eigenvalues of the coefficients matrix $[\textbf{M}]$. If the real part of at least one of the eigenvalues is positive, then the system is unstable and its response to any perturbation grows indefinitely in time. The linear formalisms of optomechanics fails to describe this phenomenon, since they always yield constant eigenvalues. Even the second-order higher-operator method with $3\times 3$ formalism, which describes the nonlinear 1-photon/1-phonon processes, fails to reproduce the correct expected stability phases. This only can be understood by employing at least the third-order $5\times 5$ operator method, which includes the nonlinear 1-photon/2-phonon processes. Hence, surprisingly enough, it is the 1-photon/2-phonon process and beyond, which contributes to the unstability of an optomechanical system.
	
	To illustrate this, we calculate the stability phases of the side-band resolved system investigated in \S{S10} of supplementary information. Illustrated in Fig. \ref{FigStability}, the stable and unstable regions of this systems across blue $\Delta<0$ and red $\Delta>0$ detunings versus input optical power $P_\text{op}$ are illustrated. The v-shaped region in violet color, maps the unstable phase, while the red and blue colors correspond to the stable operation phases. 
	
	We also have calculated the linear stability from $4\times 4$ full linear formalism, which appears on the right of Fig. \ref{FigStability}. Not surprisingly, the linear and nonlinear stability diagrams remarkably are different. The linear instability starts at moderate resonant pump powers, while it starts rapidly growing exactly over the blue detuning at a slightly higher power. 
	
	Firstly, it can be seen that across almost the entire domains of linear stability, the system is also nonlinearly stable. Secondly, by observation of the nonlinear stability in the left and linear stability on the right, it can be seen that in most of the domain of linear stability, the system is already nonlinearly stable. Hence, any attempt to drive the system within the region of linearly unstable but nonlinearly stable, ultimately results in significant growth of mechanical amplitude and therefore side-bands. Any further increase in the amplitude of side-bands become limited due to nonlinear stability. Hence, four possible stability scenarios could be expected:
	\begin{itemize}[noitemsep,nolistsep]
		\item  Linearly and Nonlinearly Stable: The intersect of the linear and nonlinear stability domains, marks a shared domain of unconditionally stable optomechanical interaction. Unless the system is influenced by thermal or other nonideal effects, the stability is always guaranteed.
		\item  Linearly Unstable, but Nonlinearly Stable: By inspection, an optomechanical system can be linearly unstable while nonlinearly stable. This corresponds to the domains where any attempt to drive the system in these regions causes immediate but limited growth in the amplitude of mechanical oscillations.
		\item Linearly and Nonlinearly Unstable: Under this scenario, the optomechanical cavity is always unstable regardless of the other nonideal effects. This happens only at remarkably high drive powers.
		\item Linear Stable, but Nonlinearly Unstable: The unlikely and surprising case of linear stability and nonlinear unstability is also possible according to the stability maps at some portions of non-resonant high drive powers. This strange behavior corresponds to the case when the system remains stable only at infinitesimal optical powers. Any fluctuation beyond tiny amplitudes shall drive the system into unstable growing and large amplitudes.
	\end{itemize}
	
	There is a threshold power $P_\text{th}$ at which instabilites start to appear. For the side-band resolved case, this happens at $P_\text{th}=5\text{mW}$ on the blue side. As it expected and in agreement to experimental observations, the unstable domain mostly covers the blue domain with $\Delta<0$. However, at higher optical powers than $P_\text{th}$, instability phase can diffuse well into the red detunings as well $\Delta>0$. For the cavity under consideration, this happens at much higher optical power of $P=14P_\text{th}=70\text{mW}$. Therefore, the general impression that instability always occurs on the blue side at every detuning above a certain threshold power is not correct. Interestingly, the boundary separating the dynamically stable and dynamically unstable phases for this side-band resolved sample with $\Omega>>\kappa>>\Gamma$ can be well estimated using
	\begin{equation}
	\label{Boundary}
	\bar{n}(\Delta)>n_\text{cr},
	\end{equation}
	\noindent
	in which $n_\text{cr}$ is a critical cavity photon number. Extensive numerical tests for cavities within deep side-band resolved $\Omega>>\kappa$, deep Doppler $\Omega>>\kappa$, and intermediate $\Omega\sim\kappa$ regimes reveal the existence of such a critical intracavity photon number limit $n_\text{cr}$, beyond which dynamical instability takes over. However, for a cavity in deep side-band resolved regime, it can be estimated in a phenomenological way, and is roughly given by
	\begin{equation}
	\label{CriticalDensity}
	n_\text{cr}\sim \frac{4}{\mathcal{C}_0}\left(\frac{\Omega}{\kappa}\right)^2. 
	\end{equation}
	\begin{figure}[ht!]
		\centering
		\includegraphics[width=6.0in]{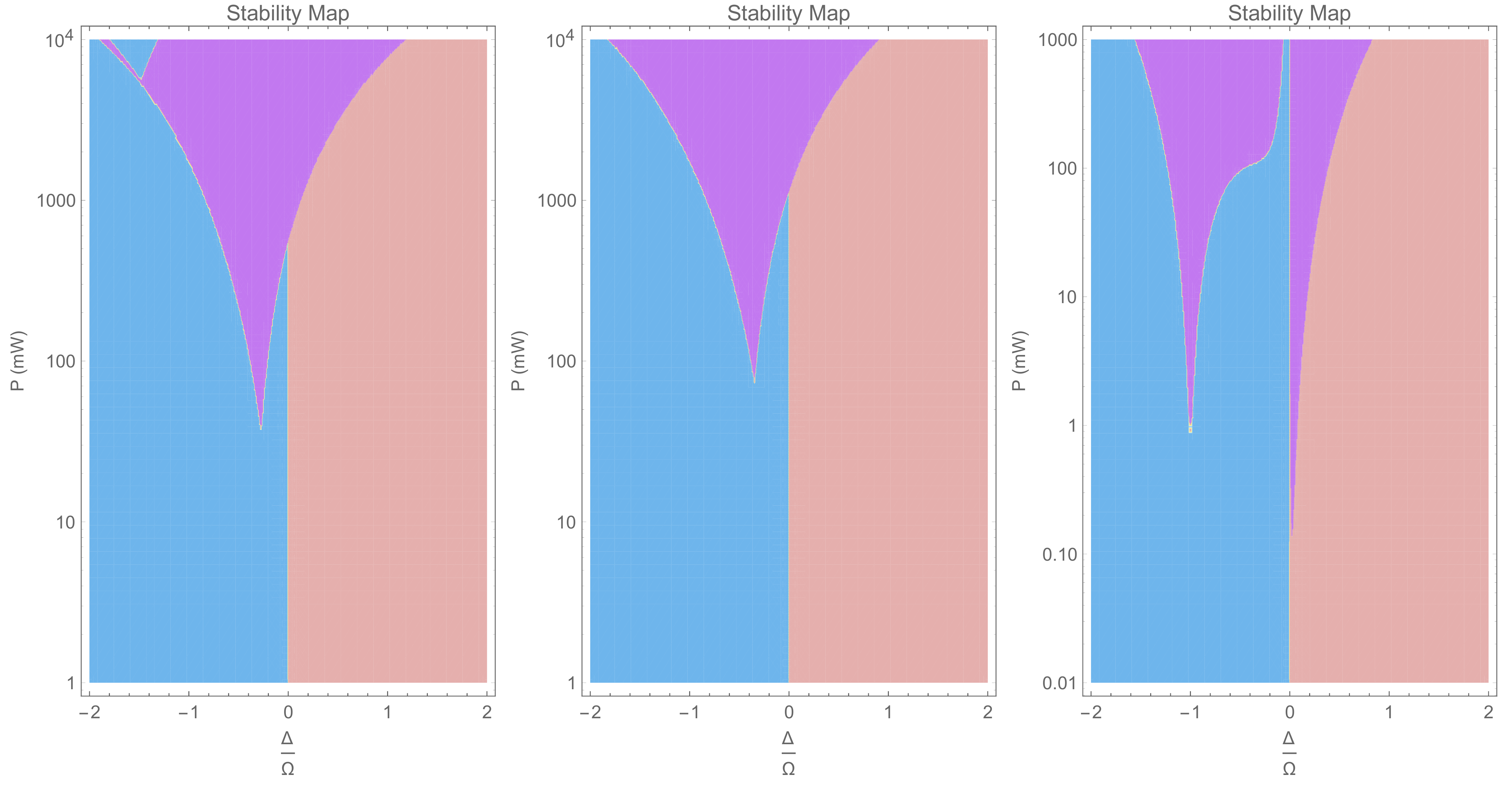}
		\caption{Dynamic stability phase map of a side-band resolved optomechanical systems studied in \S{S10} of supplementary information. Red and blue domains correspond to stable red $\Delta>0$ and blue $\Delta<0$ detunings. The violet phase is the unstable region: Full numerical simulation of nonlinear dynamics (left); Boundary marked by (\ref{Boundary}) (middle); Linear stability map for the same system (Right). \label{FigStability}}
	\end{figure}
	
	\noindent
	Here, $\mathcal{C}_0$ is the single-photon cooperativity given by
	\begin{equation}
	\label{C0}
	\mathcal{C}_0=g_0^2/\kappa\Gamma.
	\end{equation}
	If the cavity is not side-band resolved, (\ref{CriticalDensity}) cannot be used, but numerical computations can still yield the limiting number $n_\text{cr}$. 
	
	For Doppler cavities, no such dynamic instability can be observed, and therefore thermal effects should dominate over dynamical effects in driving a Doppler cavity toward instability. Meanwhile, it is the nonlinear optomechanical dynamics which seems to be dominant in driving a side-band resolved cavity into instability. As a result, the existence of such a critical maximum intracavity photon number is not related to thermal effects, but rather to the nonlinear stability. 
	
	\section*{Methods}
	
	For an extensive description of theoretical methods, refer to the Supplementary Information.
	
	\section*{Conclusions}
	
	A new analytical method was shown to solve the standard optomechanical interaction with cubic nonlinearity interaction, based on the higher-order operators. It was demonstrated that not only the higher-order operator method can reproduce the linear optomechanics, but also it can predict and provide estimates to unnoticed effects such as a new type of symmetry breaking in frequency, here referred to as side-band inequivalence, and yield new explicit expressions for quantities such as the coherent phonon population and higher-order spring effect. Corrections to the standard spring effect due to higher-order interactions have been found, and it has been shown that such corrections arise mainly because of the coherent phonons and can significantly influence measurement of single-photon optomechanical interaction rate through spring effect. A minimal basis has been defined which allows exact and explicit solution to standard nonlinear optomechanics, using the method of higher-order operators. This method can be finally used to investigate the dynamic nonlinear stability of optomechanical systems, and it has been demonstrated that at least the third-order nonlinear processes are prerequisite for occurrence of dynamic instability. We have shown that there is a reasonable correspondence between the onset of nonlinear dynamic instability and a critical intracavity photon number limit, which remains independent of thermal effects. 
	
	\section*{Additional Information}
	
	The author declares no competing financial and/or non-financial interests in this work.
	
	\section*{Acknowledgement}
	
	Discussions with Georg Arnold at IST Austria and group members of the laboratory of Quantum Foundations and Quantum Information on the Nano- and Microscale are appreciated. This paper is dedicated to the celebrated artist, Anastasia Huppmann.

	\newpage
\rfoot{\thepage/S21}
\section*{\centering Supplementary Information: Theoretical Methods}
\begin{center}
	\textbf{Method of Higher-order Operators for Quantum Optomechanics}
	
	Sina Khorasani
\end{center}

\setcounter{equation}{0}
\setcounter{page}{1}
\setcounter{table}{0}
\setcounter{section}{0}
\setcounter{figure}{0}
\makeatletter
\renewcommand{\theequation}{S\arabic{equation}}
\renewcommand{\thefigure}{S\arabic{figure}} 
\renewcommand{\thepage}{S\arabic{page}}
\renewcommand{\thetable}{S\arabic{table}}
\renewcommand{\thesection}{S\arabic{section}}
\renewcommand{\bibnumfmt}[1]{[S#1]}
\renewcommand{\citenumfont}[1]{S#1}

\section{Optomechanical Hamiltonian}

The Langevin equations for the Hamiltonian $\mathbb{H}_{\text{OM}}=\hbar \Omega\hat{m}-\hbar \Delta\hat{n}-\hbar g_0\hat{n}(\hat{b}+{\hat{b}}^{\dagger }) $ with the basis $\{A\}^\text{T}=\left\{\hat{a},\hat{b},\hat{a}\hat{b},\hat{a}{\hat{b}}^{\dagger },\hat{n},\hat{c}\right\}$ are given exactly by 
\begin{eqnarray}
\nonumber
\left[ \begin{array}{c}
\begin{array}{cccccc}
i\Delta-\frac{\kappa }{2} & 0 & ig_0 & ig_0 & 0 & 0 \\ 
0 & -(i\Omega+\frac{\Gamma}{2}) & 0 & 0 & ig_0 & 0 \\ 
ig_0\left(\hat{m}+\hat{n}+1\right) & 0 & -i\left(\Omega-\Delta-g_0 \hat{b}\right)-\frac{\gamma }{2} & 0 & 0 & 0 \\ 
ig_0\left(\hat{m}-\hat{n}\right) & 0 & 0 & i\left(\Omega+\Delta+g_0\hat{b}^\dagger\right)-\frac{\gamma }{2} & 0 & 0 \\ 
0 & 0 & 0 & 0 & -\kappa  & 0 \\ 
0 & 0 & ig_0\hat{a} & ig_0\hat{a} & 0 & 2i[\Delta+g_0(\hat{b}+\hat{b}^\dagger)]-\kappa  \end{array}
\end{array}
\right]\\ 
\label{eq4}
\times
\left\{ \begin{array}{c}
\hat{a} \\ 
\hat{b} \\ 
\hat{a}\hat{b} \\ 
\hat{a}{\hat{b}}^{\dagger } \\ 
\hat{n} \\ 
\hat{c}
\end{array}
\right\}
-\left\{ \begin{array}{c}
\sqrt{\kappa }{\hat{a}}_{\text{in}} \\ 
\sqrt{\Gamma}{\hat{b}}_{\text{in}} \\ 
\sqrt{\gamma }{\left(\hat{a}\hat{b}\right)}_{\text{in}} \\ 
\sqrt{\gamma }{\left(\hat{a}{\hat{b}}^{\dagger }\right)}_{\text{in}} \\ 
\sqrt{2\kappa }{\hat{n}}_{\text{in}} \\ 
\sqrt{2\kappa }{\hat{c}}_{\text{in}} 
\end{array}
\right\}=\frac{d}{dt}\left\{ \begin{array}{c}
\hat{a} \\ 
\hat{b} \\ 
\hat{a}\hat{b} \\ 
\hat{a}{\hat{b}}^{\dagger } \\ 
\hat{n} \\ 
\hat{c} 
\end{array}
\right\},
\end{eqnarray}
\noindent 
\noindent 
in which $\gamma =\kappa +\Gamma$. We have set $\hat{x}=\hat{a}$ in all equations except the second and third where both of the bath operators $\hat{x}=\hat{a}$ and $\hat{x}=\hat{b}$ are taken separately to construct the noise terms, $\sqrt{2}{\hat{n}}_\text{in}={\hat{a}}^{\dagger }{\hat{a}}_\text{in}+{{\hat{a}}^{\dagger }}_\text{in}\hat{a}$, $(\hat{a}\hat{b}^\dagger)_\text{in}={\hat{a}}_\text{in}\hat{b}^\dagger+\hat{a}\hat{b}_\text{in}^\dagger$, $(\hat{a}\hat{b})_\text{in}={\hat{a}}_\text{in}\hat{b}+\hat{a}{\hat{b}}_\text{in}$ and $\sqrt{2}{\hat{c}}_\text{in}=\hat{a}{\hat{a}}_\text{in}$. 

The system (\ref{eq4}) is still nonlinear and non-integrable because of the dependence of the coefficients matrix on the operators. But it can be simplified by first noting that from the fifth equation we could expect any disturbance in $\hat{n}$ would decay as $\delta\hat{n}(t)\sim\exp(-\kappa t)$ on time scales smaller than $\kappa^{-1}$. This can be further approximated as $\hat{n}\sim\bar{n}$ at steady input. Similar argument goes with $\delta\hat{m}\sim\exp(-\Gamma t)$ in response to a disturbance on time scales smaller than $\Gamma^{-1}$, which enables us to make the approximate replacement  $\hat{m}\sim\bar{m}$ at equilibrium. 

For the phononic mechanical operators $\hat{b}$ and $\hat{b}^\dagger$ appearing within the brackets, approximate decays $\delta\hat{b}(t)\sim\exp[-(i\Omega+\frac{1}{2}\Gamma) t]$ and $\delta\hat{b}^\dagger(t)\sim\exp[(i\Omega-\frac{1}{2}\Gamma) t]$ in response to disturbances hold, making the coefficients matrix time-dependent. But these can be nevertheless dropped in whole if we notice that $g_0\bar{b}<<\Omega$ which is the normal experimental condition of weakly-coupling in optomechanics. Otherwise, they can replaced by constant amplitudes $\bar{b}$ and $\bar{b}^\ast$ given below in (\ref{eq9}) on sufficiently longer time scales than $\Gamma^{-1}$ for strongly-coupled systems. 

Such types of approximations are in fact quite highly in use within the context of continuous wave standard optomechanics. Therefore, once the steady state solution to (\ref{eq4}) around the equilibrium values due to optical drive $\langle\hat{a}_\text{in}\rangle=\alpha$ is sought, the coefficients matrix can be kept time-independent, keeping only the fluctuations of input terms as the only source. The case of time dependent drive $\alpha=\alpha(t)$ for pulsed experiments shall be discussed later in the article.

Having said that, all the operators $\hat{n}$, $\hat{m}$, $\hat{b}$, and $\hat{b}^\dagger$ in the coefficients matrix can be replaced by their respective average values to proceed with the second-order accurate optomechanical system of equations as
\begin{eqnarray}
\nonumber
\left[ \begin{array}{c}
\begin{array}{cccccc}
i\Delta-\frac{\kappa }{2} & 0 & ig_0 & ig_0 & 0 & 0 \\ 
0 & -(i\Omega+\frac{\Gamma}{2}) & 0 & 0 & ig_0 & 0 \\ 
iL^+ & 0 & -i\left(\Omega-\Delta-s\right)-\frac{\gamma }{2} & 0 & 0 & 0 \\ 
iL^- & 0 & 0 & i\left(\Omega+\Delta+s^\ast\right)-\frac{\gamma }{2} & 0 & 0 \\ 
0 & 0 & 0 & 0 & -\kappa  & 0 \\ 
0 & 0 & ig & ig & 0 & 2i(\Delta+2\Re[s])-\kappa  \end{array}
\end{array}
\right]\\ 
\label{eq5}
\times
\left\{ \begin{array}{c}
\hat{a} \\ 
\hat{b} \\ 
\hat{a}\hat{b} \\ 
\hat{a}{\hat{b}}^{\dagger } \\ 
\hat{n} \\ 
\hat{c}
\end{array}
\right\}
-\left\{ \begin{array}{c}
\sqrt{\kappa }{\hat{a}}_{\text{in}} \\ 
\sqrt{\Gamma}{\hat{b}}_{\text{in}} \\ 
\sqrt{\gamma }{\left(\hat{a}\hat{b}\right)}_{\text{in}} \\ 
\sqrt{\gamma }{\left(\hat{a}{\hat{b}}^{\dagger }\right)}_{\text{in}} \\ 
\sqrt{2\kappa }{\hat{n}}_{\text{in}} \\ 
\sqrt{2\kappa }{\hat{c}}_{\text{in}} 
\end{array}
\right\}=\frac{d}{dt}\left\{ \begin{array}{c}
\hat{a} \\ 
\hat{b} \\ 
\hat{a}\hat{b} \\ 
\hat{a}{\hat{b}}^{\dagger } \\ 
\hat{n} \\ 
\hat{c} 
\end{array}
\right\},
\end{eqnarray}
\noindent 
in which $g=g_0\sqrt{\bar{n}}$, $s=g_0\bar{b}$ with $\Re[\bar{b}]=\bar{x}/2x_{\rm zp}$ and $x_{\rm zp}$ being the zero-point displacement,  $L^+=g_0(\bar{m}+\bar{n}+1)$, and $L^-=g_0(\bar{m}-\bar{n})$. These can be further approximated by $L^\pm \approx \pm g_0\bar{n}=\pm F$ under normal experimental conditions of an ultracold cavity with sufficiently high pumping. The average mirror displacement $\bar{x}$ is due to the average radiation pressure. The fact that $L^+\neq-L^-$ provides the quantum mechanical asymmetry between blue and red sidebands. 

It is easy to verify that this way of linearization decouples the state operators and reduces the space into a 3-dimensional one spanned by $\{A\}^{\rm T}=\{\hat{a},\hat{a}\hat{b},\hat{a}\hat{b}^\dagger\}$. This will be discussed in further details later.

In the absence of red-side-band optical cooling tone as well as any other interaction, the average population value is $\bar{m}={1}/{\left[\exp\left({\hbar \Omega}/{k_\text{B}T}\right)-1\right]}$, while $\bar{n}$ can be obtained from the steady state solution of the first row by replacements of input noise term $\sqrt{\kappa}{\hat{a}}_\text{in}\to \alpha+\sqrt{\kappa}{\hat{a}}_\text{in}$. Here, $\alpha $ is the input photon flux originally due to an undisplayed resonant drive term $\mathbb{H}_\text{d}=\hbar(\alpha\hat{a}+\alpha^\ast\hat{a}^\dagger)$ added to the Hamiltonian $\mathbb{H}_\text{OM}$. Furthermore, $\alpha$ has some non-zero phase taken from the cavity population $\bar{n}$ away. Now that the drive term $\mathbb{H}_\text{d}$ has been dropped from $\mathbb{H}_\text{OM}$, and $\hat{a}_\text{in}$ now only contains the fluctuations with zero-average $\langle\hat{a}_\text{in}\rangle=0$. 

As it will be shown later, the quantity $\bar{m}$ here being referred to as the coherent phonon population, can enter the optomechanical interaction processes due to higher-order effects, where its value normally needs to be fitted for cavities in the Doppler regime. Hence, the coherent phonon population $\bar{m}$ is independent of the simple thermal equilibrium value $m$, and actually represents those number of phonons who take part in the optomechanical interaction. We will observe that the negative detunings with the blue process can actually lead to a relatively constant phonon population, whereas on the red detunings it starts to decrease with the detuning.

Defining  $K=\bar{n}\kappa $ we may use the substitutions for the noise and input terms as
\begin{eqnarray}
\label{eq6}
\sqrt{\gamma }{\left(\hat{a}\hat{b}\right)}_\text{in}&\to& \sqrt{\Gamma\bar{n}}{\hat{b}}_\text{in}+\sqrt{\kappa }\bar{b}{\hat{a}}_\text{in}+\bar{b}\alpha,\\ \nonumber
{\sqrt{\gamma }\left(\hat{a}{\hat{b}}^{\dagger }\right)}_\text{in}&\to& \sqrt{\Gamma\bar{n}}{\hat{b}}_\text{in}^\dagger+\sqrt{\kappa }\bar{b}^\ast{\hat{a}}_\text{in}+\bar{b}^\ast\alpha,\\ \nonumber
\sqrt{\kappa}{\hat{n}}_\text{in}&\to& \sqrt{K}{\hat{a}}_\text{in}+\sqrt{K}{\hat{a}}_\text{in}^\dagger+2\sqrt{\bar{n}}\Re[\alpha],\\ \nonumber
\sqrt{\kappa}{\hat{c}}_{\rm in}&\to&\sqrt{K}{\hat{a}}_\text{in}+\sqrt{\bar{n}}\alpha.
\end{eqnarray}
\noindent
These substitutions follow the fact that terms such as $\hat{a}\hat{a}_{\rm in}$ which contain the interaction of a time-dependent operator $\hat{a}(t)$ and a purely white Weiner noise process with zero average $\langle\hat{a}_{\rm in}\rangle=0$, can be fairly well approximated by noting first that $\hat{a}(t)\sim\bar{a}\exp(i\Delta t)$ around the equilibrium, and then noting that shifting the noise process $\hat{a}_{\rm in}$ in frequency to the amount of $\Delta$ has essentially no effect by definition. Hence, the sinusoidal time dependence $\exp(i\Delta t)$ is irrelevant and can be dropped. Similar arguments hold for the phononic operator $\hat{b}(t)\sim\bar{b}\exp(-i\Omega t)$ and their Hermitian adjoints interacting with a white noise term with uniform spectrum.

This allows us to ultimately rewrite the Langevin equations (\ref{eq5}) as 
\begin{eqnarray}
\nonumber
\left[ \begin{array}{c}
\begin{array}{cccccc}
i\Delta-\frac{\kappa }{2} & 0 & ig_0 & ig_0 & 0 & 0 \\ 
0 & -(i\Omega+\frac{\Gamma}{2}) & 0 & 0 & ig_0 & 0 \\ 
iL^+ & 0 & -i\left(\Omega-\Delta-s\right)-\frac{\gamma }{2} & 0 & 0 & 0 \\ 
iL^- & 0 & 0 & i\left(\Omega+\Delta+s^\ast\right)-\frac{\gamma }{2} & 0 & 0 \\ 
0 & 0 & 0 & 0 & -\kappa  & 0 \\ 
0 & 0 & ig & ig & 0 & 2i(\Delta+2\Re[s])-\kappa  \end{array}
\end{array}
\right]\\ \label{eq7}
\times
\left\{ \begin{array}{c}
\hat{a} \\ 
\hat{b} \\ 
\hat{a}\hat{b} \\ 
\hat{a}{\hat{b}}^{\dagger } \\ 
\hat{n} \\ 
\hat{c}
\end{array}
\right\}
-\left[ 
\begin{array}{cccc}
\sqrt{\kappa} & 0 & 0 & 0\\
0 & 0 & \sqrt{\Gamma} & 0\\
\sqrt{\kappa}\bar{b} & 0 & \sqrt{\Gamma\bar{n}} & 0\\
\sqrt{\kappa}\bar{b}^\ast & 0 & 0 & \sqrt{\Gamma\bar{n}}\\
\sqrt{K} & \sqrt{K} & 0 & 0\\
\sqrt{K} & 0 & 0 & 0\\
\end{array}
\right]
\left\{ \begin{array}{c}
\hat{a}_\text{in} \\ 
\hat{a}_\text{in}^\dagger \\ 
\hat{b}_\text{in} \\ 
\hat{b}_\text{in}^\dagger 
\end{array}
\right\}-
\left[ 
\begin{array}{cc}
1 & 0\\
0 & 0\\
\bar{b} & 0\\
\bar{b}^\ast & 0\\
\sqrt{\bar{n}} & \sqrt{\bar{n}}\\
\sqrt{\bar{n}} & 0\\
\end{array}
\right]
\left\{ 
\begin{array}{c}
\alpha \\ 
\alpha^\ast
\end{array}
\right\}
=\frac{d}{dt}\left\{ \begin{array}{c}
\hat{a} \\ 
\hat{b} \\ 
\hat{a}\hat{b} \\ 
\hat{a}{\hat{b}}^{\dagger } \\ 
\hat{n} \\ 
\hat{c} 
\end{array}
\right\}.
\end{eqnarray}
\noindent
The second term on the right is the noise fluctuations due to the optical and mechanical fields with zero average $\langle\hat{a}_\text{in}\rangle=\langle\hat{b}_\text{in}\rangle=0$, and the last term in the above is proportional to the input photon flux $|\alpha|=\sqrt{\epsilon\kappa P/\hbar\omega}$ where $P$ is the incident radiation power and $\epsilon$ is the coupling efficiency. As it will be mentioned briefly later, the average values $\bar{n}$ and $\bar{x}$ have to be solved by setting $d/dt=0$ on the left and taking average values, which eliminates the noise fluctuations, causing the replacements $\hat{a}\to\sqrt{\bar{n}}$, $\hat{b}\to\bar{b}$, $\hat{a}\hat{b}\to\bar{b}\sqrt{\bar{n}}$, $\hat{a}\hat{b}^\dagger\to\bar{b}^\ast\sqrt{\bar{n}}$, $\hat{n}\to\bar{n}$, and $\hat{c}\to\bar{n}/2$.

Hence, the average values $\bar{a}=\sqrt{\bar{n}}$ and $\bar{b}$ get nonlinearly coupled to the input flux $\alpha$ through the system of algebraic relations as
\begin{eqnarray}
\nonumber
\left[ \begin{array}{c}
\begin{array}{cccccc}
i\Delta-\frac{\kappa }{2} & 0 & ig_0 & ig_0 & 0 & 0 \\ 
0 & -(i\Omega+\frac{\Gamma}{2}) & 0 & 0 & ig_0 & 0 \\ 
iL^+ & 0 & -i\left(\Omega-\Delta-g_0\bar{b}\right)-\frac{\gamma }{2} & 0 & 0 & 0 \\ 
iL^- & 0 & 0 & i\left(\Omega+\Delta+g_0\bar{b}^\ast\right)-\frac{\gamma }{2} & 0 & 0 \\ 
0 & 0 & 0 & 0 & -\kappa  & 0 \\ 
0 & 0 & ig & ig & 0 & i[\Delta+g_0(\bar{b}+\bar{b}^\ast)]-\frac{\kappa}{2}  \end{array}
\end{array}
\right]\\ \label{eq8} 
\times
\left\{ \begin{array}{c}
\bar{a} \\ 
\bar{b} \\ 
\bar{a}\bar{b} \\ 
\bar{a}\bar{b}^\ast \\ 
\bar{a}^2 \\ 
\bar{a}^2
\end{array}
\right\}
=\left\{ 
\begin{array}{cc}
1 & 0\\
0 & 0\\
\bar{b} & 0\\
\bar{b}^\ast & 0\\
\bar{a} & \bar{a}\\
\bar{a} & 0\\
\end{array}
\right\}
\left\{ 
\begin{array}{c}
\alpha \\ 
\alpha^\ast
\end{array}
\right\}.
\end{eqnarray}
\noindent
With a given input photon flux $|\alpha|$, this system can be now solved to obtain the phase $\angle \alpha$ in such a way that $\angle\bar{a}=0$. Then $\bar{a}$ and $\bar{b}$ can be obtained in an algebraic manner. This sets up a system of equations in terms of the total of four unknowns $\angle\alpha$, $\bar{a}=\sqrt{\bar{n}}$, $\bar{b}$, and $\bar{b}^\ast$. 

In the above system, the second equation is independent of $\alpha$, while together the fifth they yield 
\begin{eqnarray}
\label{eq9}
\bar{b}&=&\frac{ig_0}{i\Omega+\frac{1}{2}\Gamma}\bar{a}^2,\\ \nonumber
\bar{a}&=&-\frac{1}{\kappa}(\alpha+\alpha^\ast).
\end{eqnarray}
\noindent
This also already solves $\bar{b}^\ast$ in terms of $\bar{a}$. Plugging in the results into the first equation leads to the third-order algebraic equation which can be now solved. Doing this and some algebraic manipulation gives the equation
\begin{equation}
\label{eq10}
ig_0^2\frac{2\Omega}{\Omega^2+\frac{1}{4}\Gamma^2}\bar{a}^3+\left(i\Delta-\frac{\kappa}{2}\right)\bar{a}=\alpha.
\end{equation}
\noindent
This equation in general is expected to yield only real-valued $\bar{a}$. Separating the real and imaginary parts gives
\begin{eqnarray}
\label{eq11}
\Re[\alpha]&=&-\kappa\frac{\bar{a}}{2}\\
\nonumber
\Im[\alpha]&=&g_0^2\frac{2\Omega}{\Omega^2+\frac{1}{4}\Gamma^2}\bar{a}^3+\Delta\bar{a}.
\end{eqnarray}
The first of these is the same as the second of (\ref{eq9}). The above two equations can be now iteratively solved to yield $\angle\alpha$ and $\bar{a}$ for a given $|\alpha|$. One may also discard $\angle\alpha$ by combining the above two, resulting in 
\begin{equation}
\label{eq12}
|\alpha|^2=\left[\frac{\kappa^2}{4}+\left(\frac{2g_0^2\Omega}{\Omega^2+\frac{1}{4}\Gamma^2}\bar{n}+\Delta\right)^2\right]\bar{n}.
\end{equation}
\noindent
Only real and positive-valued roots of (\ref{eq12}) for $\bar{n}$ are acceptable. Sufficiently large blue-detuning with $\Delta<\Delta_\text{b}<0$ causes the well-known bistability. It is easy to find the negative blue detuning $\Delta_\text{b}<0$ at which bistability starts to appear, by looking for the only negative real root of the cubic equation
\begin{equation}
\label{eqDetuning}
-\Delta_\text{b}\left(\Delta_\text{b}^2+\frac{9}{4}\kappa^2\right)=\frac{27g_0^2\Omega}{\Omega^2+\frac{1}{4}\Gamma^2}|\alpha|^2.
\end{equation}

These two noise terms we assume have the flat shot-noise uncorrelated spectral power densities 
\begin{eqnarray}
\label{eq13}
S_{AA}(\omega)=\frac{1}{2}, \\ \nonumber
S_{BB}(\omega)=m+\frac{1}{2},
\end{eqnarray}
\noindent
which are identical on both positive and negative frequencies. The ultimate difference of noise power spectral densities will be later maintained by the asymmetry caused by $L^+ -L^-=2F+g_0 > 0$. Here, $m=1/\left[\exp(\hbar\Omega/k_\text{B}T)-1\right]$ is the population of incoherent phonons under thermal equilibrium, which contribute to the random fluctuations of thermal noise. This quantity is not to be mistaken with $\bar{m}$ which here denotes the population of coherent phonons, contributing coherently to the optomechanical interaction, and are driven by the optical radiation pressure. This shall be discussed later in \S\ref{BMWm8}.

\subsection{Perturbative Solution}

At this moment, the system of equations (\ref{eq7}) can be perturbed around equilibrium values found above. This procedure and taking a Fourier transform gives out the solution. Let us define first
\begin{equation}
\label{eq14}
\textbf{M}=\left[ \begin{array}{c}
\begin{array}{cccccc}
i\Delta-\frac{\kappa }{2} & 0 & ig_0 & ig_0 & 0 & 0 \\ 
0 & -(i\Omega+\frac{\Gamma}{2}) & 0 & 0 & ig_0 & 0 \\ 
iL^+ & 0 & -i\left(\Omega-\Delta-s\right)-\frac{\gamma }{2} & 0 & 0 & 0 \\ 
iL^- & 0 & 0 & i\left(\Omega+\Delta+s^\ast\right)-\frac{\gamma }{2} & 0 & 0 \\ 
0 & 0 & 0 & 0 & -\kappa  & 0 \\ 
0 & 0 & ig & ig & 0 & 2i(\Delta+2\Re[s])-\kappa  \end{array}
\end{array}
\right],
\end{equation}
\noindent
as well as
\begin{eqnarray}
\label{eq15}
\left\{\delta A(\omega)\right\}^{\text{T}}&=&\left\{ 
\delta\hat{a}(\omega), 
\delta\hat{b}(\omega),  
\delta(\hat{a}\hat{b})(\omega),
\delta(\hat{a}{\hat{b}}^{\dagger })(\omega), 
\delta\hat{n}(\omega), 
\delta\hat{c}(\omega) 
\right\},\\ \nonumber
\left\{A_\text{in}(\omega)\right\}^\text{T}&=&\left\{ 
\hat{a}_\text{in}(\omega) ,
\hat{a}_\text{in}^\dagger(\omega) , 
\hat{b}_\text{in}(\omega) ,
\hat{b}_\text{in}^\dagger(\omega) 
\right\},\\ \nonumber
\left[\sqrt{\Gamma}\right]&=&\left[ 
\begin{array}{cccc}
\sqrt{\kappa} & 0 & 0 & 0\\
0 & 0 & \sqrt{\Gamma} & 0\\
\sqrt{\kappa}\bar{b} & 0 & \sqrt{\Gamma\bar{n}} & 0\\
\sqrt{\kappa}\bar{b}^\ast & 0 & 0 & \sqrt{\Gamma\bar{n}}\\
\sqrt{K} & \sqrt{K} & 0 & 0\\
\sqrt{K} & 0 & 0 & 0\\
\end{array}
\right].
\end{eqnarray}

Then, taking $\textbf{I}_j$ as the $j\times j$ identity matrix, we get
\begin{eqnarray}
\label{eq16}
\left\{A_\text{out}(\omega)\right\}&=&\left\{A_\text{in}(\omega)\right\}-\left[\sqrt{\Gamma}\right]^\text{T}\left\{\delta A(\omega)\right\}=\textbf{Y}(\omega)\left\{A_\text{in}(\omega)\right\}, \\ \nonumber
\left\{\delta A(\omega)\right\}&=&\textbf{Z}(\omega)\left\{A_\text{in}(\omega)\right\}, \\ \nonumber
\textbf{Z}(\omega)&=&\left[\textbf{M}-i\omega\textbf{I}_6\right]^{-1}\left[\sqrt{\Gamma}\right], \\ \nonumber
\textbf{Y}(\omega)&=&\textbf{I}_4-\left[\sqrt{\Gamma}\right]^\text{T}\textbf{Z}(\omega).
\end{eqnarray}
\noindent
Here, $\textbf{Y}(\omega)$ is the scattering matrix connecting the input and output ports. Now, the spectral density of reflected light from the cavity can be found using (\ref{eq13}) by the expression
\begin{equation}
\label{eq17}
S(\omega)=\left[|Y_{11}(\omega)|^2+|Y_{12}(\omega)|^2\right]S_{AA}(\omega)+\left[|Y_{13}(\omega)|^2+|Y_{14}(\omega)|^2\right]S_{BB}(\omega),
\end{equation}
\noindent
as long as the noise processes of $\hat{a}_\text{in}$ and $\hat{b}_\text{in}$ have zero cross-correlation \cite{SBowen}.

\section{Linearized Optomechanics}

It is fairly easy to see that the system of equations (\ref{eq18}) when simplified and rewritten for the basis $\{\hat{a},\hat{b},\hat{b}^\dagger\}$ reproduces the widely used linearized optomechanical equations \cite{SAspel1}. To demonstrate this, we ignore the perturbation matrix $\delta\textbf{N}$, as well as $\bar{b}/\sqrt{\bar{n}}$ in the noise terms, and then employ the substitutions 
\begin{eqnarray}
\label{eq19a}
\hat{a}\hat{b}&\to&\exp\left[(i\Delta-\frac{1}{2}\kappa)t\right]\bar{a}\hat{b}=\exp\left[(i\Delta-\frac{1}{2}\kappa)t\right]\sqrt{\bar{n}}\hat{b},\\ \nonumber
\hat{a}\hat{b}^\dagger&\to&\exp\left[(i\Delta-\frac{1}{2}\kappa)t\right]\bar{a}\hat{b}^\dagger=\exp\left[(i\Delta-\frac{1}{2}\kappa)t\right]\sqrt{\bar{n}}\hat{b}^\dagger.
\end{eqnarray}
\noindent
This will immediately result in rewriting (\ref{eq18}) as
\begin{equation}
\label{eq19b}
\frac{d}{dt}
\left\{ 
\begin{array}{c}
\delta\hat{a} \\ 
\delta\hat{b} \\ 
\delta\hat{b}^{\dagger }
\end{array}
\right\}=\left[ 
\begin{array}{ccc}
i\Delta-\frac{\kappa }{2} & ig_0 & ig_0 \\ 
0 &  -i\Omega-\frac{\Gamma }{2} & 0 \\ 
0 & 0 & i\Omega-\frac{\Gamma }{2}  
\end{array}
\right]
\left\{ 
\begin{array}{c}
\delta\hat{a} \\ 
\delta\hat{b} \\ 
\delta\hat{b}^{\dagger }
\end{array}
\right\}+\left[ 
\begin{array}{ccc}
\sqrt{\kappa} & 0 & 0\\
0 & \sqrt{\Gamma} & 0\\
0 & 0 & \sqrt{\Gamma}
\end{array}
\right]\left\{ \begin{array}{c}
\hat{a}_\text{in} \\ 
\hat{b}_\text{in} \\ 
\hat{b}_\text{in}^\dagger 
\end{array}
\right\},
\end{equation}
\noindent
which is nothing but exactly the linearized state equations of optomechanics. Hence, the method of higher-order operators \cite{SPaper2} is mathematically able to reproduce the less approximate linearized optomechanics.

\section{Pulsed Drive}

Under the situation of pulsed drive, one may assume the input photon rate $\alpha(t)$ to be a function of time. If the input drive varies on a time-scale or longer than the mechanical period with $|d\alpha(t)/dt|<\Omega\alpha$, then one may assume $\bar{n}(t)$ is solved through (\ref{eq12}) at each moment with updated momentary mechanical frequency $\Omega(t)$ and linewidth $\Gamma(t)$ to yield an effective time dependent coefficients matrix $\textbf{M}(t)$. This offers the solution 
\begin{eqnarray}
\label{Pulsed}
\{A(t)\}&=&\exp\left[\int_{0}^{t}\textbf{M}(\tau)d\tau\right]\{A(0)\} +\int_{0}^{t}\exp\left[\int_{0}^{t-\tau}\textbf{M}(\nu)d\nu\right][\beta(\tau)]\left\{\alpha(\tau)\right\}d\tau \\ \nonumber
&+&\int_{0}^{t}\exp\left[\int_{0}^{t-\tau}\textbf{M}(\nu)d\nu\right][\Gamma(\tau)]\{A_\text{in}(\tau)\}d\tau\\ \nonumber
[\beta(t)]^{\text{T}}&=&\left[ 
\begin{array}{cccccc}
1 & 0 & \bar{b}(t) & \bar{b}^\ast(t) & \sqrt{\bar{n}(t)} & \sqrt{\bar{n}(t)}\\
0 & 0 & 0 & \sqrt{\bar{n}(t)} & 0
\end{array}
\right], \\ \nonumber
\{\alpha(t)\}^{\text{T}}&=&\left\{
\alpha(t), 
\alpha^\ast(t)
\right\}.
\end{eqnarray}

\section{Side-band Inequivalence}\label{Inequivalence}

Let us go back to the set of equations (\ref{eq14}) and only retain the first, third, and fourth equations.  This reduction gives a $3\times 3$ system of equations, identical to (\ref{eq16}) with the redefinitions 
\begin{eqnarray}
\label{eq18}
\textbf{M}&=&\textbf{N}+\delta\textbf{N}\\ \nonumber
\textbf{N}&=&\left[ 
\begin{array}{ccc}
i\Delta-\frac{\kappa }{2} & ig_0 & ig_0 \\ 
iF &  -i(\Omega-\Delta)-\frac{\gamma }{2} & 0 \\ 
-iF & 0 & i(\Omega+\Delta)-\frac{\gamma }{2}  
\end{array}
\right],\\ \nonumber
\left\{\delta A(\omega)\right\}^{\text{T}}&=&\left\{ \delta\hat{a}(\omega), \delta(\hat{a}\hat{b})(\omega), \delta(\hat{a}{\hat{b}}^{\dagger })(\omega) \right\},\\ \nonumber
\left\{A_\text{in}(\omega)\right\}^{\text{T}}&=&\left\{ \hat{a}_\text{in}(\omega),\hat{b}_\text{in}(\omega), 
\hat{b}_\text{in}^\dagger(\omega) \right\},\\ \nonumber
\left[\sqrt{\Gamma}\right]&=&\left[ 
\begin{array}{ccc}
\sqrt{\kappa} & 0 & 0\\
\sqrt{\kappa}\bar{b} & \sqrt{\Gamma\bar{n}} & 0\\
\sqrt{\kappa}\bar{b}^\ast & 0 & \sqrt{\Gamma\bar{n}}
\end{array}
\right].
\end{eqnarray}
\noindent
Here, the perturbation matrix $\delta\textbf{N}$ is defined through the relation
\begin{equation}
\label{eq19}
\delta\textbf{N}=\left[ 
\begin{array}{ccc}
0 & 0 & 0 \\ 
if^+ & is & 0 \\ 
if^- & 0 & is^\ast  
\end{array}
\right],
\end{equation}
in which $f^+=g_0(\bar{m}+1)$ and $f^-=g_0\bar{m}$. It is quite apparent that the second and third rows of $\textbf{N}$ in (\ref{eq18}) are complex conjugates. 

By setting $\Delta=0$ in (\ref{eq18}) one would expect identically displaced sidebands at $\pm\Omega$. However, this is contingent on the fact that the eigenvalues of $\textbf{N}$ be either complex conjugates as $\Im[\eta]=\pm\Omega$ corresponding to the frequencies of the sidebands, or $\Im[\eta]=0$ corresponding to the resonant pump. However, the presence of perturbation matrix $\delta\textbf{N}$ breaks this symmetry between the sidebands. This causes a very tiny displacement of sidebands so that $\Delta_\text{r}+\Delta_\text{b}\neq0$. First figure of the main article illustrates the side-band asymmetry for various intracavity photon numbers $\bar{n}=(g/g_0)^2$ and coherent phonon numbers $\bar{m}$, when $g_0/\Omega=10^{-3}$. This effect is actually due to the higher-order optomechanical spring effect analyzed in the following. 

It has to be noticed that the horizontal axes are nonlinear functions of the incident light intensity and therefore $\alpha$. Typically, an inequivalence would be observable in a heterodyne side-band resolved experiment if the effect is large enough to allow clear and measurable motion of side-bands. This condition requires $|\Delta_\text{r}+\Delta_\text{b}|>\Gamma=\Omega/Q_\text{m}$, in which $Q_\text{m}$ is the mechanical quality factor. If $Q_\text{m}>10^5$, then an intracavity occupation number of $\bar{n}>10^4$ should be sufficient to detect any such inequivalence.

The side-band inequivalence should not be mistaken with the fundamental energy conservation and time reversal symmetry. Firstly according to these, the spectral density on the negative frequencies of the spectrum should be mirror symmetric with respect to the positive frequencies. Normally, the actual optical frequency $\omega$ is much larger than the mechanical frequency $\Omega$, so that the observed red- and blue-detuned sidebands within the range $\Delta\in(-\Omega,+\Omega)$ actually entirely correspond to the positive absolute frequencies. So the speculation that $\delta\Delta$ could be non-zero has nothing to do with the time-reversal symmetry. Secondly, side-band inequivalence is a purely nonlinear effect and is therefore strictly forbidden in any linearized approximation of the optomechanical Hamiltonian. 

\section{Resonance Shift}
The contribution of the terms $\pm iF+if^{\pm}$ to the mechanical frequency $\Omega$ in the second and third equations of (\ref{eq19}), can be held responsible for the so-called optomechanical spring effect \cite{SKip1,SAspel1,SBowen,SQuad9,SSpring1,SSpring2,SSpring3,SSpring4,SSpring5}. As the result of optomechanical interaction, both of the optical and mechanical resonance frequencies and damping rates undergo shifts. Even at the limit of zero input optical power $\alpha=0$ and therefore zero cavity photon number $\bar{n}=0$, it is possible to show that there is a temperature-dependent shift in the mechanical resonance frequency, markedly different from the lattice-expansion dependent effect. This effect is solely due to the optomechanical interaction with virtual cavity photons, which completely vanishes when $g_0=0$. In close relationship to the shift of resonances, we can also study the optomechanical spring effect with the corrections from higher-order interactions included.

The analysis of spring effect is normally done by consideration of the effective optomechanical force acting upon the damped mechanical oscillator, thus obtaining a shift in squared mechanical frequency $\delta(\Omega^2)$, whose real and imaginary parts give expressions for $\delta\Omega$ and $\delta\Gamma$. Corrections to these two terms due to higher-order interactions are discussed in the main article. Here, we demonstrate that the analysis using higher-order operator algebra can recover some important lost information regarding the optical and mechanical resonances when the analysis is done on the linearized basis $\{A\}^\text{T}=\{\hat{a},\hat{a}^\dagger,\hat{b},\hat{b}^\dagger\}$. 

To proceed, we consider finding eigenvalues of the matrix $\textbf{M}$ as defined in (\ref{eq18}). Ignoring all higher-order nonlinear effects beyond the basis $\{A\}^\text{T}=\{\hat{a},\hat{a}\hat{b},\hat{a}\hat{b}^\dagger\}$, we set $s=0$. This enables us to search for the eigenvalues of the coefficients matrix $\textbf{M}$ as
\begin{eqnarray}
\label{eqSpring1}
\text{eig}\{\textbf{M}\}&=&\text{eig}\left[ 
\begin{array}{ccc}
i\Delta-\frac{\kappa }{2} & ig_0 & ig_0 \\ 
i(G+f^+) &  -i(\Omega-\Delta)-\frac{\gamma }{2} & 0 \\ 
-i(G-f^-) & 0 & i(\Omega+\Delta)-\frac{\gamma }{2}  
\end{array}
\right]\\ \nonumber
&=&i\left\{
\begin{array}{c}
\Delta+\lambda_1(\Delta,T)+i\gamma_1(\Delta,T) \\
\Delta+\lambda_2(\Delta,T)+i\gamma_2(\Delta,T) \\
\Delta+\lambda_3(\Delta,T)+i\gamma_3(\Delta,T) 
\end{array}
\right\}\\ \nonumber
&=&i\left\{
\begin{array}{c}
\Delta+\eta_1(\Delta,T) \\
\Delta+\eta_2(\Delta,T) \\
\Delta+\eta_3(\Delta,T) 
\end{array}
\right\},
\end{eqnarray}
\noindent
in which $\lambda_j=\Re[\eta_j]$ and $\gamma_j=\Im[\eta_j]$ with $j=1,2,3$ are real valued functions of $\Delta$ and bath temperature $T$. The temperature $T$ determines $\bar{m}$ while $\bar{n}$ is a function of $\Delta$ as well as input photon rate $\alpha$. 

In general, the three eigenvalues $\eta_j=\lambda_j(\Delta,T)+i\gamma_j(\Delta,T), j=1,2,3$ are expected to be deviate from the three free-running values $\psi_1=i\frac{1}{2}\kappa$, $\psi_2=-\Omega+i\frac{1}{2}\gamma$, and $\psi_3=\Omega+i\frac{1}{2}\gamma$, as $\eta_j\approx \psi_j-\Delta$ because of non-zero $g_0$. Solving the three equations therefore gives the values of shifted optical and mechanical frequencies and their damping rates compared to the bare values in absence of optomechanical interactions with $g_0=0$ 
\begin{eqnarray}
\label{eqSpring2}
\delta\Omega&=&-\frac{1}{2}\Re[\eta_2-\eta_3]-\Omega, \\ \nonumber
\delta\omega&=&-\frac{1}{2}\Re[\eta_2+\eta_3], \\ \nonumber
\delta\Gamma&=&\Im[-2\eta_1+\eta_2+\eta_3]-\Gamma, \\ \nonumber
\delta\kappa&=&2\Im[\eta_1]-\kappa. 
\end{eqnarray}
\noindent
This  method to calculate the alteration of resonances, does not regard the strength of the optomechanical interaction or any of the damping rates. In contrast, the known methods to analyze this phenomenon normally require $g<<\kappa$ and $\Gamma+\delta\Gamma<<\kappa$ \cite{SAspel1}.

The above values can be calculated numerically for a typical optomechanical cavity, whose parameters are displayed in Table \ref{Table1}. The selected values of the four example optomechanical set ups result in very different configurations. System A is very strongly coupled with $g/(\kappa/2)=1.2$ and in far Doppler limit $(\kappa/2)/\Omega=15$. Meanwhile, Systems B and C with $(\kappa/2)/\Omega=0.15$, and System D with $(\kappa/2)/\Omega=2.2\times 10^{-6}$ are all in the resolved-side band regime. System B is ultrastrongly coupled with $g/(\kappa/2)=1.07\times10^4$, while for Systems C and D we have respectively $g/(\kappa/2)=1.2$ and $g/(\kappa/2)=3.48$.

\begin{table}[ht!]
	\centering
	\caption{Parameters of the simulated optomechanical example; $P_\text{op}$: input optical power; $\lambda$: optical wavelength; $\Omega_m$: mechanical angular frequency; $Q$: optical quality factor; $Q_m$: mechanical quality factor; $g_0$: single-photon optomechanical interaction rate; $T$: absolute temperature; $\mathcal{C}_0=4g_0^2/\kappa\Gamma$: single-photon cooperativity \cite{SAspel1}; $\mathcal{C}=\bar{n}\mathcal{C}_0$: multi-photon cooperativity. System A is strongly coupled and in Doppler regime. Systems B and C are side-band resolved but strongly coupled. System D is the one used in experiment \cite{SAspel3,SAspel4}. System E is used for study of nonlinear transduction.}
	\label{Table1}
	\begin{tabular}{cccccccccc}
		\hline\hline
		System & $P_\text{op}$ & $\lambda$ & $\Omega_m/2\pi$ & $Q$ & $Q_m$  & $g_0/2\pi$ & $T$ & $\mathcal{C}_0$ & $\mathcal{C}$\\ \hline
		A & $2\mu\text{W}$ & $1\mu\text{m}$ & $1\text{GHz}$ & $10^4$ & $10^3$ & $160\text{kHz}$ & 1K & $3.38\times 10^{-4}$ & $3.81$\\ 
		B & $2\mu\text{W}$ & $1\mu\text{m}$ & $1\text{GHz}$ & $10^6$ & $10^4$ & $16\text{kHz}$ & 1K & $3.38\times 10^{-5}$ & $3.41\times 10^{3}$\\  
		C & $20\text{nW}$ & $1\mu\text{m}$ & $1\text{GHz}$ & $10^6$ & $10^4$ & $16\text{kHz}$ & 1K & $3.38\times 10^{-5}$ & 0.381\\
		D & $450\text{nW}$ & $1.55\mu\text{m}$ & $5.3\text{GHz}$ & $2.3\times10^5$ & $3.8\times 10^5$ & $869\text{kHz}$ & 35mK & $1.5\times 10^{-2}$ & $2.8\times 10^5$\\
		E & $-$ & $1\mu\text{m}$ & $1\text{GHz}$ & $10^6$ & $10^4$ & $400\text{kHz}$ & $-$ & $2.21\times 10^{-2}$ & $-$\\
		\hline\hline
	\end{tabular}
\end{table}

A very simple way to estimate the shift in eigenvalues is by separating the real and imaginary parts of the optomechanical interaction, as $\Omega\to\Omega-\Re[s]$ and $\gamma\to\gamma+2\Im[s]$. These shifts in mechanical frequency and damping rates can be approximated using (\ref{eq9}) as
\begin{eqnarray}
\label{eq20}
\delta\Omega+\delta\omega&\approx&-g_0\Re[\bar{b}]=-\frac{g^2\Omega}{\Omega^2+\frac{1}{4}\Gamma^2},\\ \nonumber
\delta\Gamma+\delta\kappa&\approx&g_0\Im[\bar{b}]=\frac{g^2\Gamma}{2(\Omega^2+\frac{1}{4}\Gamma^2)},
\end{eqnarray}
\noindent
where $g$ has been defined under (\ref{eq5}). This approximation requires the optomechanical processes $\{\hat{a}\hat{b},\hat{a}\hat{b}^\dagger\}$ being independent of the other state variables. Since this decoupling is not exact, relations (\ref{eq20}) also will remain approximate. However, the accuracy of these are still quite remarkable as is demonstrated here. 

\begin{figure}[ht!]
	\centering
	\includegraphics[width=2.1in]{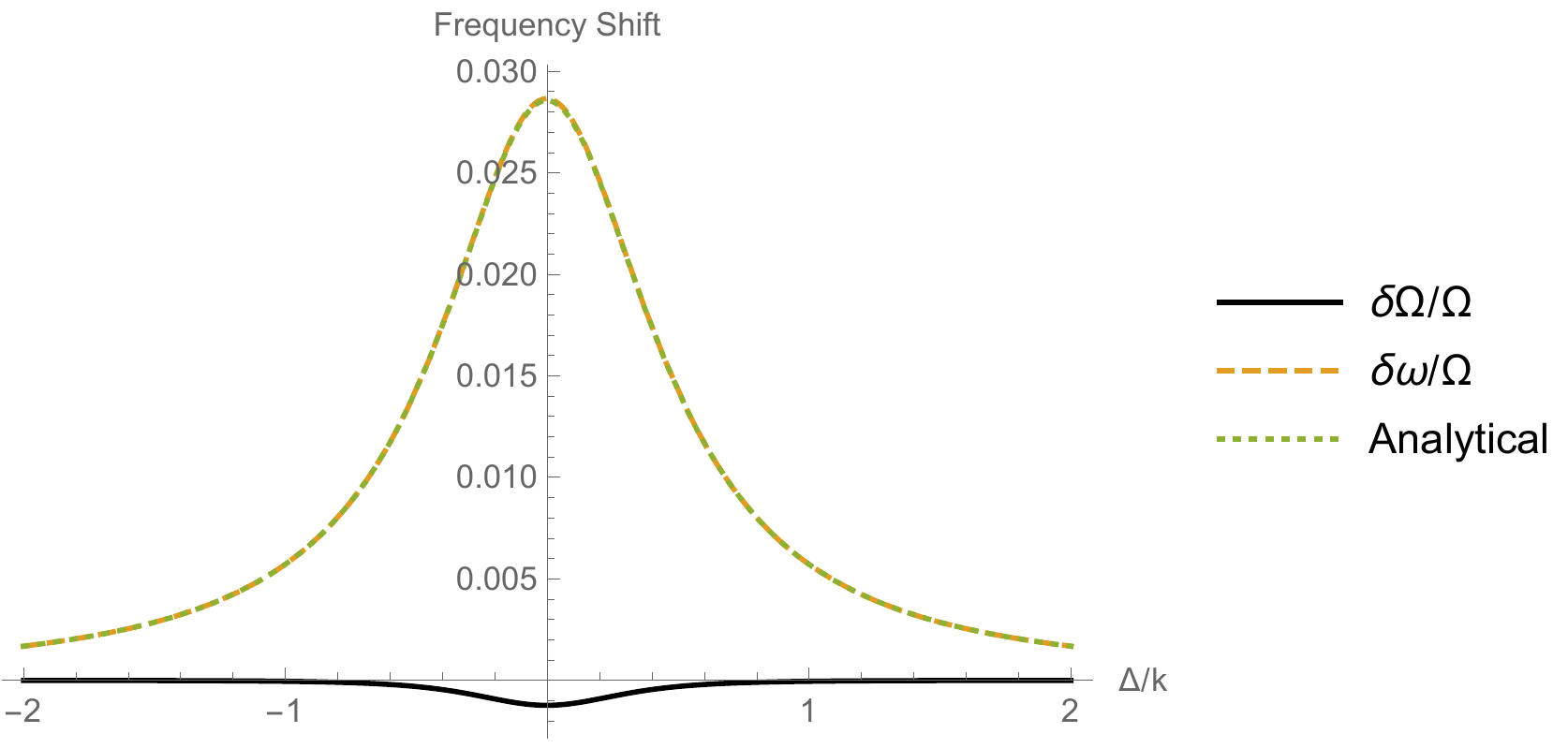}\includegraphics[width=2.1in]{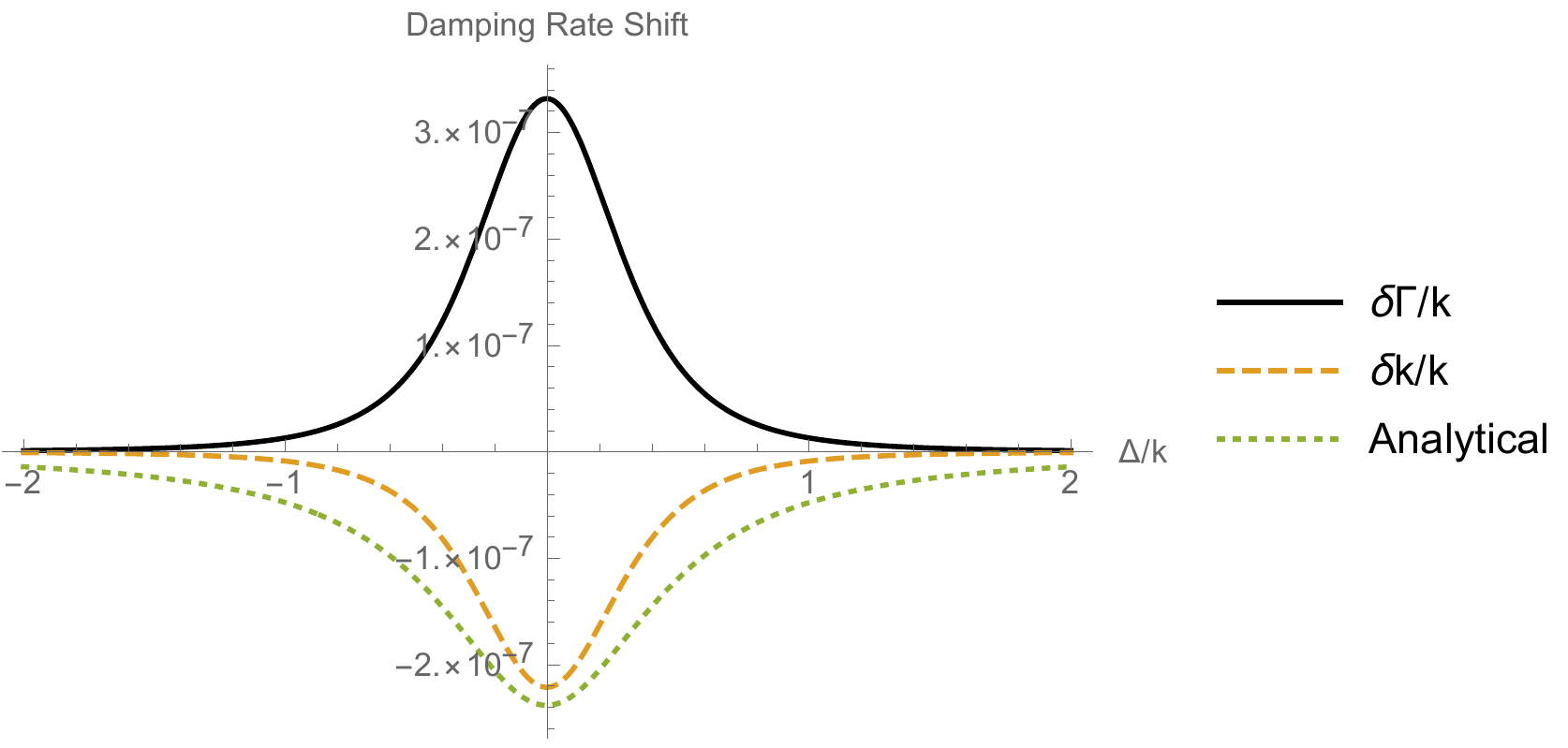}
	\caption{Shift in frequency and damping rates of optical and mechanical partitions due to optomechanical interaction for System A\label{Fig1}.}
\end{figure}

\begin{figure}[ht!]
	\centering
	\includegraphics[width=2.1in]{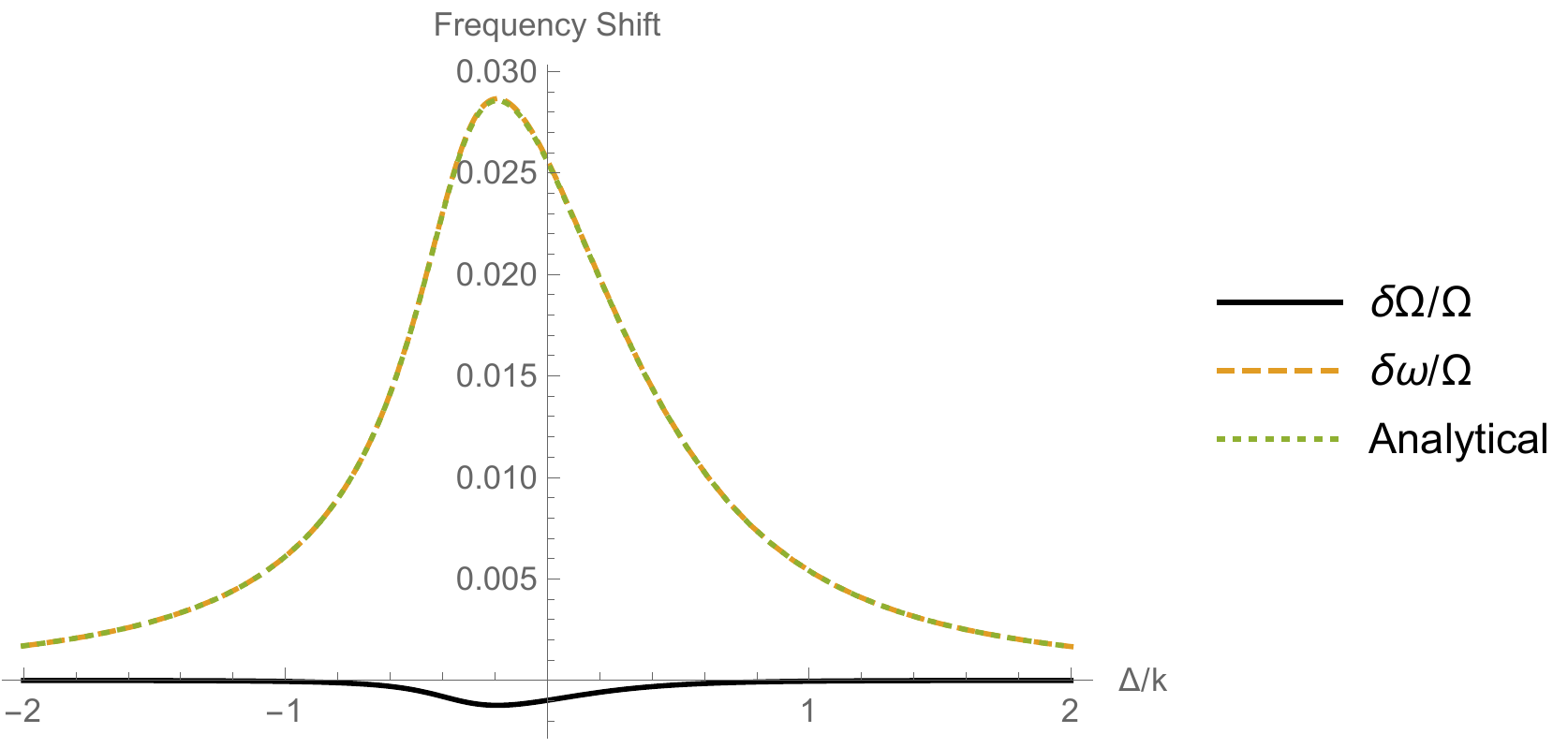}\includegraphics[width=2.1in]{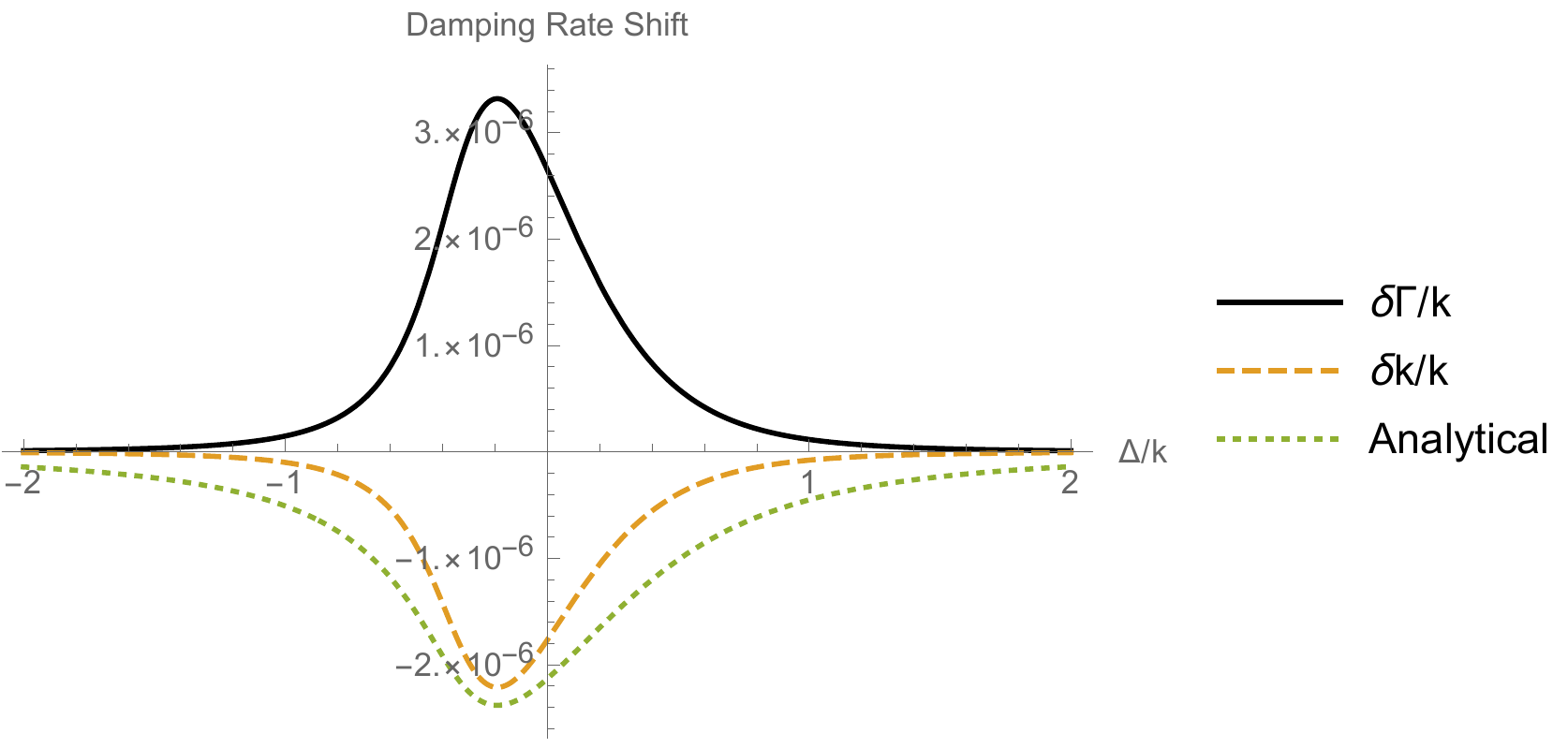}
	\caption{Shift in frequency and damping rates of optical and mechanical partitions due to optomechanical interaction for System B\label{Fig2}. Large cooperativity causes asymmetric behavior of frequency and damping shifts.}
\end{figure}

\begin{figure}[ht!]
	\centering
	\includegraphics[width=2.1in]{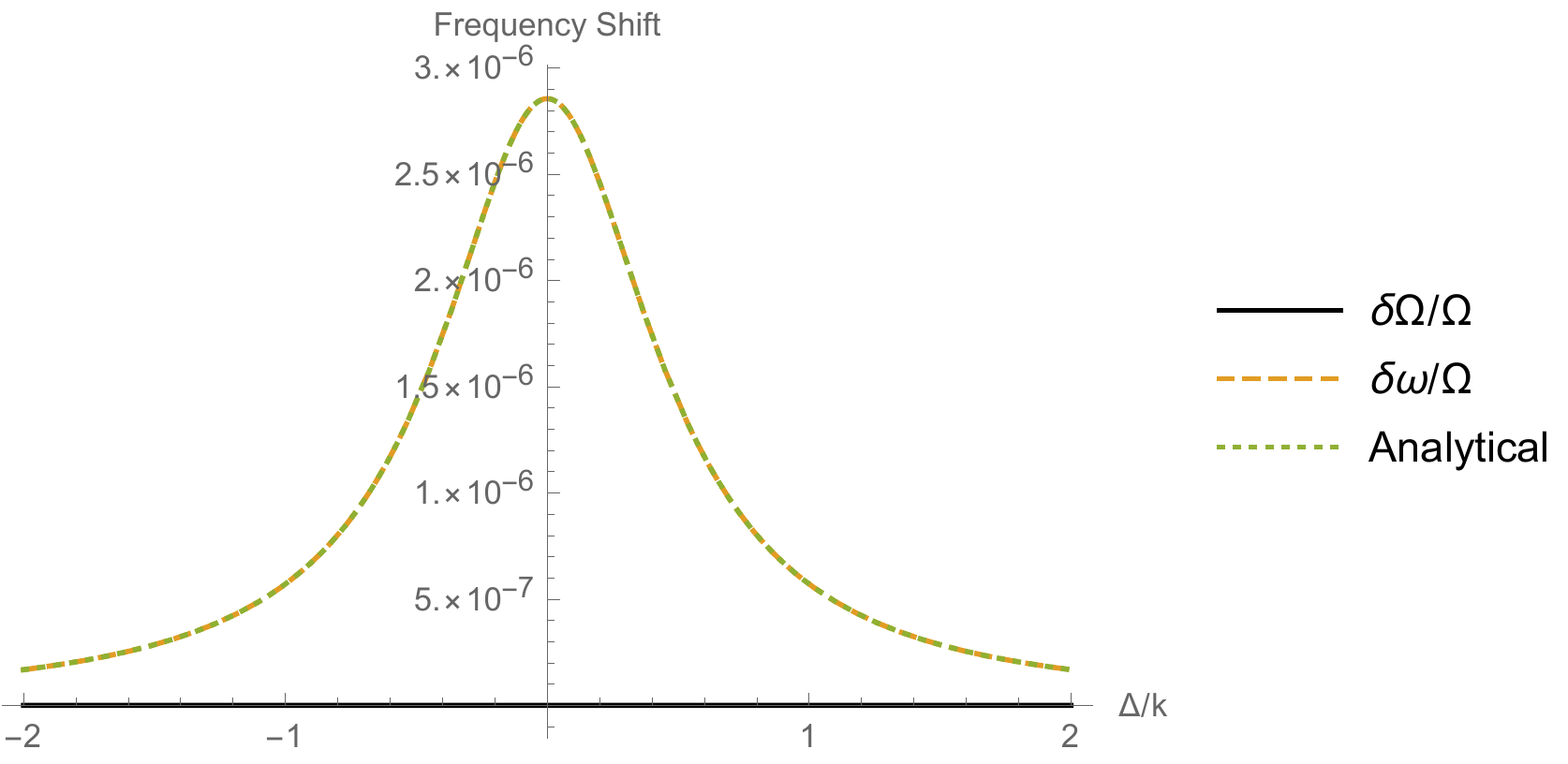}\includegraphics[width=2.1in]{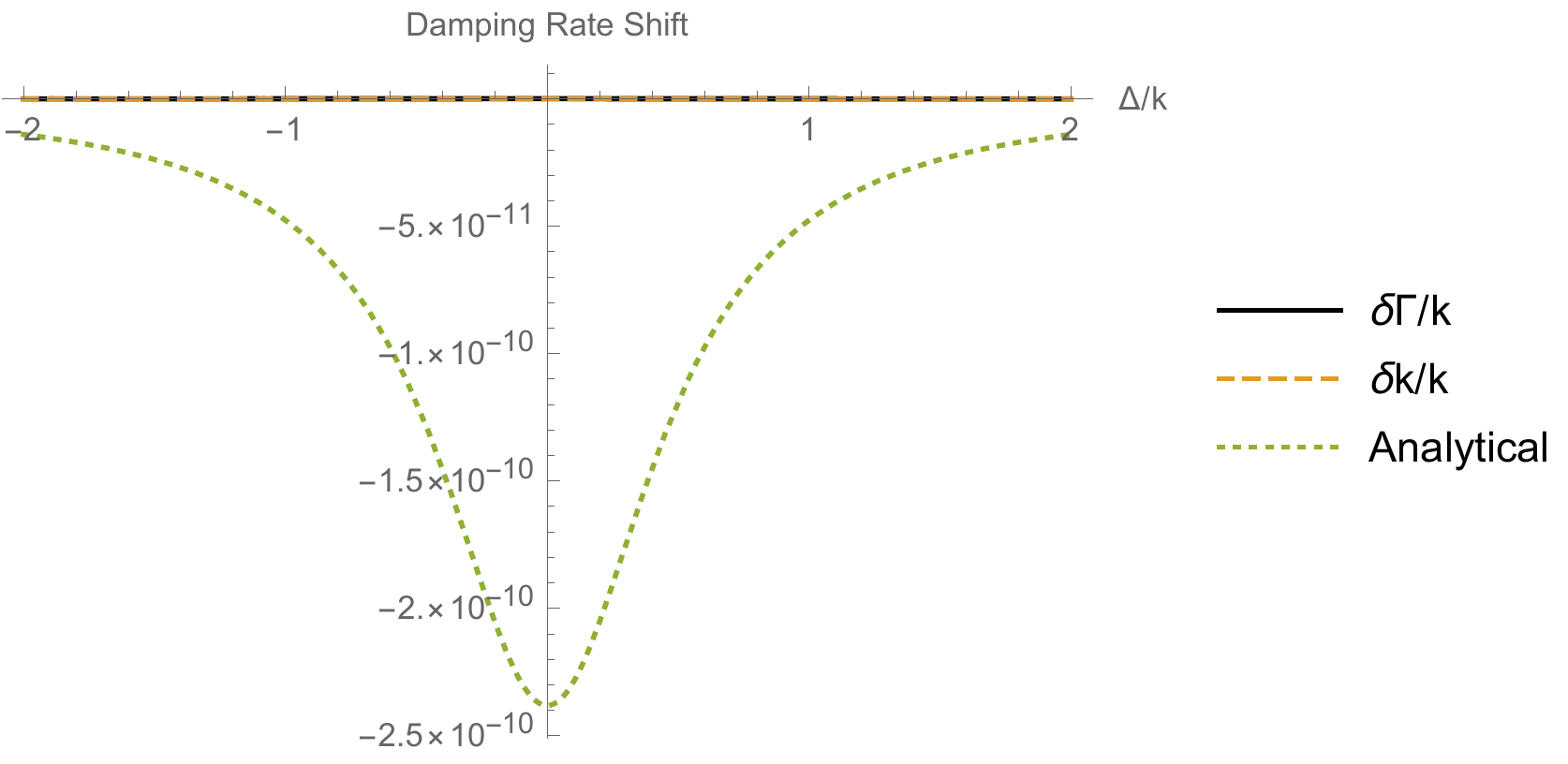}
	\caption{Shift in frequency and damping rates of optical and mechanical partitions due to optomechanical interaction for System C\label{Fig3}.}
\end{figure}

\begin{figure}[ht!]
	\centering
	\includegraphics[width=2.1in]{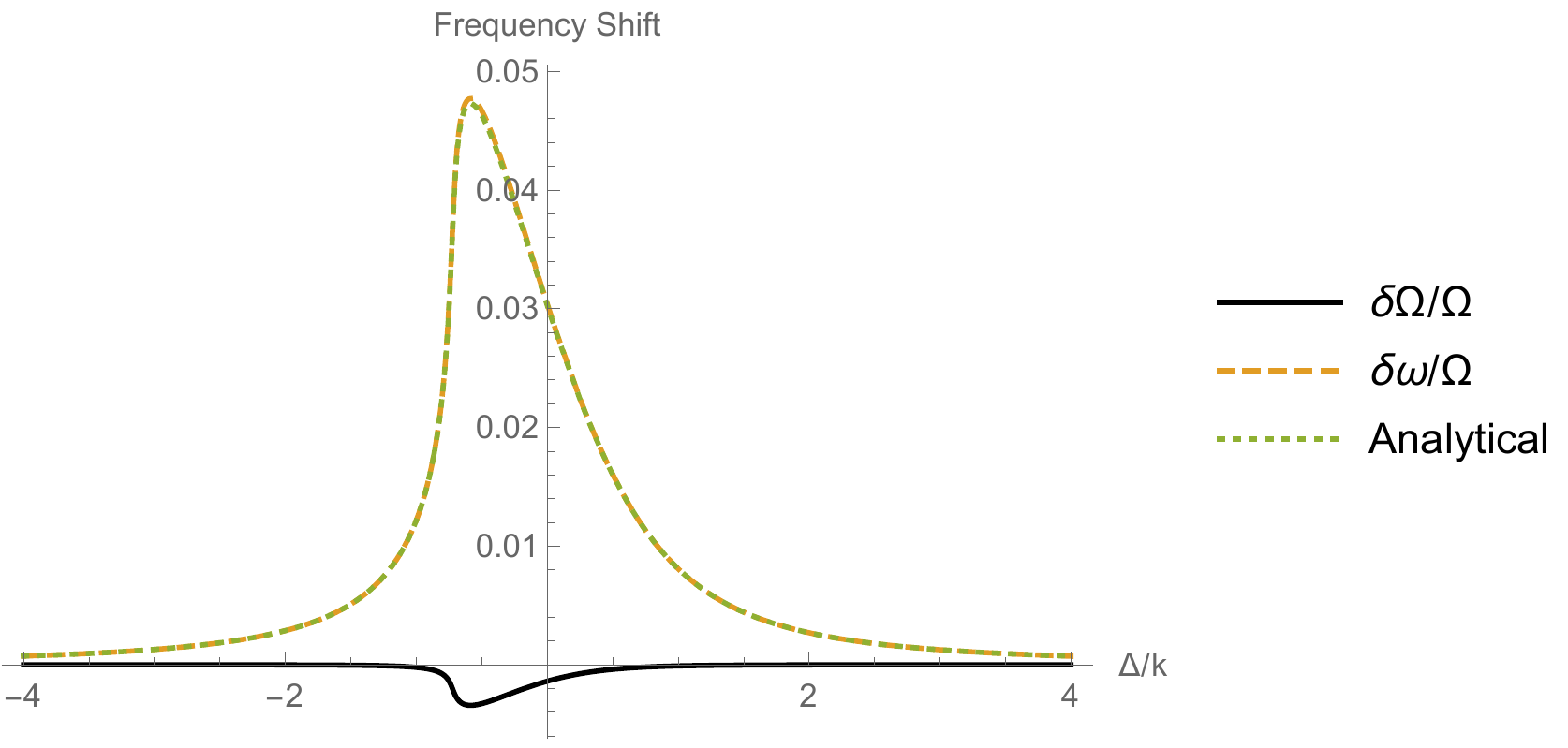}\includegraphics[width=2.1in]{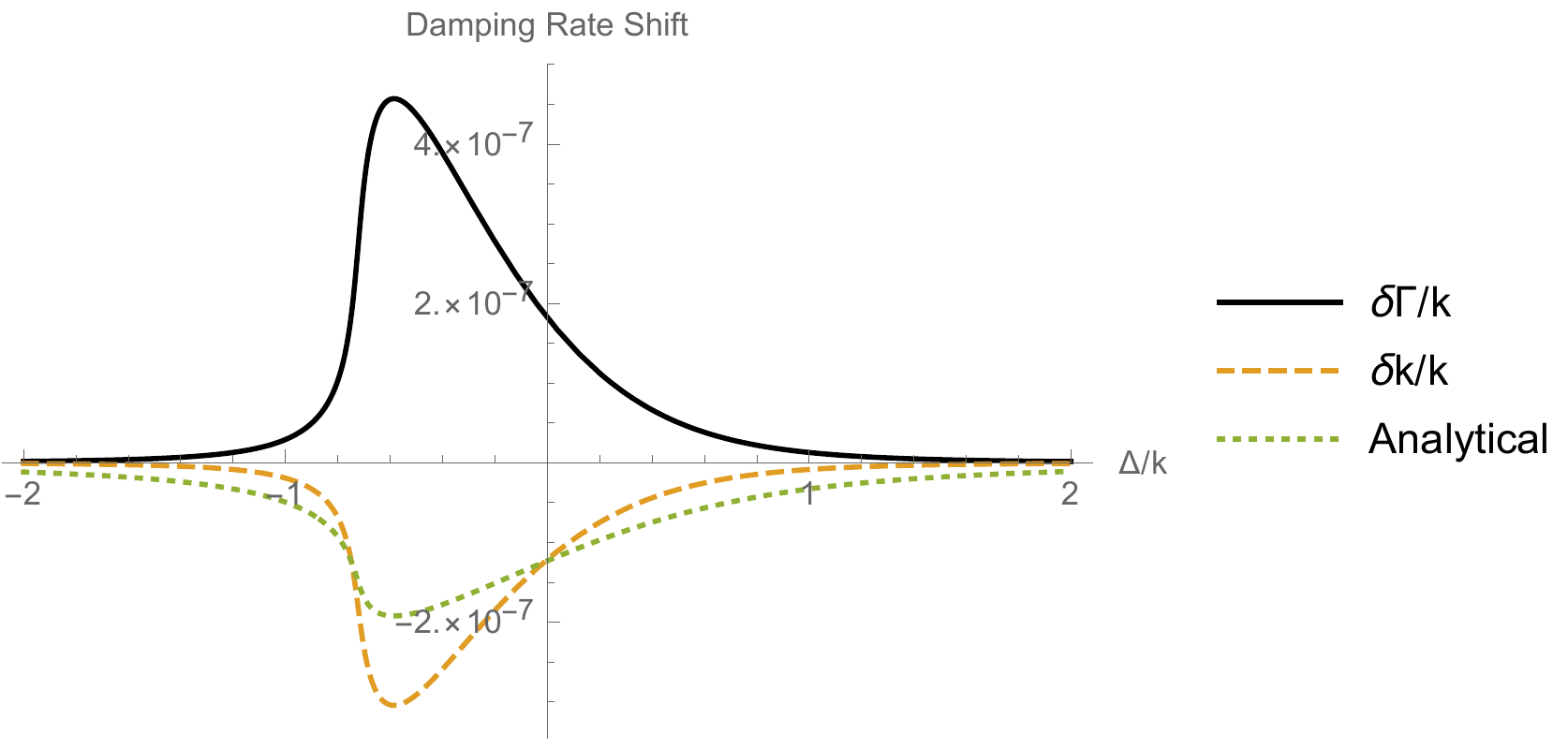}
	\caption{Shift in frequency and damping rates of optical and mechanical partitions due to optomechanical interaction for System D\label{Fig4}. }
\end{figure}

First of all, it is noticed through extensive numerical computations that the shifts in optical and mechanical frequencies take place primarily in the optical part.  That implies the resonance shift is typically much stronger in the optical partition of the system instead of the mechanical partition, leading to marked change in reflection spectra of optomechanical cavities. We calculate and plot each of the four individual components of (\ref{eqSpring2}) along with the analytical expressions (\ref{eq20}) for the four systems A, B, C and D described in Table \ref{Table1}, respectively illustrated in Figs. \ref{Fig1}, \ref{Fig2}, \ref{Fig3} and \ref{Fig4} for the shifts in mechanical and optical frequencies and dissipation-decay rates. 

Extensive numerical calculations for various configurations establish the fact that it is actually the optical resonance frequency which receives the optomechanical interaction effect. The asymmetry of this shift in cavity optical frequency across the zero-detuning $\Delta=0$, is well exhibited in Fig. \ref{Fig2} for System B, and in Fig. \ref{Fig4} for System D, both of which are taken to have relatively large intracavity photon numbers around $10^{7}$ to $10^{8}$. This clear numerical signature underlines the fact that the well-known asymmetry of cavity optical response at high intensities should be actually a result of this higher-order spring effect, rather than thermally induced instabilities. 

\begin{figure}[ht!]
	\centering
	\includegraphics[width=2.1in]{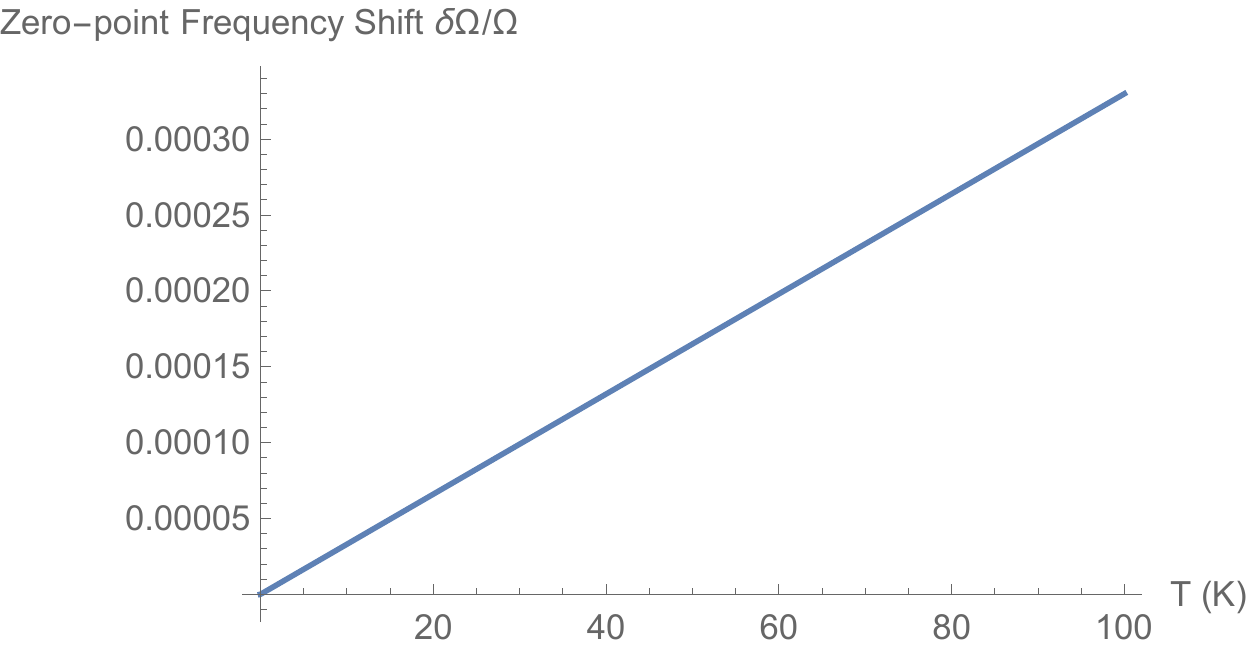}\includegraphics[width=2.1in]{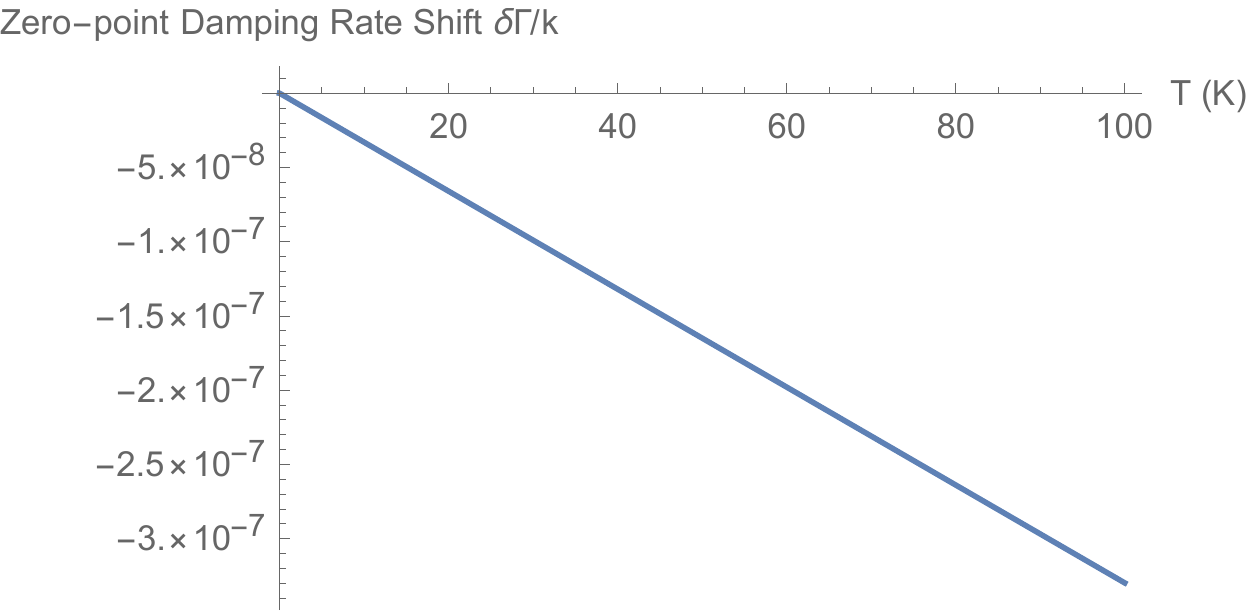}
	\caption{Temperature dependence of mechanical frequency shift due to optomechanical interaction with zero-point radiation field for System E. This temperature-dependent shift amounts to $3.3\text{kHz/K}$\label{Fig5}.}
\end{figure}

Summarizing, any such higher-order resonance shift will cause a change in mechanical frequency $\delta\Omega$ and decay rate $\delta\Gamma$, as well as optical detuning $\delta\omega=-\delta\Delta$ and decay rate $\delta\kappa$. While all these four components are non-zero, it is $\delta\Delta$ which is ultimately dominant over the three others in the bistability relation (\ref{eq12}). This will make the cavity response to follow the bistability and therefore appear to be asymmetric at high illumination drive intensities.

As a result of higher-order spring effect and $\delta\Delta$, a shift in intracavity photon number follows $\delta\bar{n}$, which immediately shifts the higher-order spring effect and therefore $\delta\Delta$. The infinite cycle of shifts in intracavity photon number and optical resonance frequency establishes a deterministic chaotic behavior, which is also a well known experimental observation in the community.

Furthermore, for System E, the zero-point optical field can change the mechanical frequency as large as $3.3\text{kHz/K}$ as illustrated in Fig. \ref{Fig5}. Being in close relationship with the Dynamical Casimir effect \cite{SMacri}, this value could be in principle measured if temperature-induced expansion and the resulting change of mechanical frequency is much smaller. The thermal expansion coefficient of Silicon is roughly $2.6\times10^{-6}\text{K}^{-1}$, roughly equivalent to $2.6\text{kHz/K}$. The contribution of zero-point field can be therefore larger or at least within the same order of magnitude. Here, we have assumed that the thermal expansion coefficient of Silicon is independent of temperature and also $\Omega$ shifts linearly with temperature. 

The same calculation for System D gives out a value of $0.57\text{kHz/K}$ which is much less than the temperature expansion drift of $13.7\text{kHz/K}$ for the same structure. This phenomenon has been also noticed and referred to as the Nonlinear Transduction \cite{STransduction} where the photon-phonon coupling can induce a temperature-dependent change in the resonance frequency of the cavity, even on the order of cavity linewidth.

\section{Coherent Phonon Population}\label{BMWm8}
It is here shown that the method of higher-order operators allows one to find an explicit expression for $\bar{m}$. In order to do this, we need to write down the $3\times 3$ reduced set of higher-order optomechanical equations with the fluctuations terms dropped, which reads
\begin{equation}
\label{m1}
\frac{d}{dt}\left\{\begin{array}{c}
\hat{a}\\
\hat{a}\hat{b}\\
\hat{a}\hat{b}^\dagger
\end{array}
\right\}=\left[\begin{array}{ccc}
i\Delta-\frac{1}{2}\kappa & ig_0 & ig_0 \\
ig_0(\bar{m}+\bar{n}+1) & -i(\Omega-\Delta)-\frac{1}{2}\gamma & 0\\
ig_0(\bar{m}-\bar{n}) & 0 & i(\Omega+\Delta)-\frac{1}{2}\gamma
\end{array}\right]\left\{\begin{array}{c}
\hat{a}\\
\hat{a}\hat{b}\\
\hat{a}\hat{b}^\dagger
\end{array}
\right\}-\left\{\begin{array}{c}
\alpha \\ 
\bar{b}\alpha \\
\bar{b}^\ast\alpha^\ast
\end{array}
\right\}.
\end{equation}
Here, $\alpha$ should be taken as a complex number from (\ref{eq11}), $\bar{b}$ is substituted from (\ref{eq9}) in terms of $\bar{a}$, where $\bar{a}=\sqrt{\bar{n}}$ is taken as a real number and $\bar{n}$ can already by found from the solution of the third-order algebraic equation (\ref{eq12}). 

We are here interested in the steady state solutions, so that the time derivative $d/dt=0$ can be set to zero. Then we arrive at the system of equations
\begin{equation}
\label{m2}
\left[\begin{array}{ccc}
i\Delta-\frac{1}{2}\kappa & ig_0 & ig_0 \\
ig_0(\bar{m}+\bar{n}+1) & -i(\Omega-\Delta)-\frac{1}{2}\gamma & 0\\
ig_0(\bar{m}-\bar{n}) & 0 & i(\Omega+\Delta)-\frac{1}{2}\gamma
\end{array}\right]\left\{\begin{array}{c}
\bar{a}\\
\overline{ab}\\
\overline{ab^\ast}
\end{array}
\right\}=\left\{\begin{array}{c}
\alpha \\ 
\bar{b}\alpha \\
\bar{b}^\ast\alpha^\ast
\end{array}
\right\}.
\end{equation}
In the above system of equations, $\overline{ab}$ corresponds to the time-average of the operators $\braket{\hat{a}\hat{b}}$, while $\overline{ab^\ast}$ corresponds to the time-average of the operators $\braket{\hat{a}\hat{b}^\dagger}$. Quite obviously, $\overline{ab}\approx\bar{a}\bar{b}$ and  $\overline{ab^\ast}\approx\bar{a}\bar{b}^\ast$ can approximately hold based on the mean-field approximation. We shall here furthermore observe that this approximation does not any longer hold for the coherent phonons as $\bar{m}=\braket{\hat{b}^\dagger\hat{b}}\neq\bar{b}^\ast\bar{b}$ for the reasons discussed below.

We now can rearrange (\ref{m2}) in terms of the unknown quantities $\bar{m}$, $\overline{ab}$, and $\overline{ab^\ast}$ as 
\begin{equation}
\label{m3}
\left[\begin{array}{ccc}
0 & ig_0 & ig_0 \\
ig_0\bar{a} & -i(\Omega-\Delta)-\frac{1}{2}\gamma & 0\\
ig_0\bar{a} & 0 & i(\Omega+\Delta)-\frac{1}{2}\gamma
\end{array}\right]\left\{\begin{array}{c}
\bar{m}\\
\overline{ab}\\
\overline{ab^\ast}
\end{array}
\right\}=\left\{\begin{array}{c}
\alpha-\left(i\Delta-\frac{1}{2}\kappa\right)\bar{a} \\ 
\bar{b}\alpha-ig_0\left(\bar{n}+1\right)\bar{a} \\
\bar{b}^\ast\alpha^\ast+ig_0 \bar{n}\bar{a}
\end{array}
\right\}.
\end{equation}
This linear system of equations after appropriate substitutions from (\ref{eq9}) and (\ref{eq11}) now can be solved to find
\begin{equation}
\label{m4}
\bar{m}(\Delta)=\frac{64 g_0^2 \Omega ^2 \bar{n}^2(\Delta ) \left(\gamma ^2+\gamma  \Gamma +4 \Delta ^2\right)-\left[(\Gamma ^2+4 \Omega ^2\right)^2 \left(\gamma ^2+4 \Delta  (\Delta +\Omega )\right]}{2 \left(\gamma ^2+4 \Delta ^2\right) \left(\Gamma ^2+4 \Omega ^2\right)^2}.
\end{equation}
Here, a small imaginary part remains which has to be dropped and results from the inexactness of (\ref{eq9}) and (\ref{eq11}) coming from linearized optomechanics, and not being in complete consistency with the higher-order formalism. 

In a similar manner, one may find
\begin{eqnarray}
\label{m5}
\overline{ab}&=&
\frac{i \sqrt{\bar{n}} \left[g_0^2 \left(8 \bar{n}+4\right)+(2 \Delta +i \kappa ) (2 (\Delta +\Omega )+i \gamma )\right]-2 i \alpha  \left[\frac{8 \Gamma  g_0^2 \bar{n}}{\Gamma ^2+4 \Omega ^2}+\gamma -2 i (\Delta +\Omega )\right]}{4 g_0 (\gamma -2 i \Delta )}, \\ \nonumber
\overline{ab^\ast}&=&\frac{2 \alpha  \left[\frac{8 i \Gamma  g_0^2 \bar{n}}{\Gamma ^2+4 \Omega ^2}-i \gamma -2 \Delta +2 \Omega \right]-\sqrt{\bar{n}} \left[4 i g_0^2 \left(2 \bar{n}+1\right)+(\kappa -2 i \Delta ) (i \gamma +2 \Delta -2 \Omega )\right]}{4 g_0 (\gamma -2 i \Delta )}.
\end{eqnarray}

The expression (\ref{m4}) for $\bar{m}$ is accurate within half a quanta $\pm\frac{1}{2}$, so that in order to satisfy the zero limits at infinite detuning 
\begin{equation}
\label{m6}
\lim_{\Delta\rightarrow\infty}\bar{m}(\Delta)=0, 
\end{equation}
a half-quanta must be added to (\ref{m4}). Then it will read
\begin{eqnarray}
\label{m8} 
\bar{m}(\Delta)&=&\frac{32 g_0^2 \Omega ^2 \left(\gamma ^2+\gamma  \Gamma +4 \Delta ^2\right)}{\left(\gamma ^2+4 \Delta ^2\right) \left(\Gamma ^2+4 \Omega ^2\right)^2}\bar{n}^2(\Delta ) -\frac{2 \Delta  \Omega  }{\gamma ^2+4 \Delta ^2 }\pm\frac{1}{2}\\ \nonumber
&\approx&\frac{32 g_0^2 \Omega ^2 \left(\gamma ^2+\gamma  \Gamma +4 \Delta ^2\right)}{\left(\gamma ^2+4 \Delta ^2\right) \left(\Gamma ^2+4 \Omega ^2\right)^2}\bar{n}^2(\Delta ) \\ \nonumber
&=&g_0^2\zeta(\Delta)\bar{n}^2(\Delta).
\end{eqnarray}
The approximation holds well if $\bar{n}$ is well above unity. Hence, we can infer from (\ref{m8}) that $\bar{m}\propto\bar{n}^2$. In the lossless limit, where $\gamma\approx 0$ and $\Gamma\approx 0$, one may even further simplify (\ref{m8}) to obtain the simple expression $\bar{m}\approx 2g_0^2\bar{n}^2/\Omega^2$ which is typically accurate within 10\% of the actual value or better. It is not difficult to check the resonant coherent phonon number $\bar{m}(0)$. In the practical limit of $\kappa>>\Gamma$, it is easy to verify that (\ref{m8}) actually simplifies to $\bar{m}(0)\approx 32 [g_0 Q_\text{m}\bar{n}(0)/\Gamma]^2$ with $Q_\text{m}=\Omega/\Gamma$ being the mechanical quality factor. In the next section, we point out a straightforward method to measure this quantity through experiment on the well-known optical spring effect.

In practice, the expression (\ref{m8}) is sensitive to the choice of optomechanical parameters and in particular $g_0$. A slight variation in the basic optomechanical parameters $\left\{g_0,\omega,\kappa,\Omega,\Gamma\right\}$ with $\gamma=\kappa+\Gamma$ as small as few percent can make a pronounced effect in expected behavior in $\bar{m}$.

For the side-band resolved systems in the lossless limit, it is within 10\% of the relationship $\bar{m}\propto 2|\bar{b}|^2$, meanwhile for Doppler cavities, the agreement is roughly within 3\% or better. This result perfectly agrees to the large-amplitude oscillation limit of $\bar{b}(t)\approx\delta\hat{b}(t)+[\bar{b}+\bar{b}\exp(-i\Omega t)]$ where $\delta\hat{b}$ represents the random fluctuations in the mechanical field with $\braket{\delta\hat{b}}=0$. This also tells that the coherent oscillations of the mechanical field are not differential in amplitude, and can vary in the range $(0,2|\bar{b}|)$. So, the amplitude of coherent mechanical oscillations is just as big as their average. This large-amplitude coherent mechanical wave is driven and waked by the optical coherent field inside the cavity, through optomechanical interactions. The mean field approximations $\overline{ab}\approx\bar{a}\bar{b}$ and $\overline{ab^\ast}\approx\bar{a}\bar{b}^\ast$ seem however to always hold better than 0.1\% for Doppler cavities. This accuracy breaks down for side-band resolved cavities.

If there are more than one mechanical fields available $j=1,2,\cdots$, the coherent phonon population of each mode $\bar{m}_j$ shall be determined with the corresponding sets of optomechanical parameters $\left\{g_{0,j},\omega_j,\kappa_j,\Omega_j,\Gamma_j\right\}$, with the expected approximate result $\bar{m}_j\approx 2g_{0,j}^2\bar{n}_j^2/\Omega_j^2$ as long as the mechanical modes are almost uncorrelated. The case of coherent phonon numbers of two or more correlated mechanical modes needs a separate study.

It is here again stressed out that the oscillations of the mechanical field can be decomposed into the incoherent and coherent parts. The incoherent part results from random thermal fluctuations with the thermal occupancy $m$, as well as half a quanta contributing from the quantum noise of the coherent part, while the coherent oscillations correspond to the coherent phonon number $\bar{m}$. The same also should be true for the optical field, however, the random fluctuations of a coherent light is only half a quanta, and the thermal optical occupancy $n$ of optomechanical cavity is normally negligible under practical considerations and working temperatures. 

\section{Higher-order Spring Effect}\label{Spring}

It is possible to calculate the optomechanical spring effect due to the standard linearized and higher-order interactions. In order to do this, we start from the matrix $[\textbf{M}]$ given in (\ref{eq18}), and after dropping the noise and drive input terms we notice the expansion of first Langevin equation for the operator $\hat{a}$. That reads
\begin{equation}
\label{eqA1}
\frac{d}{dt}\hat{a}=(i\Delta-\frac{1}{2}\kappa)\hat{a}+ig_0\hat{a}(\hat{b}+\hat{b}^\dagger).
\end{equation}
From the second and third equations we get
\begin{eqnarray}
\label{eqA2}
\hat{a}\frac{d}{dt}\hat{b}+\left[(i\Delta-\frac{1}{2}\kappa)\hat{a}+ig_0\hat{a}(\hat{b}+\hat{b}^\dagger)\right]\hat{b}&=&i(f^+ +F)\hat{a}-[i(\Omega-\Delta-g_0\hat{b})+\frac{1}{2}\gamma]\hat{a}\hat{b}, \\ \nonumber
\hat{a}\frac{d}{dt}\hat{b}^\dagger+\left[(i\Delta-\frac{1}{2}\kappa)\hat{a}+ig_0\hat{a}(\hat{b}+\hat{b}^\dagger)\right]\hat{b}^\dagger&=&i(f^- -F)\hat{a}+[i(\Omega+\Delta+g_0\hat{b})-\frac{1}{2}\gamma]\hat{a}\hat{b}^\dagger.
\end{eqnarray}
This is equivalent to 
\begin{eqnarray}
\label{eqA3}
\frac{d}{dt}\hat{b}&=&i(f^+ +F+\hat{b}^\dagger\hat{b})-(i\Omega+\frac{1}{2}\Gamma)\hat{b}, \\ \nonumber
\frac{d}{dt}\hat{b}^\dagger&=&i(f^- -F+\hat{b}\hat{b}^\dagger)+(i\Omega-\frac{1}{2}\Gamma)\hat{b}^\dagger.
\end{eqnarray}
These two equations can be now combined after dropping the nonlinear terms by addition and subtraction, and then taking the Fourier transform to yield the system
\begin{equation}
\label{eqA4}
\left[
\begin{array}{cc}
-iw+\frac{1}{2}\Gamma & i\Omega \\ 
i\Omega & -iw+\frac{1}{2}\Gamma
\end{array}
\right]\left\{
\begin{array}{c}
\hat{b}+\hat{b}^\dagger \\
\hat{b}-\hat{b}^\dagger
\end{array}
\right\}=2ig_0\left\{
\begin{array}{c}
\bar{m}+\frac{1}{2} \\
\bar{n}+\frac{1}{2}
\end{array}
\right\}.
\end{equation}
This can be solved now to yield the expression for $\delta\hat{x}=x_\text{zp}(\hat{b}+\hat{b}^\dagger)$ as 
\begin{equation}
\label{eqA5}
\delta\hat{x}(w)=2ix_\text{zp}g_0\frac{(-iw+\frac{1}{2}\Gamma)(\bar{m}+\frac{1}{2})-i\Omega(\bar{n}+\frac{1}{2})}{(-iw+\frac{1}{2}\Gamma)^2-(-i\Omega)^2}.
\end{equation}
A rearrangement of this expression yields
\begin{equation}
\label{eqA6}
\delta\hat{x}(w)=\frac{2x_\text{zp}g_0\Omega}{-(w+i\frac{1}{2}\Gamma)^2+\Omega^2}\left\{\bar{n}+\left[\left(\frac{w}{\Omega}+i\frac{\Gamma}{2\Omega}\right)\left(\bar{m}+\frac{1}{2}\right)+\frac{1}{2}\right]\right\}.
\end{equation}
It is straightforward now to see that the term within brackets contributes to the necessary corrections to the spring effect. This will change the mechanical response function $\Sigma(w)$ \cite{SKip1,SAspel1,SBowen,SKip2} as
\begin{equation}
\label{eqA7}
\Sigma(w,\Delta)=2\Omega g_0^2 \left[\frac{1}{(\Delta+w)+\frac{i}{2}\kappa}+\frac{1}{(\Delta-w)-\frac{i}{2}\kappa}\right][\bar{n}+\mu(w)],
\end{equation}
where a term with the dimension of mass in the numerator, which in the following calculation ultimately cancels out, and is equal to the effective motion mass $m_\text{eff}$, is not shown for simplicity. We have also
\begin{equation}
\label{eqA8}
\mu(w)=\frac{1}{\Omega}\left(w+\frac{i}{2}\Gamma\right)\left(\bar{m}+\frac{1}{2}\right)+\frac{1}{2},
\end{equation}
represents corrections to yield the effective cavity photon number $\bar{n}_\text{eff}=\bar{n}+\mu(w)$ because of higher-order interactions. This corrections is easy to see that are important if the pump level is not too high. Typically, for $\bar{n}<10^2$ higher-order spring effects are quite significant, and when $\bar{n}>10^3$ the higher-order effects are suppressed by the standard spring effect.

The spring effect modifies the measured mechanical frequency $\Omega$ and linewidth $\Gamma$ as
\begin{eqnarray}
\label{eqA9}
\delta\Omega(w,\Delta)&=&\frac{1}{2w}\Re[\Sigma(w,\Delta)], \\ \nonumber
\delta\Gamma(w,\Delta)&=&-\frac{1}{w}\Im[\Sigma(w,\Delta)].
\end{eqnarray}
Put together combined, we get
\begin{eqnarray}
\label{eqA10}
\delta\Omega(w,\Delta)&=& \frac{g_0^2\bar{n}\Omega}{w}\left[\frac{\Delta+w}{(\Delta+w)^2+\frac{1}{4}\kappa^2}+\frac{\Delta-w}{(\Delta-w)^2+\frac{1}{4}\kappa^2}\right]\\ \nonumber
&+&\frac{g_0^2\Re[\mu(w)]\Omega}{w}\left[\frac{\Delta+w}{(\Delta+w)^2+\frac{1}{4}\kappa^2}+\frac{\Delta-w}{(\Delta-w)^2+\frac{1}{4}\kappa^2}\right] \\ \nonumber
&+&\frac{g_0^2\Im[\mu(w)]\Omega}{w}\left[\frac{\kappa}{(\Delta+w)^2+\frac{1}{4}\kappa^2}-\frac{\kappa}{(\Delta-w)^2+\frac{1}{4}\kappa^2}\right],\\ \nonumber
\delta\Gamma(w,\Delta)&=&\frac{g_0^2\bar{n}\Omega}{w}\left[\frac{\kappa}{(\Delta+w)^2+\frac{1}{4}\kappa^2}-\frac{\kappa}{(\Delta-w)^2+\frac{1}{4}\kappa^2}\right]\\ \nonumber
&+&\frac{g_0^2\Re[\mu(w)]\Omega}{w}\left[\frac{\kappa}{(\Delta+w)^2+\frac{1}{4}\kappa^2}-\frac{\kappa}{(\Delta-w)^2+\frac{1}{4}\kappa^2}\right]\\ \nonumber
&-&\frac{g_0^2\Im[\mu(w)]\Omega}{w}\left[\frac{\Delta+w}{(\Delta+w)^2+\frac{1}{4}\kappa^2}+\frac{\Delta-w}{(\Delta-w)^2+\frac{1}{4}\kappa^2}\right].
\end{eqnarray}
Here, the second and third terms on the rights hand sides of both equations are corrections to the spring effect due to the higher-order interactions, resulting from the temperature-dependent expressions 
\begin{eqnarray}
\label{eqA11}
\Re[\mu(w)]&=&\frac{w}{\Omega}\left(\bar{m}+\frac{1}{2}\right)+\frac{1}{2}, \\ \nonumber
\Im[\mu(w)]&=&\frac{\Gamma}{2\Omega}\left(\bar{m}+\frac{1}{2}\right).
\end{eqnarray}
The temperature-dependence of (\ref{eqA11}) causes dependence of the spring effect on temperature as well.

The uncorrected standard expressions read \cite{SAspel1,SKip2}
\begin{eqnarray}
\label{eqA12}
\delta\Omega(w,\Delta)&=& \frac{g_0^2\bar{n}\Omega}{w}\left[\frac{\Delta+w}{(\Delta+w)^2+\frac{1}{4}\kappa^2}+\frac{\Delta-w}{(\Delta-w)^2+\frac{1}{4}\kappa^2}\right],\\ \nonumber
\delta\Gamma(w,\Delta)&=&\frac{g_0^2\bar{n}\Omega}{w}\left[\frac{\kappa}{(\Delta+w)^2+\frac{1}{4}\kappa^2}-\frac{\kappa}{(\Delta-w)^2+\frac{1}{4}\kappa^2}\right],
\end{eqnarray}
from which we may observe
\begin{eqnarray}
\label{eqA13}
\delta\Omega(w,\Delta)&=&-\delta\Omega(w,-\Delta),\\ \nonumber
\delta\Omega(-\Delta,\Delta)&=&\frac{g_0^2\bar{n}\Omega}{2}\frac{\Delta}{\Delta^2+\frac{1}{16}\kappa^2},
\end{eqnarray}
which do hold for the standard spring effect at sufficiently high optical powers. 

\subsection{Examples}

As an example, we first assume a side-band resolved cavity with optical resonance wavelength of $\lambda=1\mu\text{m}$ and quality factor $Q=10^6$, a mechanical resonance frequency of $\Omega=2\pi\times 1\text{GHz}$ and quality factor of $Q_\text{m}=10^5$, with the optomechanical interaction rate $g_0=2\pi\times 160\text{kHz}$. Pumping resonantly at the rates $\alpha=10^{10}\text{s}^{-1}$ and $\alpha=10^{11}\text{s}^{-1}$ correspond to $\bar{n}=1.1\times 10^2$ and $\bar{n}=1.1\times 10^4$.

\begin{figure}[ht!]
	\centering
	\includegraphics[width=2.1in]{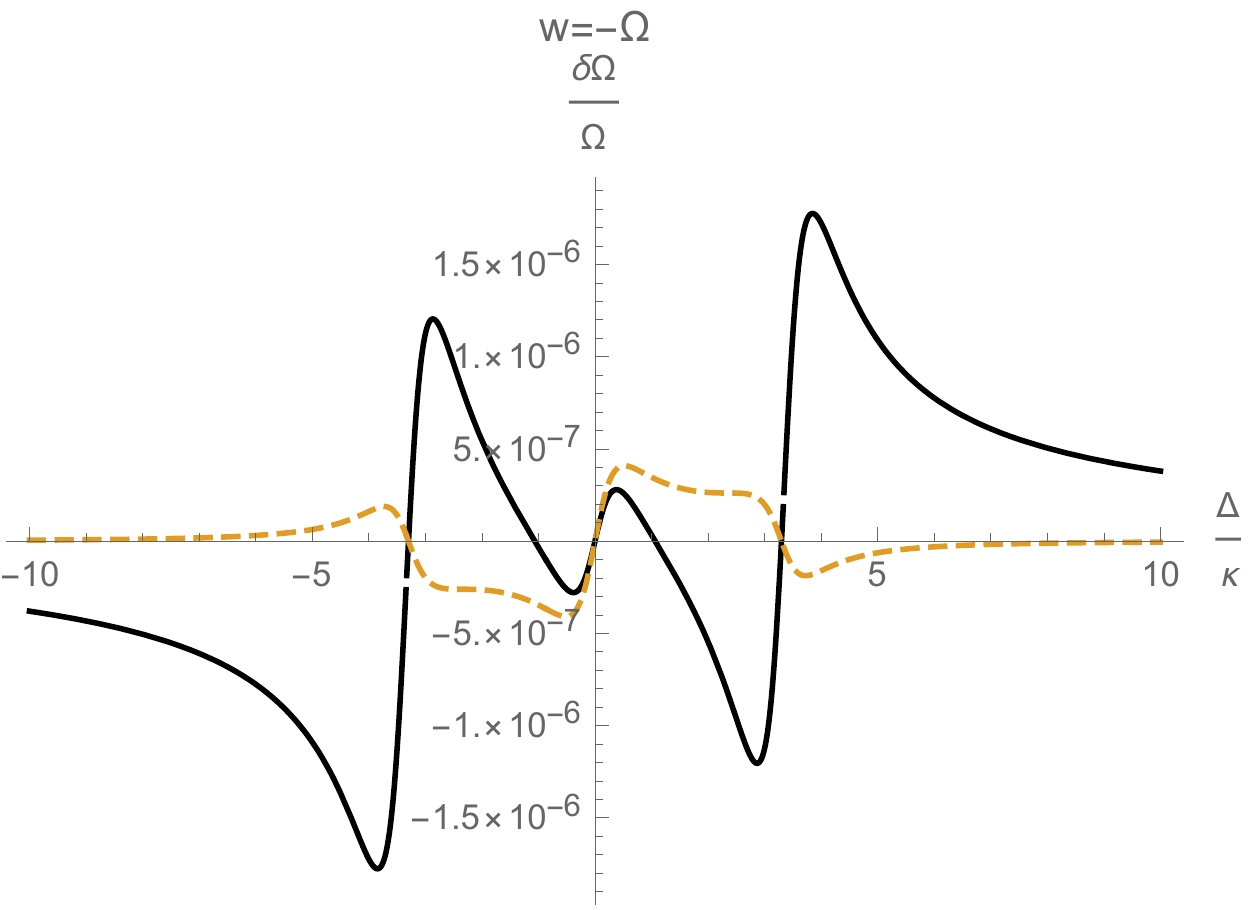}
	\includegraphics[width=2.1in]{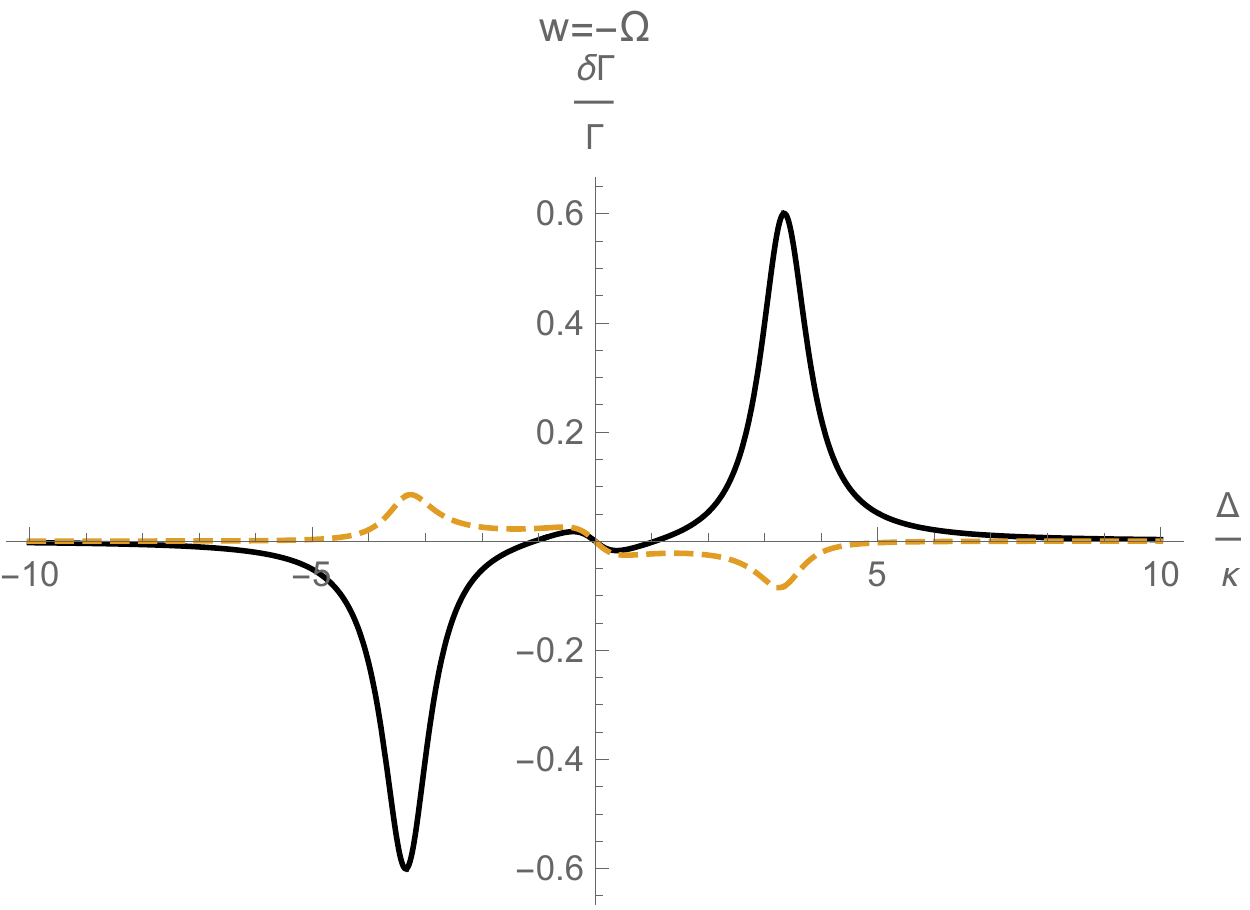} \\
	\includegraphics[width=2.1in]{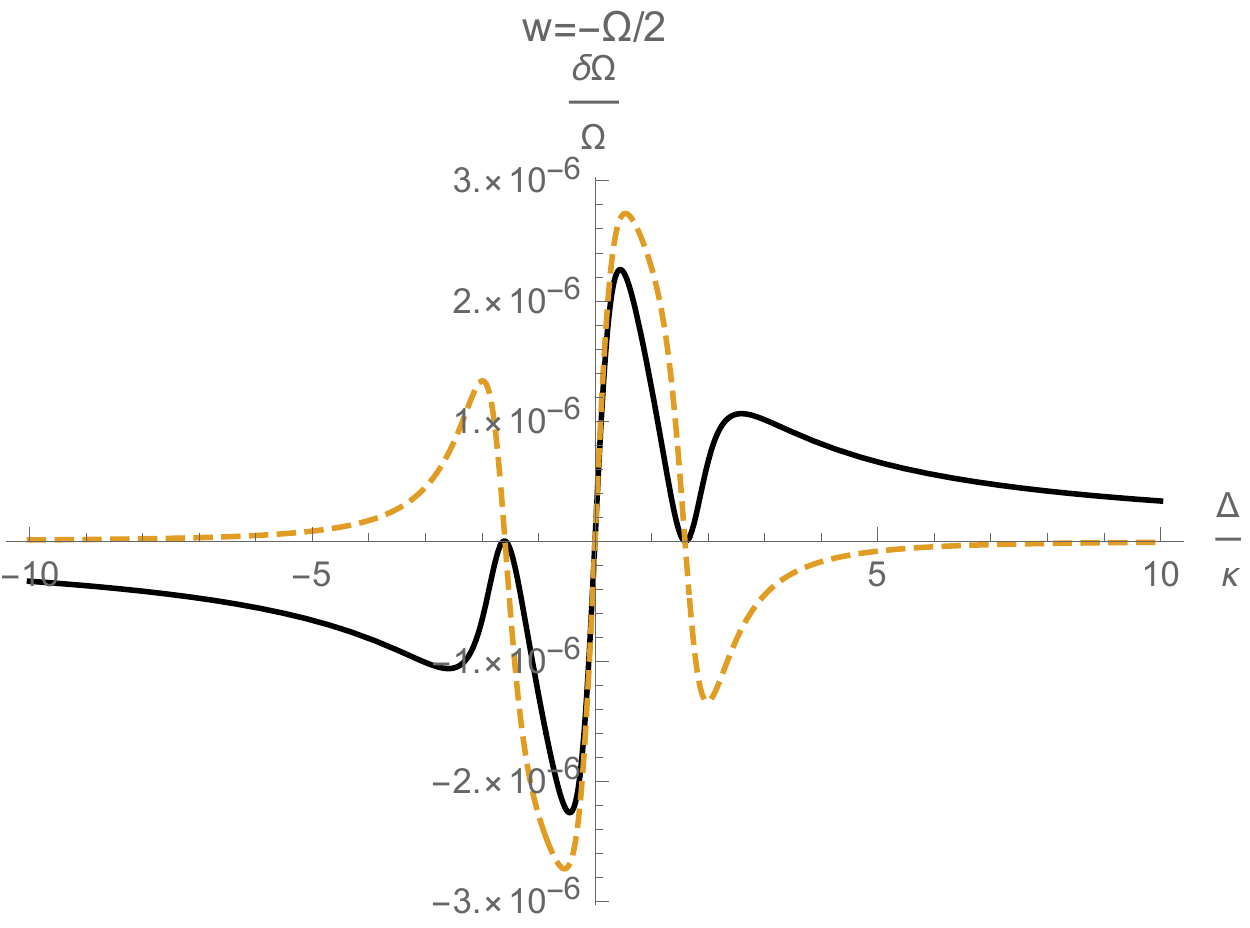} 
	\includegraphics[width=2.1in]{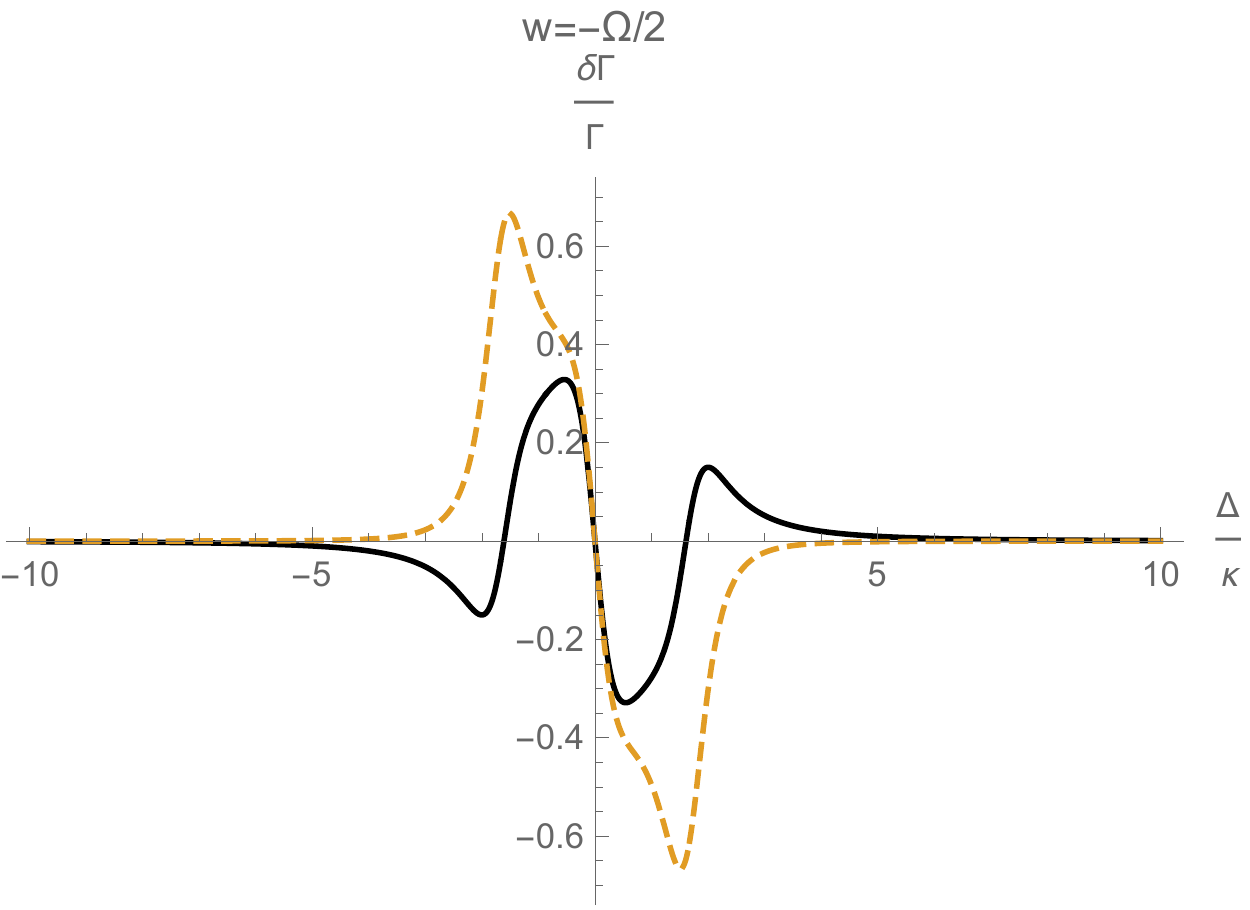} \\
	\includegraphics[width=2.1in]{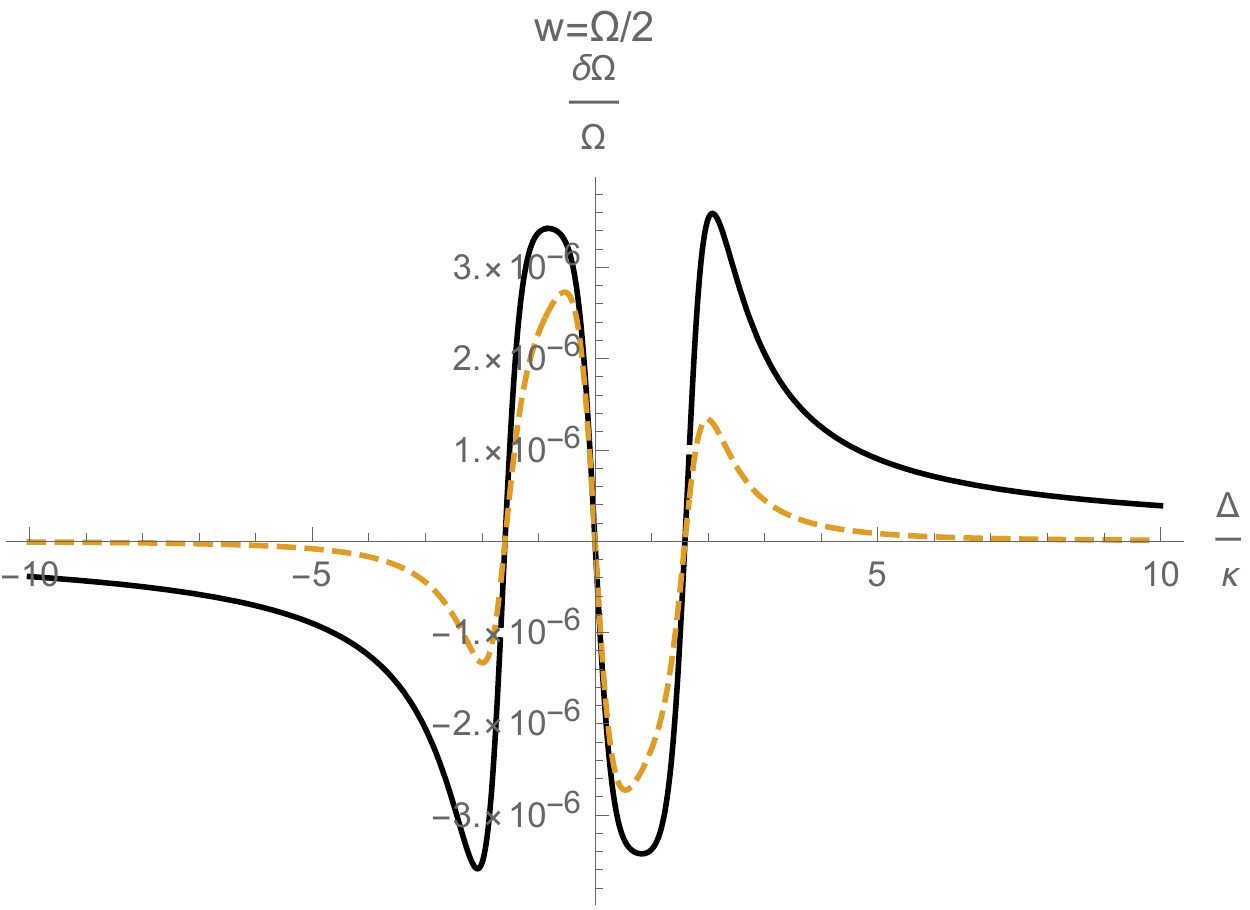} 
	\includegraphics[width=2.1in]{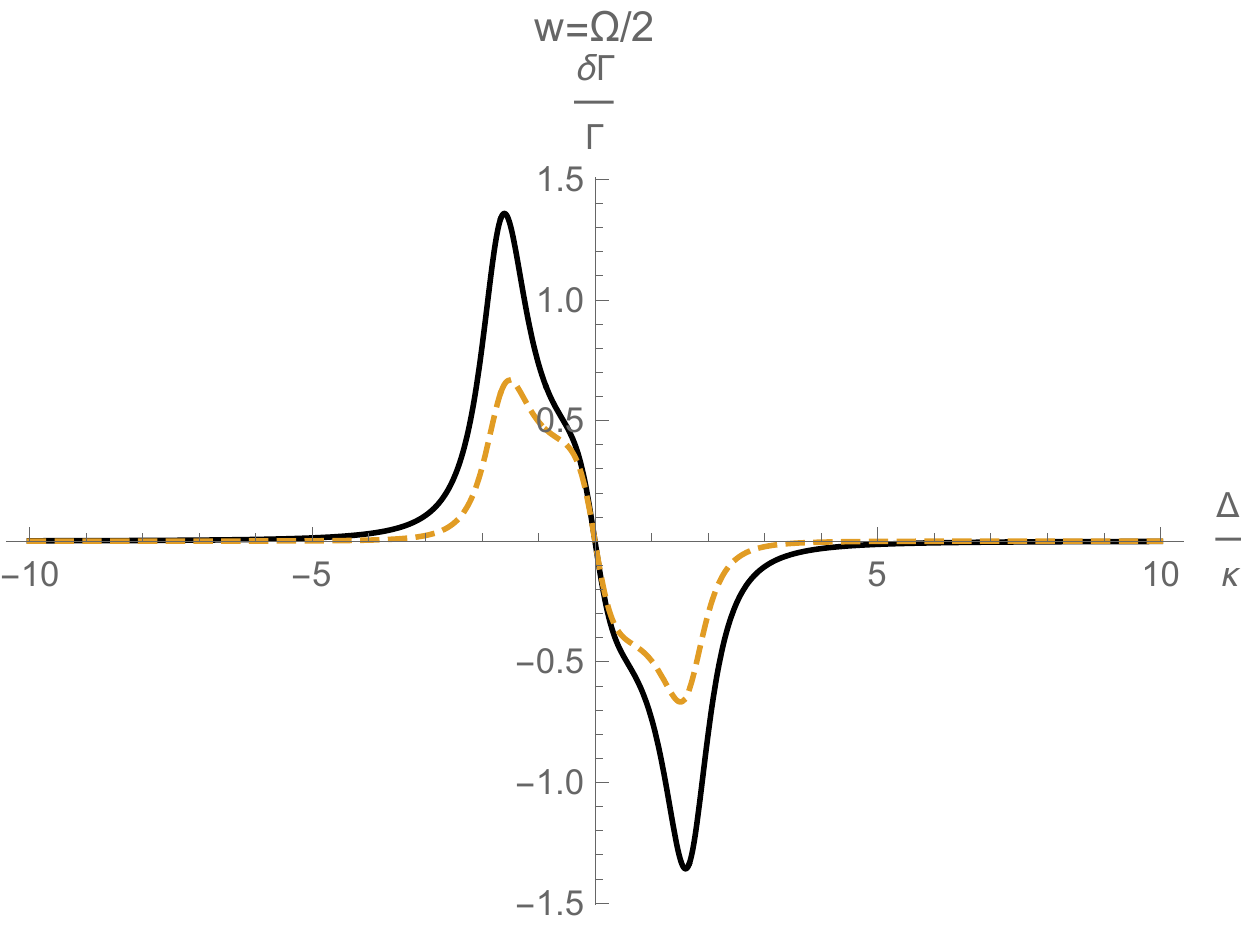} \\
	\includegraphics[width=2.1in]{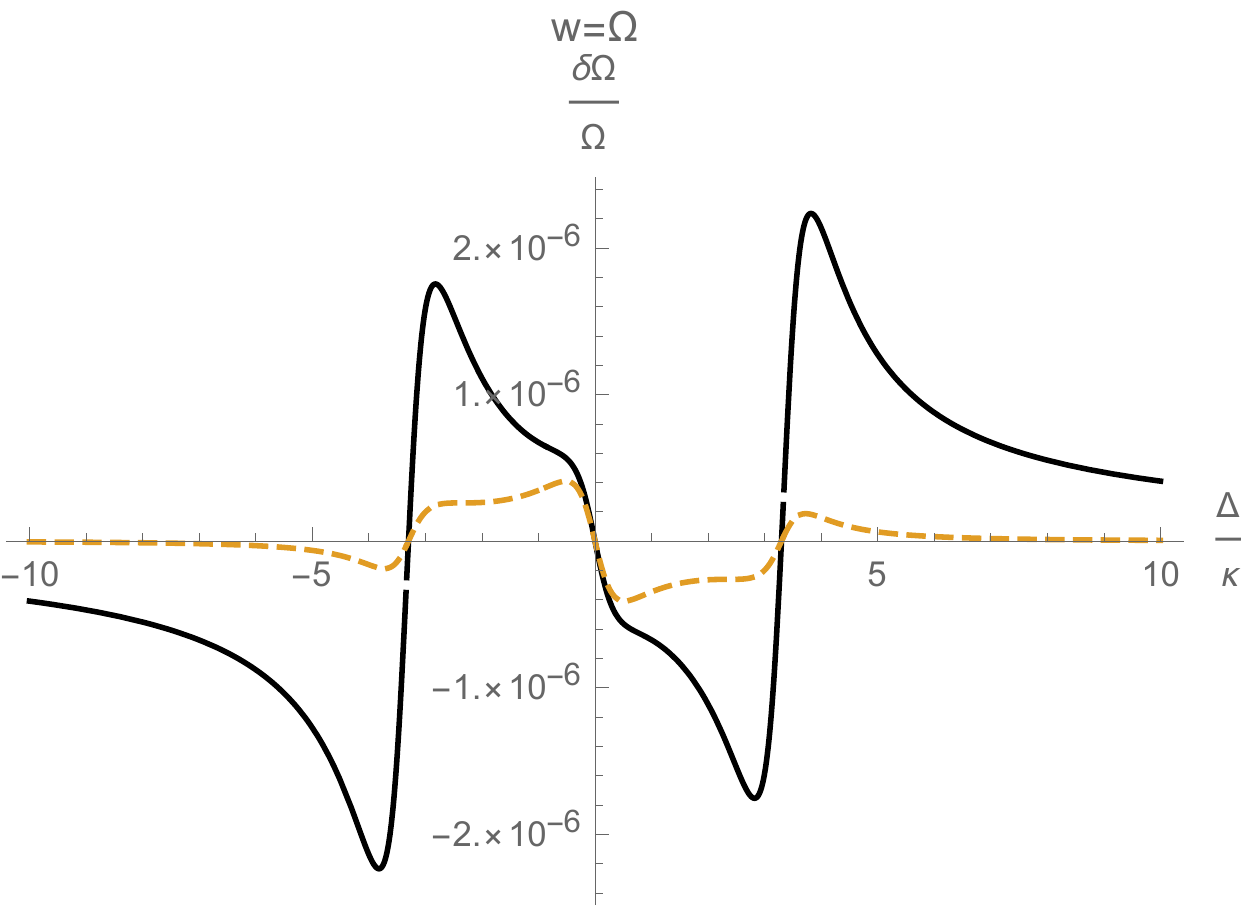} 
	\includegraphics[width=2.1in]{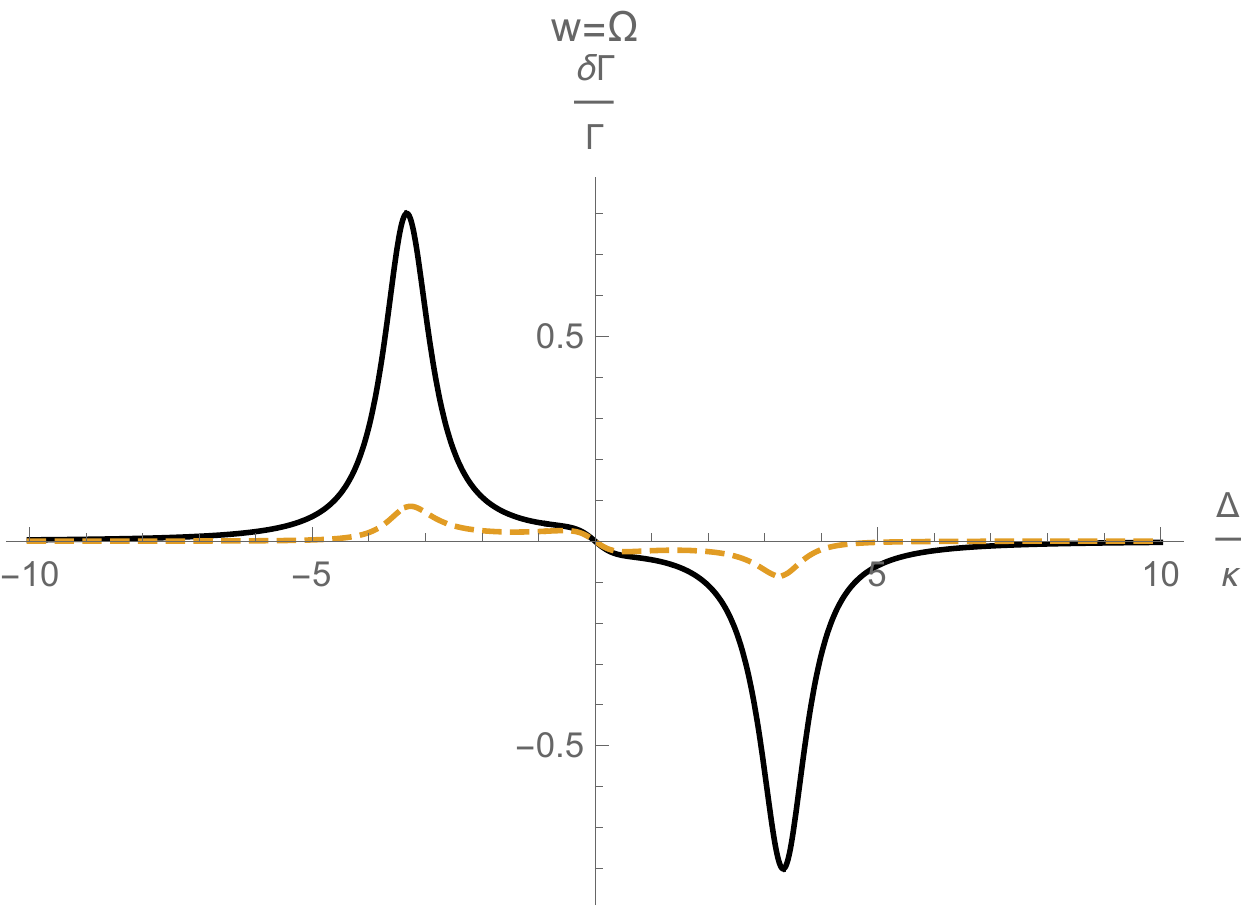} 
	\caption{Optical spring effect due to the standard (dashed) and higher-order interactions (solid black) for a two-beam measurement. From top to the bottom: $w=-\Omega$, $w=-\frac{1}{2}\Omega$, $w=\frac{1}{2}\Omega$, and $w=-\Omega$. Left column corresponds to the change in mechanical frequency $\delta\Omega$ while the right column corresponds to the change in linewidth $\delta\Gamma$. This cavity is side-band resolved and $\alpha=10^{10}\text{s}^{-1}$. \label{FigA1}}
\end{figure}

\begin{figure}[ht!]
	\centering
	\includegraphics[width=2.1in]{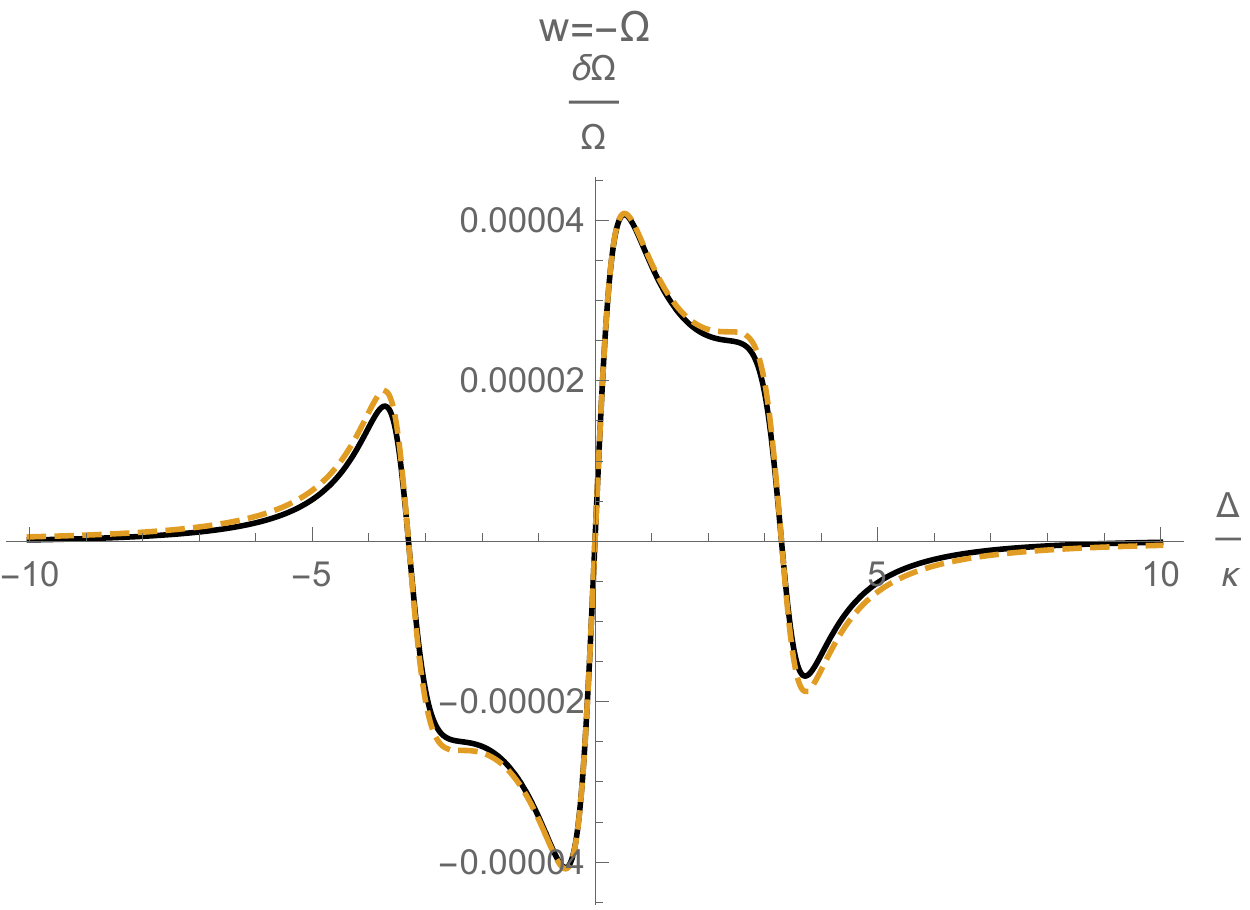}
	\includegraphics[width=2.1in]{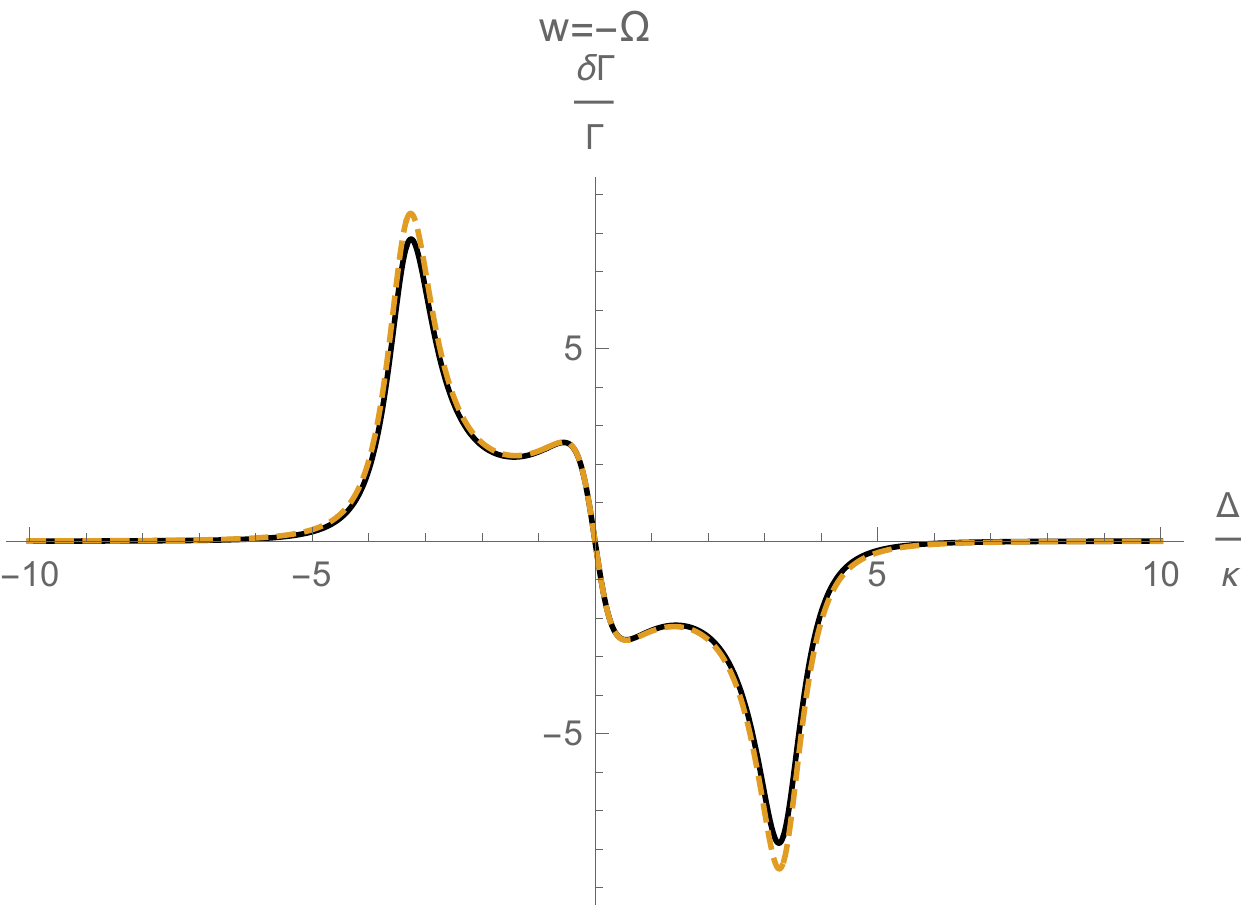} \\
	\includegraphics[width=2.1in]{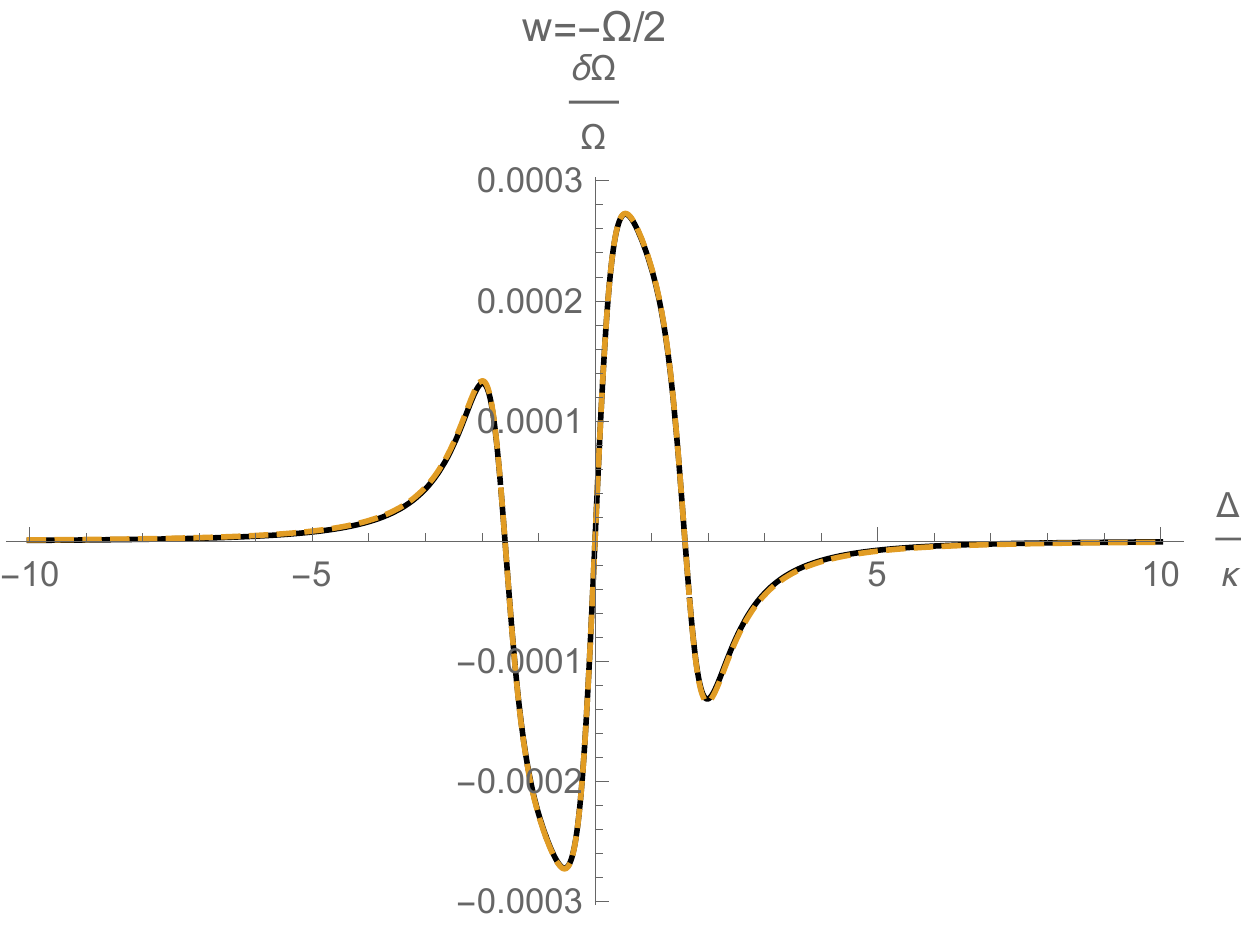} 
	\includegraphics[width=2.1in]{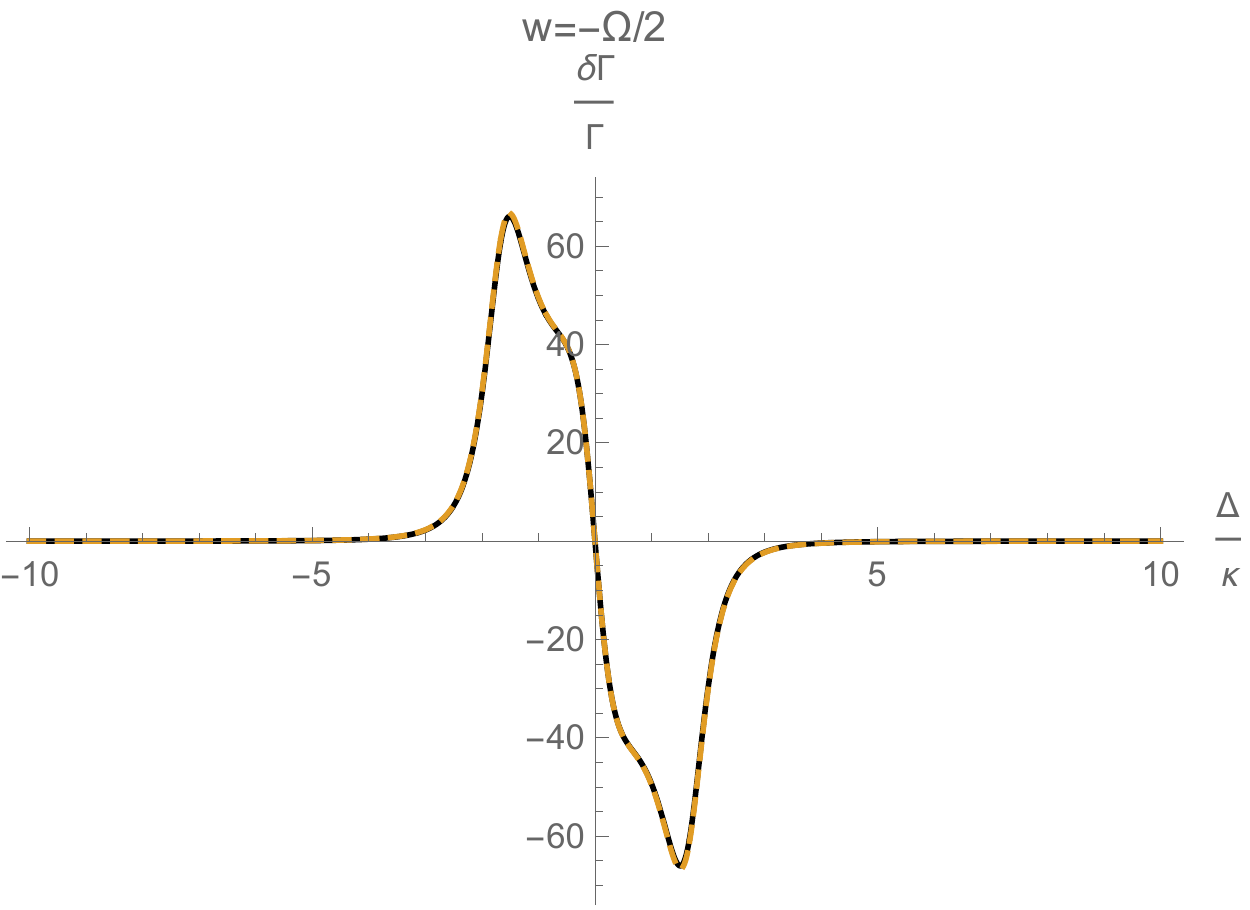} \\
	\includegraphics[width=2.1in]{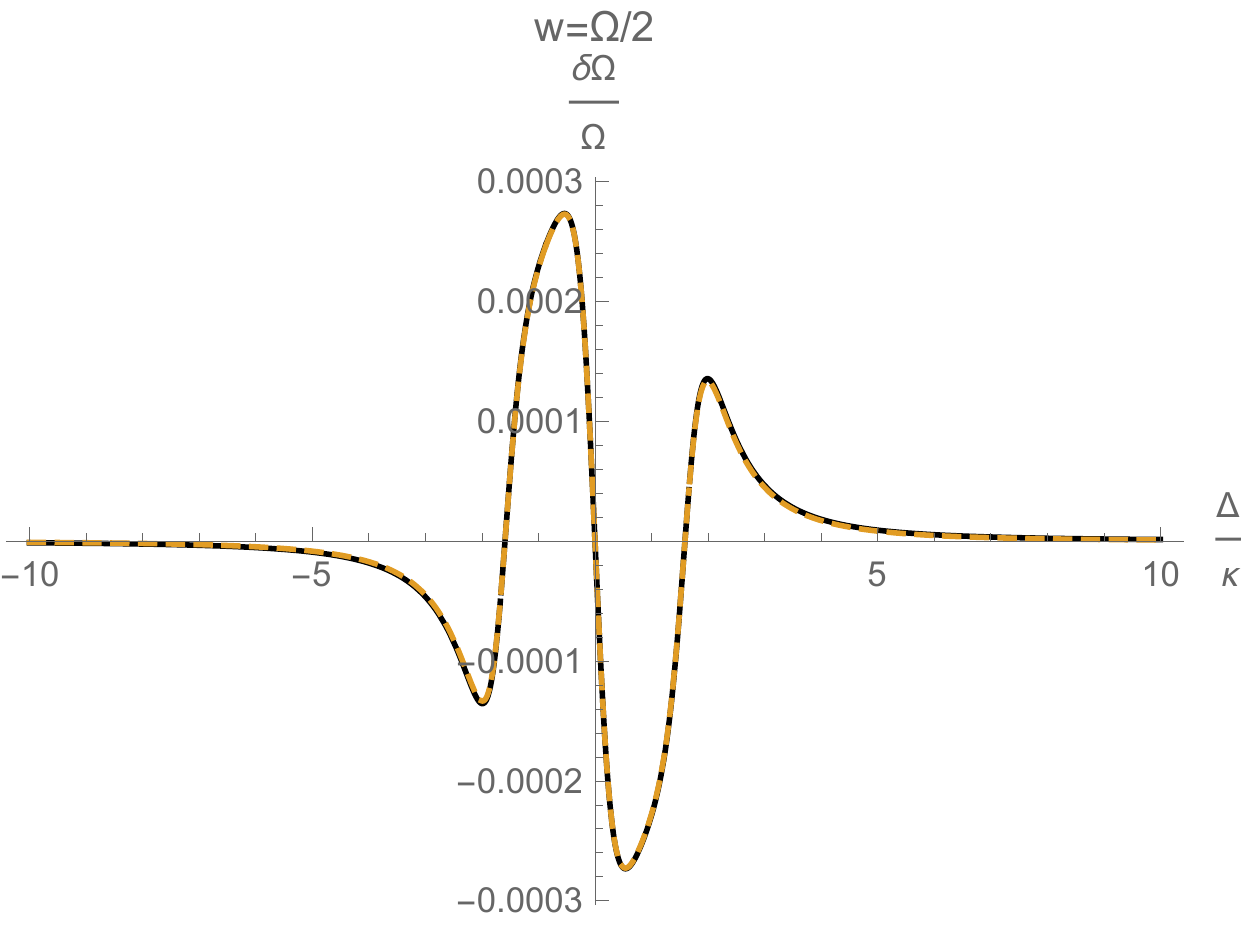} 
	\includegraphics[width=2.1in]{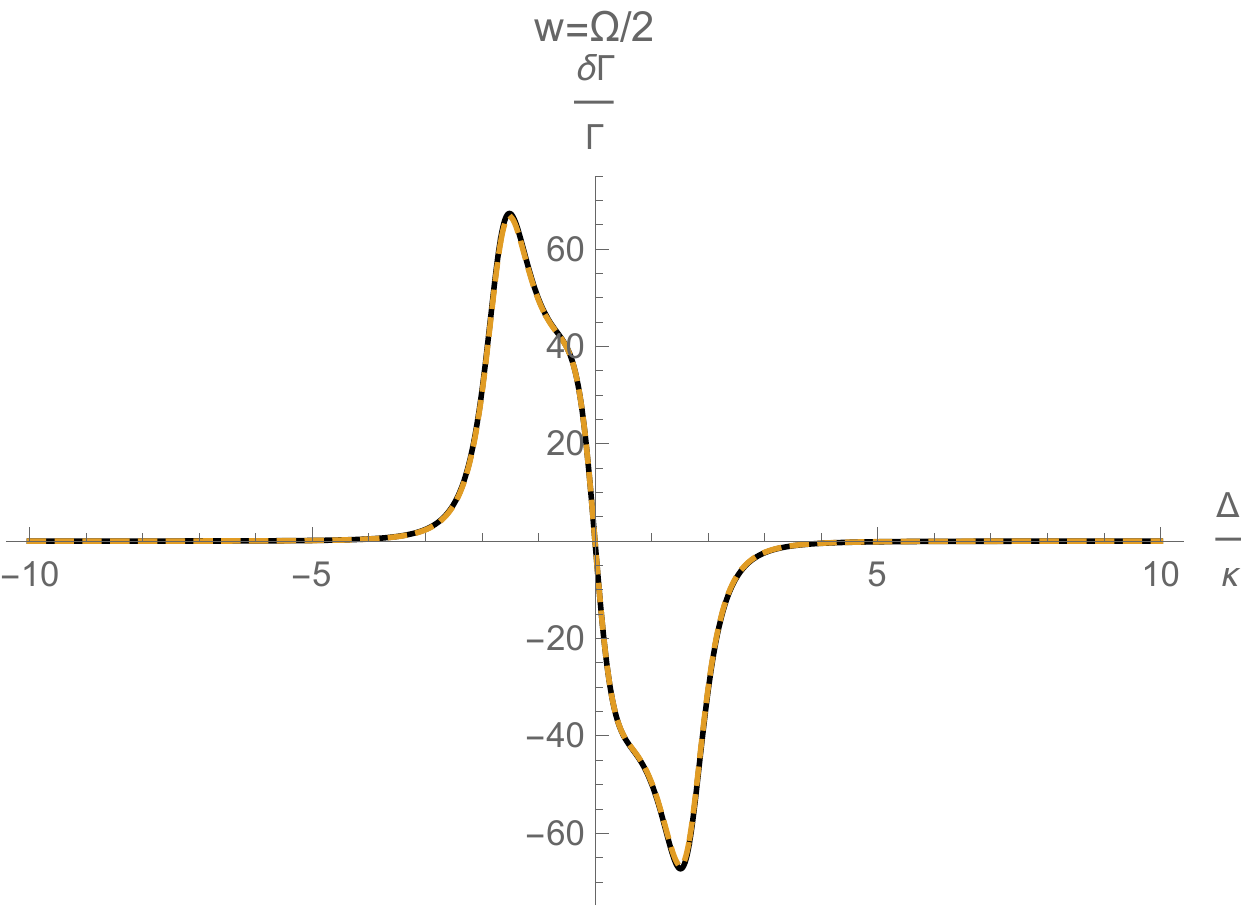} \\
	\includegraphics[width=2.1in]{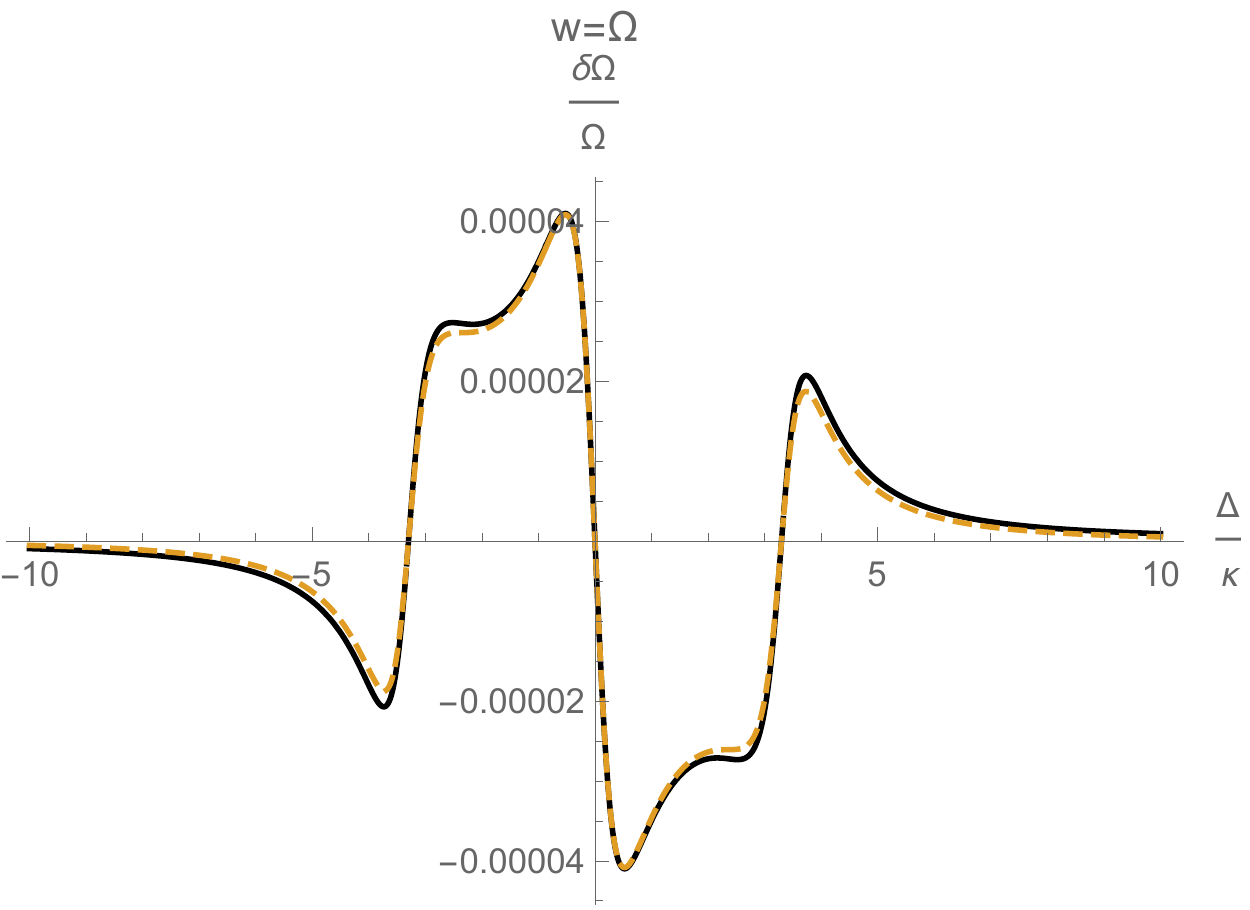} 
	\includegraphics[width=2.1in]{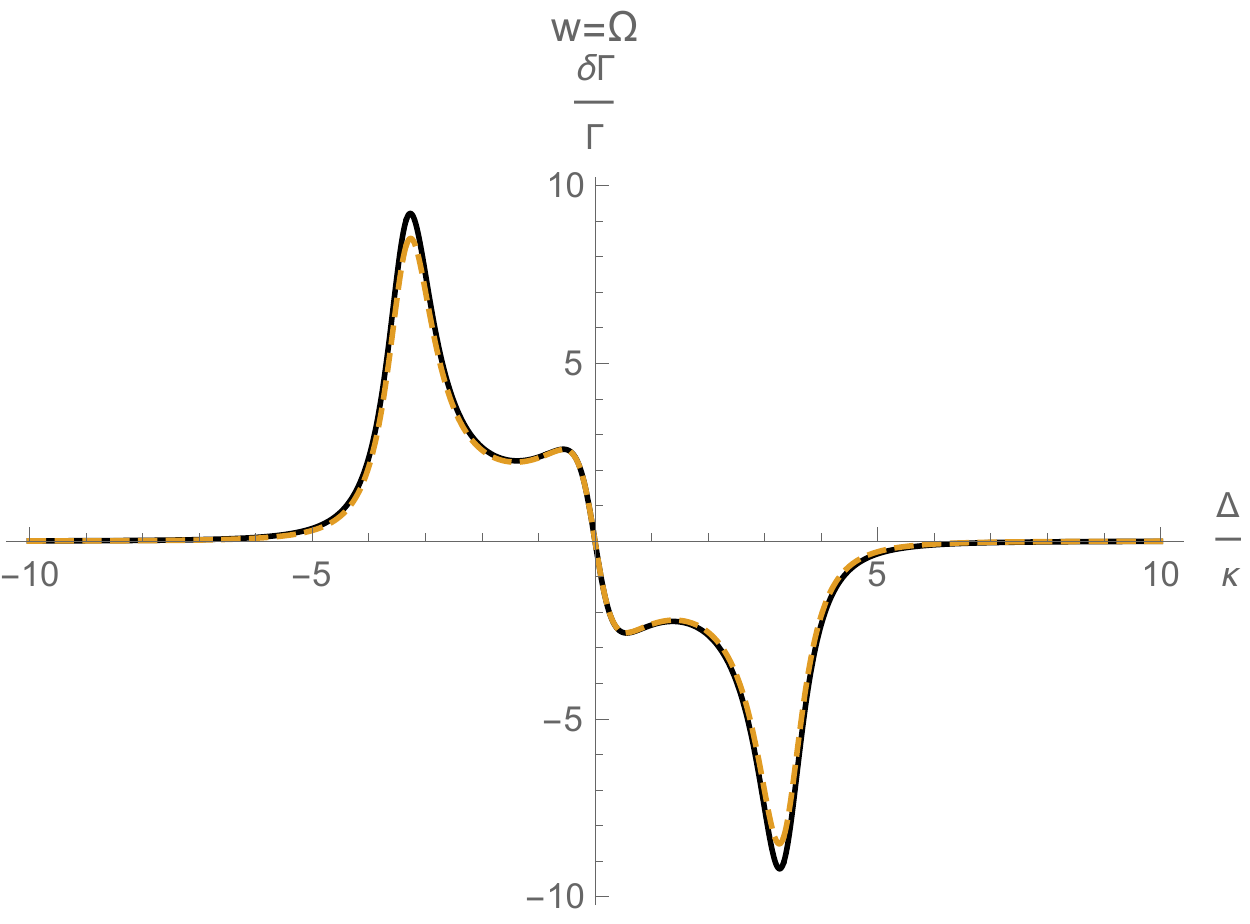} 
	\caption{Optical spring effect due to the standard (dashed) and higher-order interactions (solid black) for a two-beam measurement. From top to the bottom: $w=-\Omega$, $w=-\frac{1}{2}\Omega$, $w=\frac{1}{2}\Omega$, and $w=-\Omega$. Left column corresponds to the change in mechanical frequency $\delta\Omega$ while the right column corresponds to the change in linewidth $\delta\Gamma$. This cavity is side-band resolved and $\alpha=10^{11}\text{s}^{-1}$. \label{FigA2}}
\end{figure}

In Figs. \ref{FigA1} and \ref{FigA2}, optical spring effect has been illustrated at the two above different pumping rates, respectively, for a side-band resolved cavity. It has been supposed that a two-beam pump-probe experiment is undertaken, where the pump frequency is having the detuning $\Delta$ and the relative probe frequency is $w$. 

It can be seen that when the cavity photon number $\bar{n}$ is increased while coherent phonon number $\bar{m}$ is kept constant, the higher-order corrections to the spring effect are just negligible and can be ignored, as shown in Fig. \ref{FigA2}. However, at lower pumping rates where $\bar{n}$ is no longer much larger than $\bar{m}$, the higher-order corrections become important as shown in Fig. \ref{FigA1}. Ultimately, for a few or very low cavity photon number $\bar{n}<10$, the higher-order corrections could be orders of magnitude stronger than the standard effects. 

\begin{figure}[ht!]
	\centering
	\includegraphics[width=2.1in]{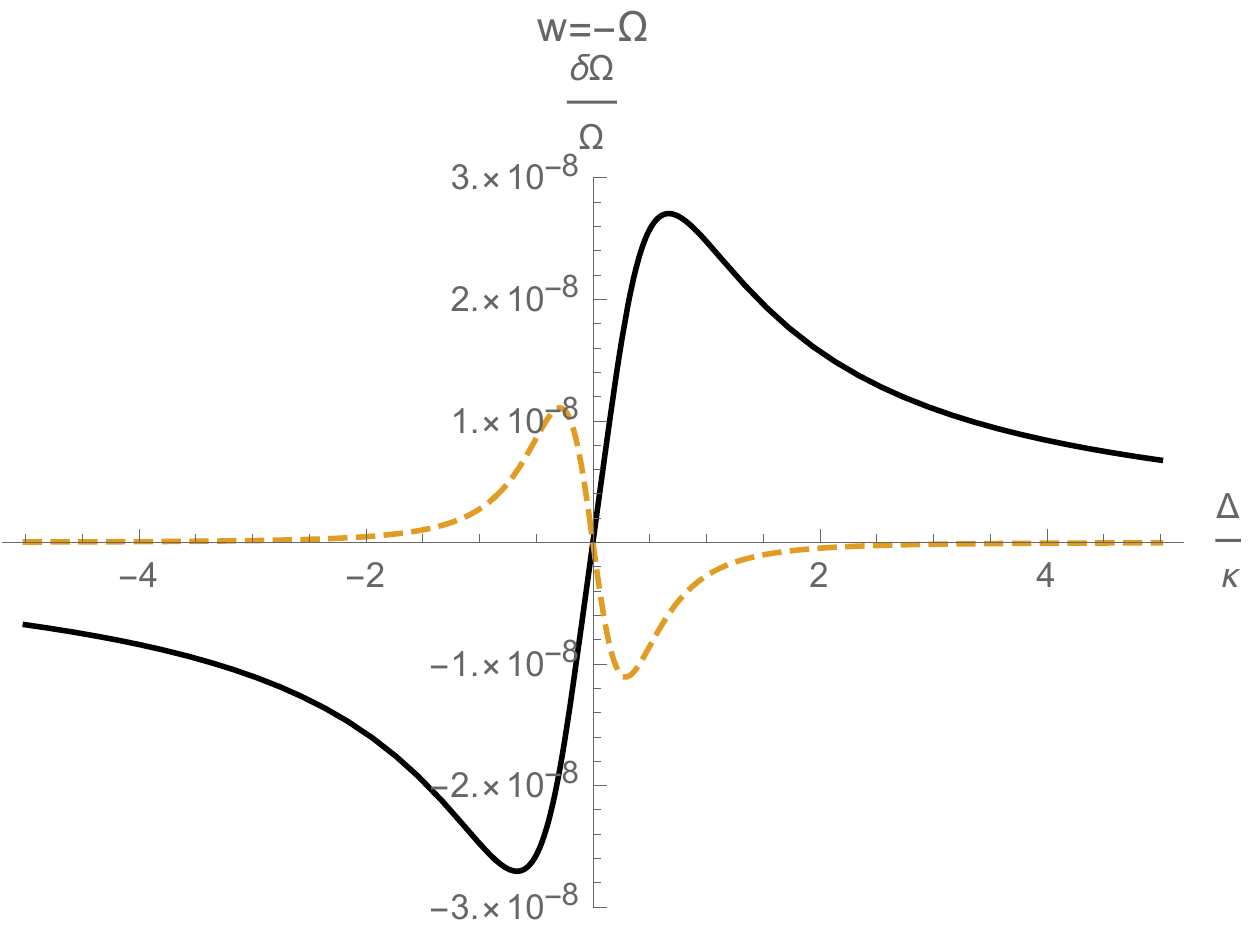}
	\includegraphics[width=2.1in]{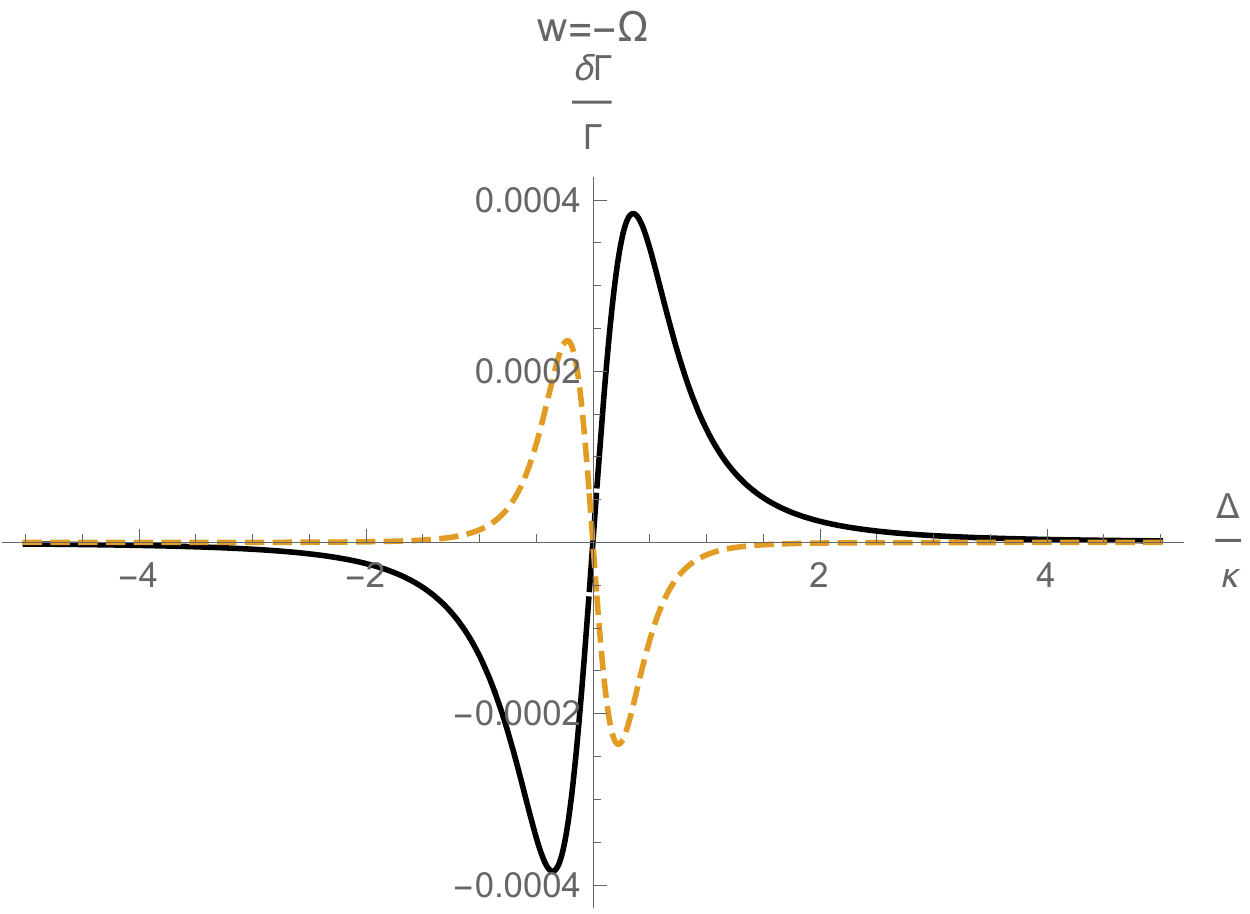} \\
	\includegraphics[width=2.1in]{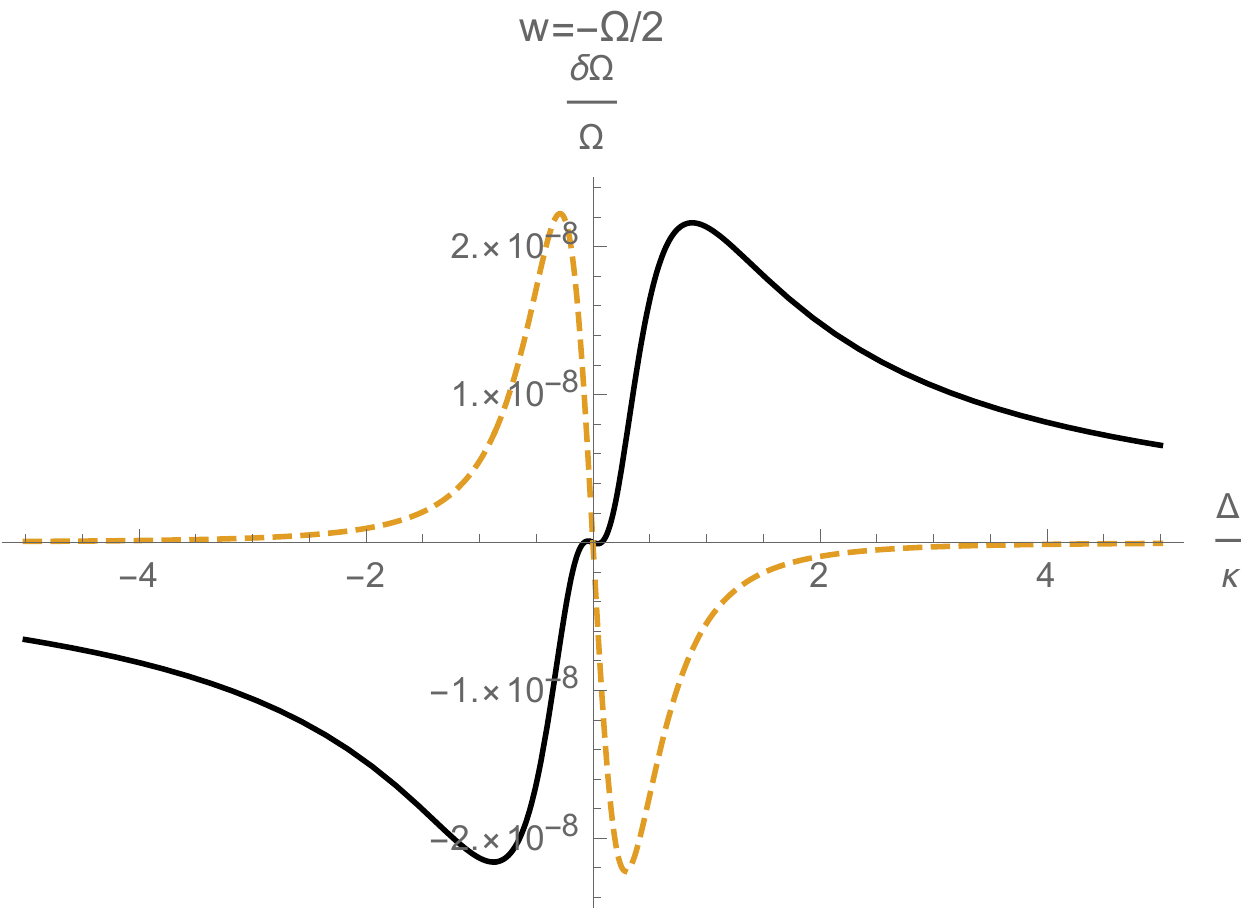} 
	\includegraphics[width=2.1in]{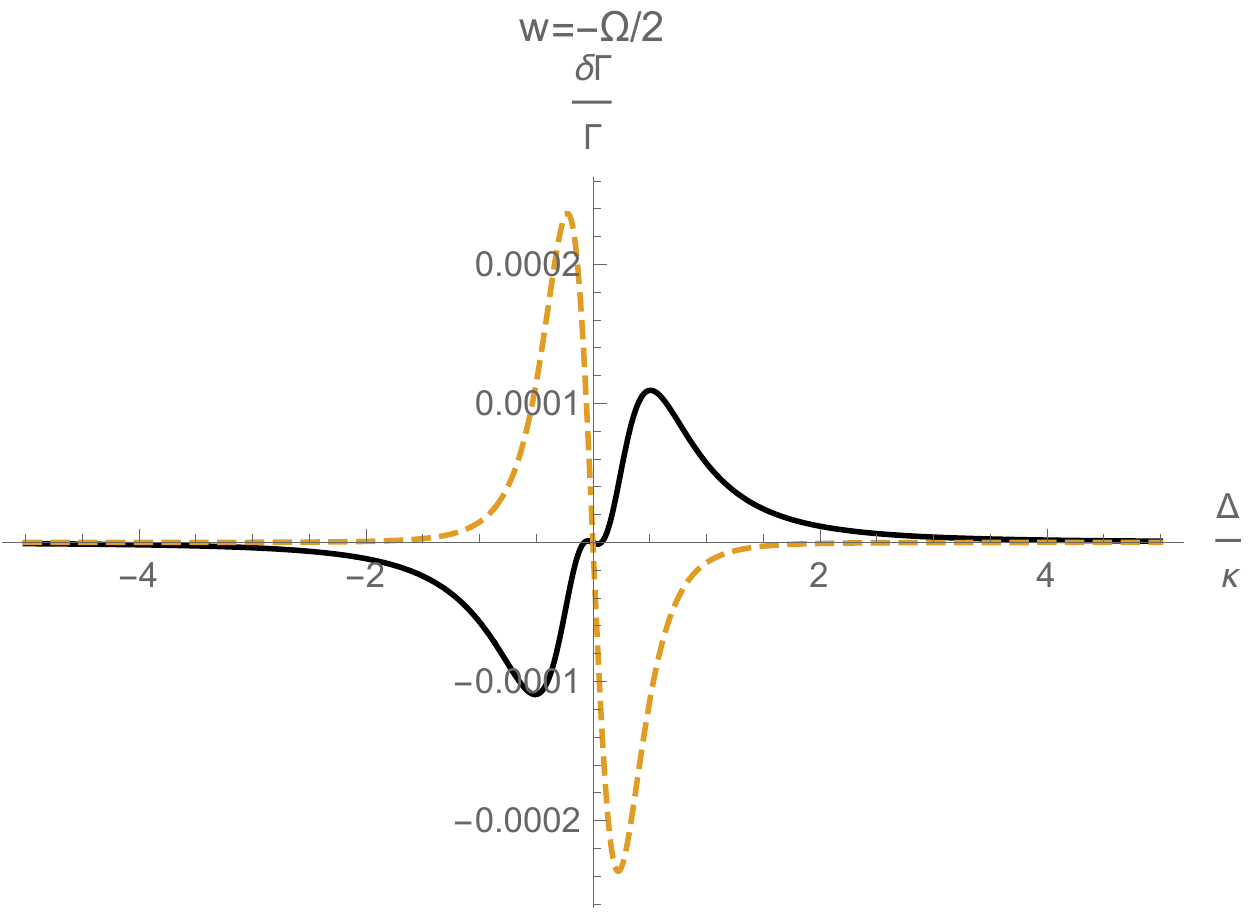} \\
	\includegraphics[width=2.1in]{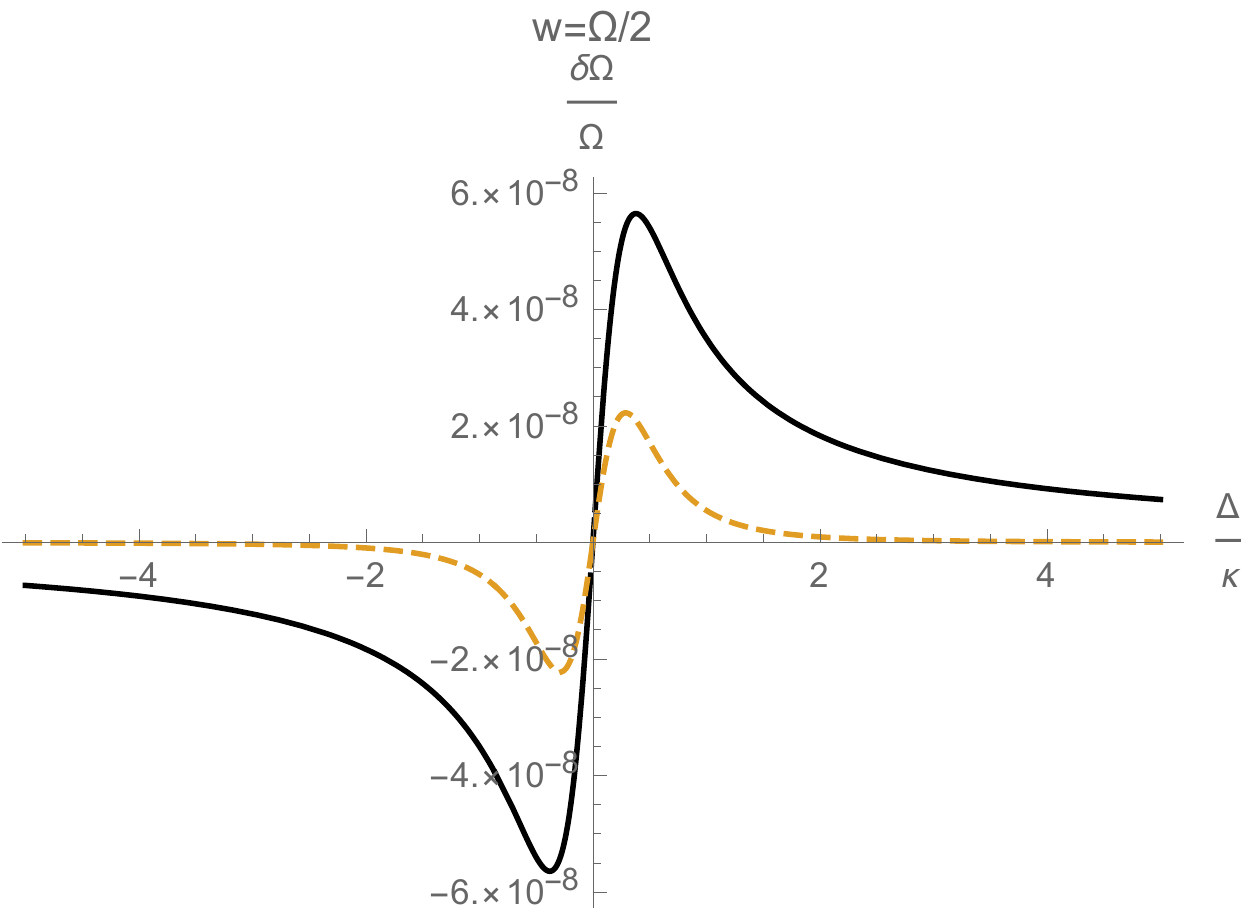} 
	\includegraphics[width=2.1in]{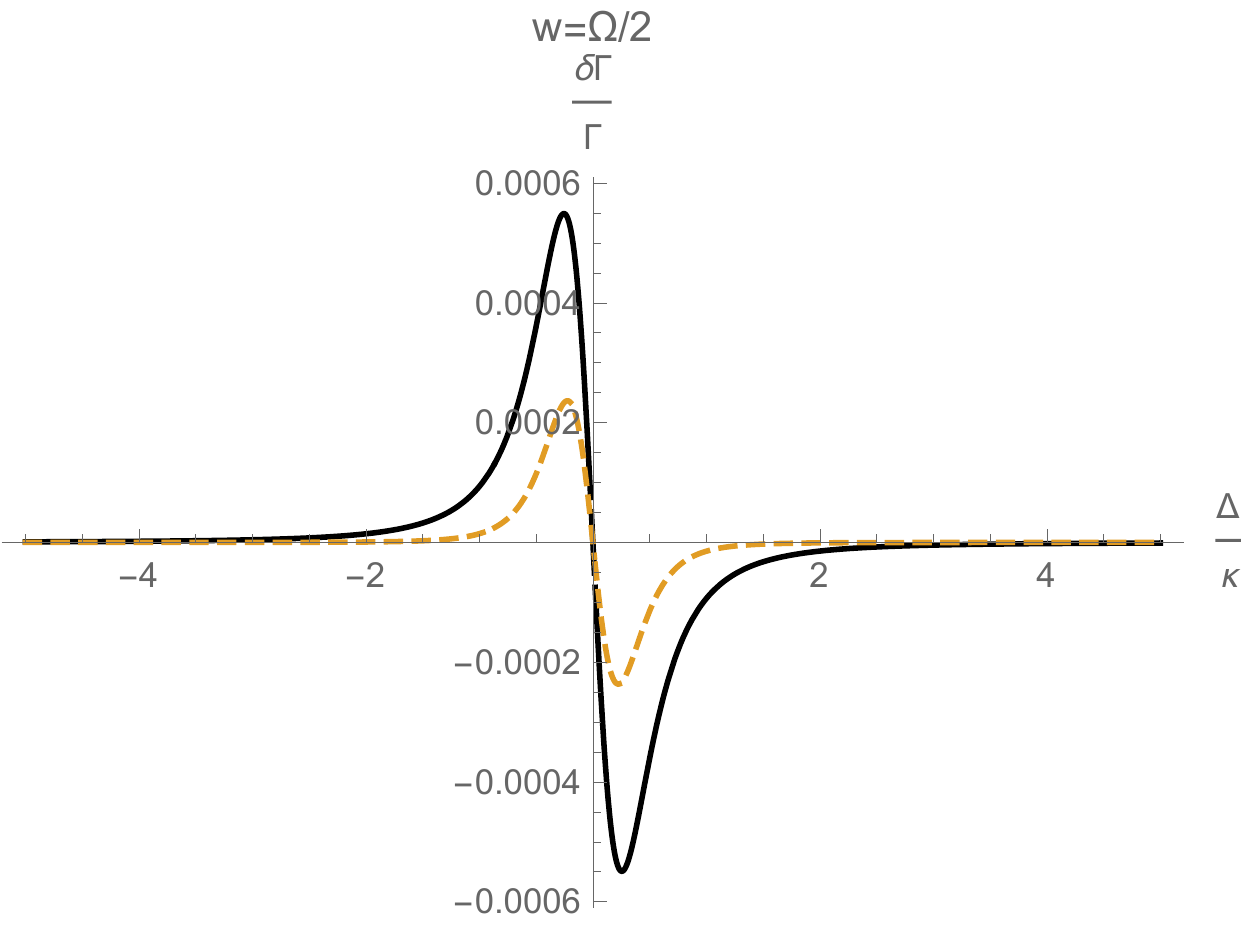} \\
	\includegraphics[width=2.1in]{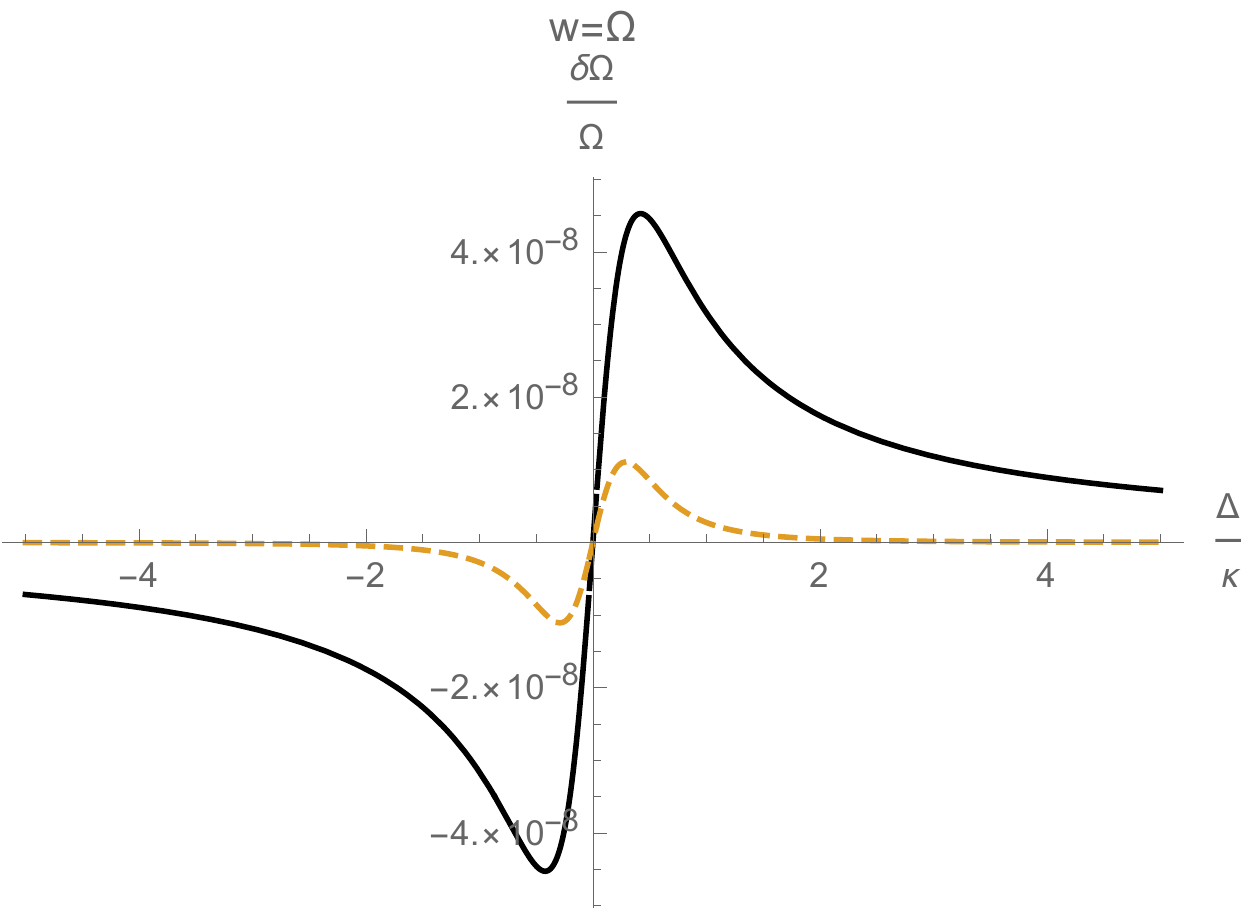} 
	\includegraphics[width=2.1in]{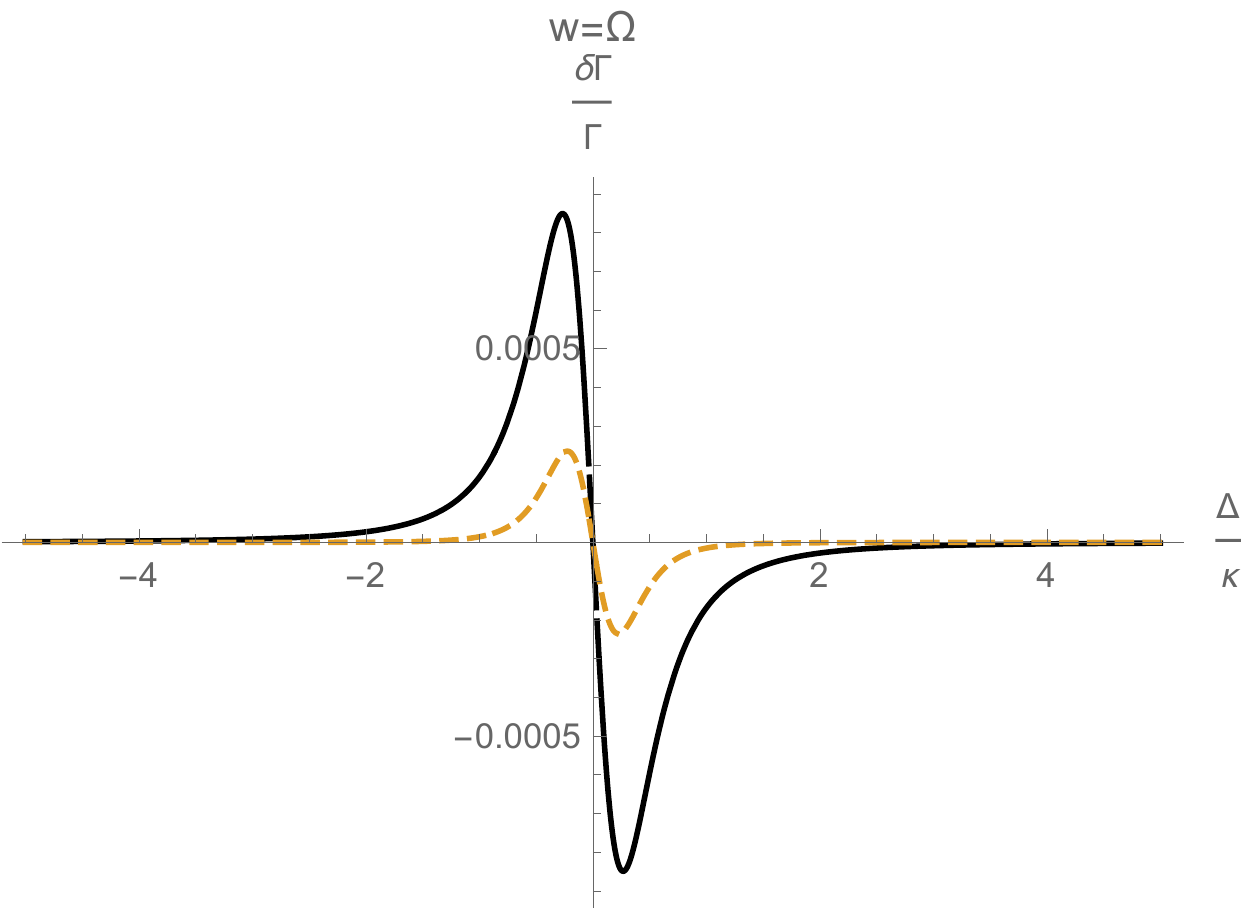} 
	\caption{Optical spring effect due to the standard (dashed) and higher-order interactions (solid black) for a two-beam measurement. From top to the bottom: $w=-\Omega$, $w=-\frac{1}{2}\Omega$, $w=\frac{1}{2}\Omega$, and $w=-\Omega$. Left column corresponds to the change in mechanical frequency $\delta\Omega$ while the right column corresponds to the change in linewidth $\delta\Gamma$. This cavity is in strong Doppler regime and $\alpha=3\times 10^{11}\text{s}^{-1}$. \label{FigA3}}
\end{figure}

In Fig. \ref{FigA3} the same cavity parameters are employed with a lowered optical quality factor of $Q=10^4$. This choice puts the cavity in the Doppler regime. It is easy to see again that in the same manner of the previous examples, sufficiently high pump levels entirely masks the higher-order effects. Taking only the case of $\alpha=3\times 10^{11}\text{s}^{-1}$ here leads to $\bar{n}=1.01\times 10^1$, while at smaller pump levels only the higher-order optomechanical effects survive. At this cavity photon number around $\bar{n}\approx 10$, differences between the standard and higher-order corrected responses are quite prominent and visible.

\section{Minimal Basis}\label{Minimal}

Complete solution of optomechanical interaction $\mathbb{H}_\text{OM}$ can be attained analytically using the minimal basis $\{A\}^{\rm T}=\{\hat{n}^2,\hat{n}\hat{b},\hat{n}\hat{b}^\dagger\}=\{\hat{N},\hat{B},\hat{B}^\dagger\}$. Construction of Langevin equations leads to the system
\begin{equation}
\label{eq21}
\frac{d}{dt}\left\{\begin{array}{c}
\hat{N}\\
\hat{B}\\
\hat{B}^\dagger
\end{array}\right\}=
\left[
\begin{array}{ccc}
-2\kappa & 0 & 0\\
ig_0 & -i\Omega-\frac{\gamma}{2} & 0\\
ig_0 & 0 & i\Omega-\frac{\gamma}{2}
\end{array}
\right]
\left\{\begin{array}{c}
\hat{N}\\
\hat{B}\\
\hat{B}^\dagger
\end{array}\right\}-\left\{\begin{array}{c}
\sqrt{4\kappa}\hat{N}_{\rm in}\\
\sqrt{\gamma}\hat{B}_{\rm in}\\
\sqrt{\gamma}\hat{B}_{\rm in}^\dagger
\end{array}\right\}.
\end{equation}
\noindent
Here, the multiplicative noise terms are defined as 
\begin{eqnarray}
\label{eq22}
\sqrt{4\kappa}\hat{N}_\text{in}&=&2\sqrt{\kappa}\left(\hat{n}\hat{a}^\dagger\hat{a}_\text{in}+\hat{a}_\text{in}^\dagger\hat{a}\hat{n}\right),\\ \nonumber
\sqrt{\gamma}\hat{B}_\text{in}&=&\sqrt{2\kappa}\hat{b}\hat{n}_\text{in}+\sqrt{\Gamma}\hat{n}\hat{b}_\text{in},
\end{eqnarray} 
\noindent
where $\hat{n}_\text{in}$ is already defined under (\ref{eq4}), and the spectral density of which can be estimated using the method described elsewhere \cite{SLoo}. A very effective method to deal with multiplicative noise is to be discussed in \S\ref{Noise}. This can be immediately noticed to be reducible as 
\begin{equation}
\label{eq23}
\frac{d}{dt}\left\{\begin{array}{c}
\hat{N}\\
\hat{B}
\end{array}\right\}=
\left[
\begin{array}{cc}
-2\kappa & 0 \\
ig_0 & -i\Omega-\frac{\gamma}{2}
\end{array}
\right]
\left\{\begin{array}{c}
\hat{N}\\
\hat{B}
\end{array}\right\}-\left\{\begin{array}{c}
\sqrt{4\kappa}\hat{N}_{\rm in}\\
\sqrt{\gamma}\hat{B}_{\rm in}
\end{array}\right\}.
\end{equation}
\noindent
These will make the evaluation of spectral densities $S_{NN}(\omega)$ and $S_{BB}(\omega)$ possible. Interestingly, (\ref{eq23}) is actually decoupled, since the equation for $\hat{N}$ is already independent of $\hat{B}$, which admits the solution
\begin{equation}
\label{eq24}
\hat{N}(t)=\hat{N}(0)e^{-2\kappa t}-2\sqrt{\kappa}e^{-2\kappa t}\int_{0}^{t}\hat{N}_\text{in}(\tau)e^{2\kappa \tau}d\tau.
\end{equation}
\noindent
We can be now plug (\ref{eq24}) in the second equation of (\ref{eq23}) to solve exactly for $\hat{B}$. We define $\vartheta=i\Omega+\frac{\gamma}{2}$ and may write down
\begin{equation}
\label{eq25}
\hat{B}(t)=\hat{B}(0)e^{-\vartheta t}-e^{-\vartheta t}\int_{0}^{t}e^{\vartheta \tau}\left[ig_0\hat{N}(\tau)+\sqrt{\gamma}\hat{B}_\text{in}(\tau)\right]d\tau.
\end{equation}

The treatment of multiplicative noise terms (\ref{eq22}) can be quite difficult in the most general form, especially that they demand prior knowledge of photonic and phononic ladder operators. However, assuming that the extra ladder operators can be replaced by their mean values, we can do the zeroth order approximations 
\begin{eqnarray}
\label{eq26}
\sqrt{4\kappa}\hat{N}_\text{in}&\approx&\sqrt{\kappa\bar{n}}\bar{n}\left(\hat{a}_\text{in}+\hat{a}_\text{in}^\dagger\right)\to 2\sqrt{\kappa\bar{n}}\bar{n}\check{a}_\text{in},\\ \nonumber
\sqrt{\gamma}\hat{B}_\text{in}&\approx&\sqrt{\kappa\bar{n}}\bar{b}\left(\hat{a}_\text{in}+\hat{a}_\text{in}^\dagger\right)+\sqrt{\Gamma}\bar{n}\hat{b}_\text{in}\to 2\sqrt{\kappa\bar{n}}\bar{b}\check{a}_\text{in}+\sqrt{\Gamma}\bar{n}\hat{b}_\text{in}.
\end{eqnarray} 
\noindent
Here, the real-valued Weiner process $\check{a}_\text{in}(t)$ with the \textit{symmetrized} classical spectral density $\check{a}_\text{in}(\omega)$ is obtained as
\begin{eqnarray}
\label{eq27}
\check{a}_\text{in}(t)&=&\frac{\hat{a}_\text{in}(t)+\hat{a}_\text{in}^\dagger(t)}{2},\\ \nonumber
\check{a}_\text{in}(\omega)&=&\Re[\hat{a}_\text{in}(\omega)].
\end{eqnarray}

While this type of approximations in multiplicative noise could be useful for many cases, there are some phenomena which cannot be reproduced without correct treatment of multiplicative noise. This shall be discussed in details in \S\ref{Noise}. Nevertheless, it is also instructive the take the expectation values of (\ref{eq21}) to obtain the classical system
\begin{equation}
\label{eq28}
\frac{d}{dt}\left\{\begin{array}{c}
N(t)\\
B(t)
\end{array}\right\}=
\left[
\begin{array}{cc}
-2\kappa & 0 \\
ig_0 & -i\Omega-\frac{\gamma}{2}
\end{array}
\right]
\left\{\begin{array}{c}
N(t)\\
B(t)\\
\end{array}\right\}+2\sqrt{\bar{n}}\left\{\begin{array}{c}
\bar{n}\\
\bar{b}
\end{array}\right\}\Re[\alpha].
\end{equation}
\noindent
Together with (\ref{eq9},\ref{eq12}), and setting the time-derivative on the left of the above to zero, makes the evaluation of steady-state values $N(\infty)=\bar{n}^2$ and $B(\infty)=\overline{nb}\approx\bar{n}\bar{b}$ readily possible. Doing this and solving for $\bar{n}$ and $\bar{b}$ precisely gives back (\ref{eq9}). This not only is in agreement with the equilibrium equation (\ref{eq12}), but also confirms the general finding that the equilibrium intracavity photon population $\bar{n}(\Delta)$ is independent of the coherent phonon population $\bar{m}(\Delta)$. However, the opposite is not correct, and as it was shown in the previous sections, $\bar{m}(\Delta)$ can actually be either determined from $\bar{n}(\Delta)$ and fitting to the experimental data, or directly estimated using the expression (\ref{m8}) in \S\ref{BMWm8}. 

Existence of such an exact transformation which puts the optomechanical interaction into exactly linear form should be connected to the polaron transformation \cite{SAspel1} which leaves behind a Kerr nonlinear term as $\hat{n}^2$ in the transformed optomechanical Hamiltonian. It furthermore highlights the fact that usage of higher-order operators ultimately can reach a fully linear system at which convergence of this method to the exact solution is evident.

\section{Higher-Order Sidebands}

When the optical frequency is much larger than the mechanical frequency, apart from the mechanical sidebands which are roughly placed at $\Delta^{(1)}\approx\pm\Omega$, there exist higher-order sidebands such as $\Delta^{(2)}\approx\pm2\Omega$ and so on. The occurrence of these higher-order sidebands, which are observable for sideband-resolved experiments, is obviously stringent on the existence of two- and multi-phonon processes. Normally, one would expect that these could be studied by constructing the Langevin equations for the operators $\hat{b}^2$, $\hat{b}^{\dagger 2}$ and so on. But this guess turns out to be incorrect, since the corresponding Langevin equations would be totally independent of the one for $\hat{a}$, implying that the second- and higher-order sidebands could not be reconstructed via the fully linearized Langevin equations. This has already been shown to be a nonlinear process which does not naturally appear in the Hamiltonian of the fully linearized optomechanics \cite{SGirvin}. But this difficulty can be appropriately addressed by the method of Higher-order Operators, too.

In order to investigate this phenomenon, let us restrict the case only to the second-order sidebands roughly located at $\Delta^{(2)}\approx\pm2\Omega$. In order to study these, it is sufficient to extend the basis $\{A\}^\text{T}=\left\{\hat{a},\hat{a}\hat{b},\hat{a}{\hat{b}}^{\dagger }\right\}$ to 
\begin{equation}
\label{eq54}
\{A\}^{\rm T}=\{\hat{a},\hat{a}\hat{b},\hat{a}\hat{b}^\dagger,\hat{a}\hat{b}^2,\hat{a}\hat{b}^{\dagger 2} \},
\end{equation}
where the third-rank higher-order operators $\hat{a}\hat{b}^2$ and $\hat{a}\hat{b}^{\dagger 2}$ take care of the one-photon two-phonon processes, ultimately leading to formation of second-order sidebands at $\Delta^{(2)}\approx\pm2\Omega$. The Langevin equations for this basis within the zeroth-order approximation of multiplicative noise reads
\begin{eqnarray}
\nonumber
\left[ 
\begin{array}{ccc|cc}
i\Delta-\frac{\kappa }{2} & ig_0 & ig_0 & 0 & 0 \\ 
ig_0(\bar{m}+\bar{n}+1) & -i(\Omega-\Delta)-\frac{\gamma }{2} & 0 & ig_0 & 0 \\ 
ig_0(\bar{m}-\bar{n}) & 0 & i(\Omega+\Delta)-\frac{\gamma }{2} & 0 & ig_0 \\ \hline
0 & ig_0(\bar{m}+2\bar{n}+2) & 0 & -i(2\Omega-\Delta)-\frac{\theta}{2} & 0 \\ 
0 & 0 & ig_0(\bar{m}-2\bar{n}-1) & 0 & i(2\Omega+\Delta)-\frac{\theta}{2}  
\end{array}
\right]\\ \label{eq55}
\times
\left\{ \begin{array}{c}
\hat{a} \\ 
\hat{a}\hat{b} \\ 
\hat{a}{\hat{b}}^{\dagger } \\ \hline
\hat{a}\hat{b}^2 \\
\hat{a}\hat{b}^{\dagger 2}
\end{array}
\right\}-\left[ 
\begin{array}{ccc}
\sqrt{\kappa} & 0 & 0\\
\sqrt{\frac{1}{2}\kappa\bar{m}} &  \sqrt{\Gamma\bar{n}} & 0\\
\sqrt{\frac{1}{2}\kappa\bar{m}} &  0 & \sqrt{\Gamma\bar{n}}\\ \hline
\frac{1}{2}\sqrt{\kappa}\bar{m} & \sqrt{\Gamma\bar{n}\bar{m}} & 0\\
\frac{1}{2}\sqrt{\kappa}\bar{m} & 0 & \sqrt{\Gamma\bar{n}\bar{m}}
\end{array}
\right]
\left\{ \begin{array}{c}
\hat{a}_\text{in} \\ 
\hat{b}_\text{in} \\ 
\hat{b}_\text{in}^\dagger 
\end{array}
\right\}+
\left[ 
\begin{array}{cc}
1 & 0\\
\bar{b} & 0\\
\bar{b}^\ast & 0\\ \hline
\bar{b}^2 & 0\\
\bar{b}^{\ast 2} & 0\\
\end{array}
\right]
\left\{ 
\begin{array}{c}
\alpha \\ 
\alpha^\ast
\end{array}
\right\} 
=\frac{d}{dt}\left\{ \begin{array}{c}
\hat{a} \\ 
\hat{a}\hat{b} \\ 
\hat{a}{\hat{b}}^{\dagger } \\  \hline
\hat{a}\hat{b}^{2} \\ 
\hat{a}\hat{b}^{\dagger 2} 
\end{array}
\right\}.
\end{eqnarray}
\noindent
Here, $\theta=\kappa+2\Gamma$ is the decay rate associated with the third-rank one-photon two-phonon processes $\hat{a}\hat{b}^2$ and $\hat{a}\hat{b}^{\dagger 2}$. The approximation $2|\bar{b}|^2\approx\bar{m}$ is used following the discussions in \S\ref{BMWm8}. We do observe that this treatment of multiplicative noise causes non-negligible error in some cases, and is due to be discussed later in \S\ref{Noise}.

\begin{figure}[ht!]
	\centering
	\includegraphics[width=2.1in]{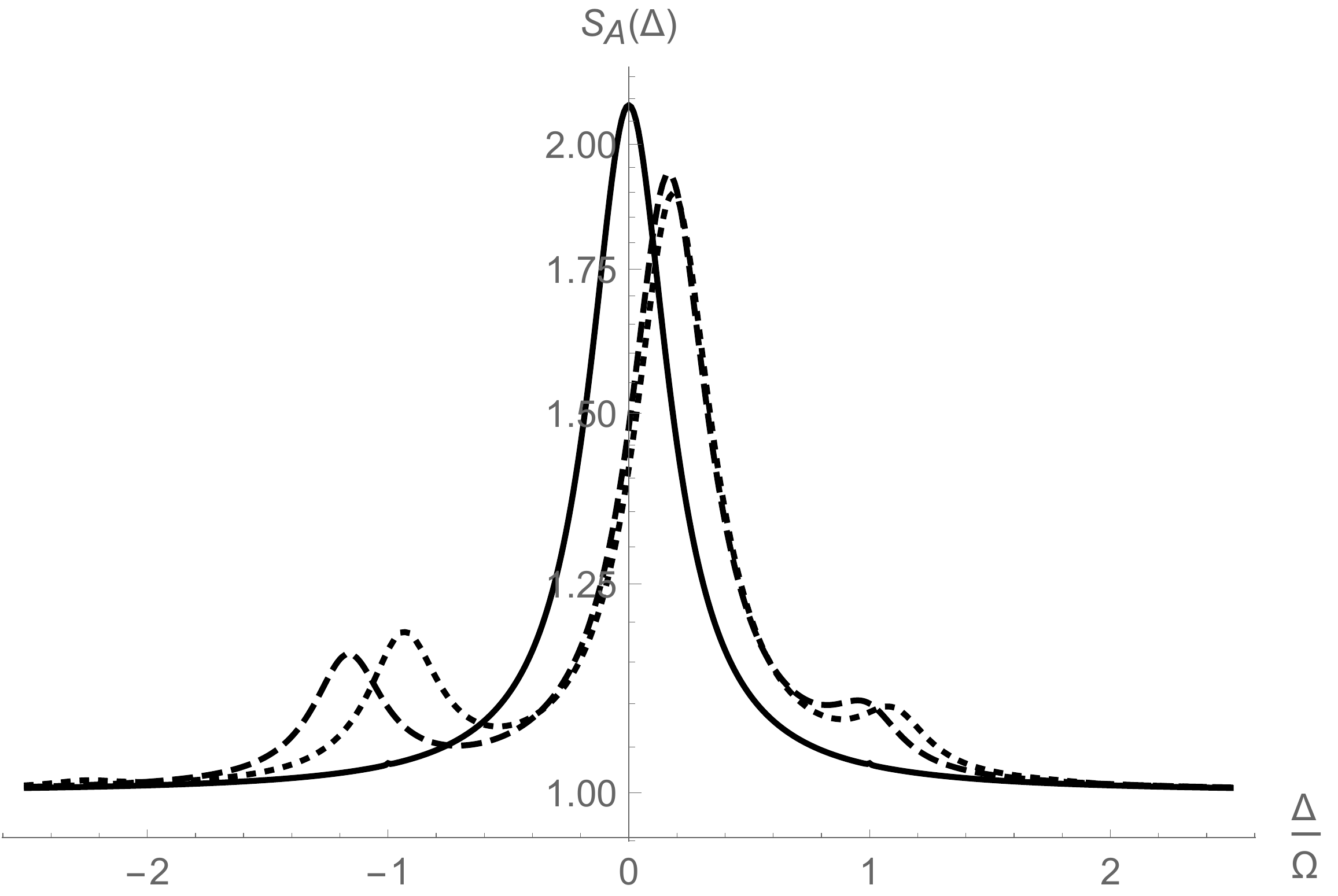}
	\includegraphics[width=2.1in]{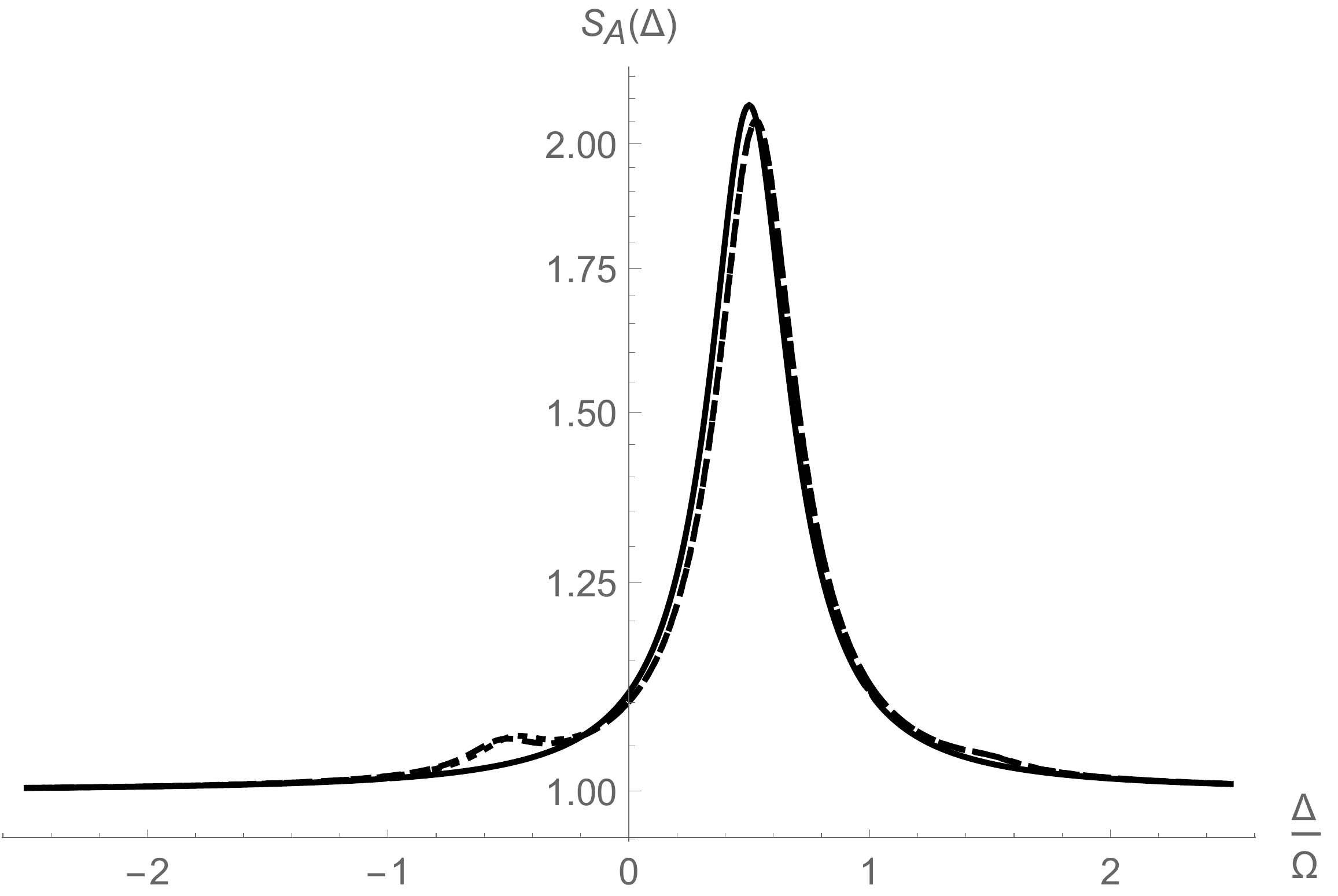}
	\caption{Estimated noise spectrum of a side-band resolved optomechanical cavity under strong pump according to the fully linearized optomechanics shown in solid black, and higher-order optomechanics with the bases $\{A\}^\text{T}=\left\{\hat{a},\hat{a}\hat{b},\hat{a}{\hat{b}}^{\dagger }\right\}$ and (\ref{eq54}) respectively shown in dashed and dotted curves: resonant pump (left); pump on the red mechanical sideband (right).\label{Fig15}}
\end{figure}

Vertical and horizontal partitions separate the single-phonon $\{\hat{a}\hat{b},\hat{a}\hat{b}^\dagger\}$ and two-phonon processes $\{\hat{a}\hat{b}^2,\hat{a}\hat{b}^{\dagger^2}\}$. So by retaining only the first $3\times 3$ blocks and first $3$ rows what remains is nothing but the equations of first-order optomechanics in terms of the second-rank single-phonon operator basis $\{\hat{a},\hat{a}\hat{b},\hat{a}\hat{b}^\dagger\}$. 

The first-order $\delta\Delta^{(1)}$ and second-order $\delta\Delta^{(2)}$ sideband inequivalences take on similar expansions as
\begin{equation}
\label{eq56}
\frac{\delta\Delta^{(1)}}{\Omega}\approx-\frac{\delta\Delta^{(2)}}{2\Omega}\approx \left(\frac{g_0}{\Omega}\right)^2\left(\bar{n}+\frac{1}{2}\right)-2\left(\frac{g_0}{\Omega}\right)^4\left(\bar{n}+\frac{1}{2}\right)\left(\bar{m}+\frac{1}{2}\right) \approx\mathcal{G}^2-4\mathcal{G}^4\mathcal{G}_0^2.
\end{equation}
Here, $\mathcal{G}_0=g_0/\Omega$ and $\mathcal{G}=g/\Omega$ are normalized interaction rates with respect to the mechanical frequency, where $g=g_0\sqrt{\bar{n}}$ is the enhanced optomechanical interaction rate. Furthermore, $\bar{m}$ is approximated from (\ref{m8}) in the above.

Results of noise spectrum calculations using the fully linearized and higher-order formulations of optomechanics is shown in Fig. \ref{Fig15}. The single-photon optomechanical interaction rate $g_0=1.68\times 10^{-3}\Omega$ and the enhanced optomechanical interaction on resonance satisfies $g=0.31\Omega$, corresponding to the strong coupling regime and optical power of $P_\text{op}=47.2\mu\text{W}$ at $T=3\text{K}$. Cavity is side-band resolved with the parameters given elsewhere \cite{SKip2,SKip3}. The input power is high-enough to cause the cavity to exhibit asymmetric reflectivity, a very clear hallmark of bistability seen easily in experiments. 

\begin{figure}[ht!]
	\centering
	\includegraphics[width=2.1in]{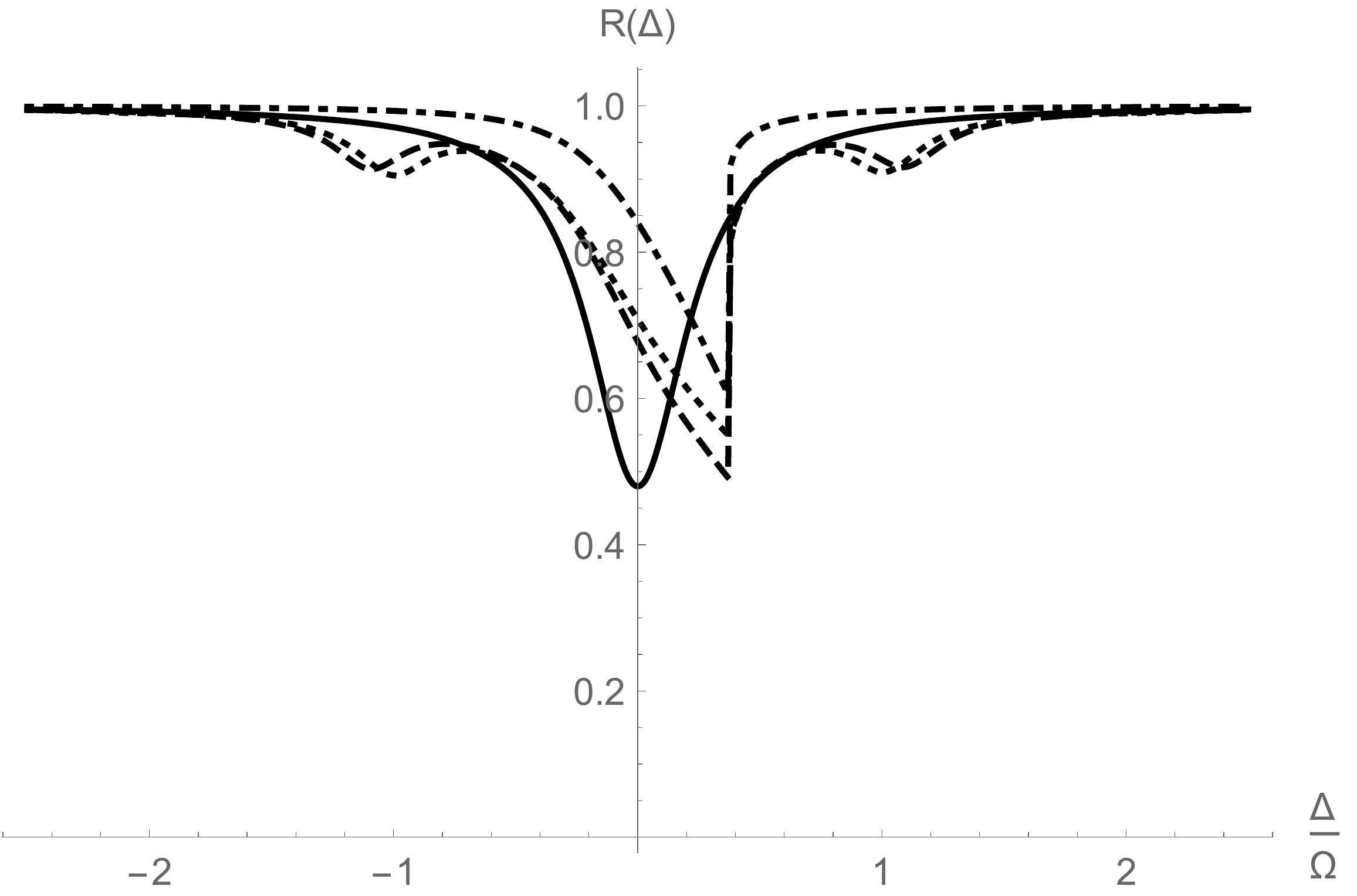}
	\caption{Estimated reflectivity $|Y_{11}(\omega)|^2$ of a side-band resolved optomechanical cavity under single pump with varying frequency according to the fully linearized optomechanics in solid black, higher-order optomechanics with the bases $\{A\}^\text{T}=\left\{\hat{a},\hat{a}\hat{b},\hat{a}{\hat{b}}^{\dagger }\right\}$ and (\ref{eq54}) respectively shown in dashed and dotted curves, and approximate semi-classical calculation using (\ref{eq9}), $\hat{b}\rightarrow\bar{b}$, $\hat{a}\rightarrow\bar{a}$ and the Langevin equation for $\hat{a}$.\label{Fig16}}
\end{figure}

This has been calculated and illustrated in Fig. \ref{Fig16} for various linear and higher-order formulations resulting from simulating a scanning pump experiment. It can be seen that the fully-linearized optomechanics cannot reasonably reproduce the highly asymmetric and non-Lorentzian lineshape of cavity under strong pump. As a simple measure of reflectivity, one may use the Langevin equation for photons $\hat{a}$ with the semi-classical substitutions  $\hat{a}\rightarrow\bar{a}$ and $\hat{b}\rightarrow\bar{b}$, where $\bar{b}$ is correspondingly given from (\ref{eq9}), and $\bar{n}$ can be nonlinearly solved from (\ref{eq12}). For a side-coupled cavity where the reflectivity is not identity, and external coupling rate $\kappa_\text{ex}$ is known, we have $\eta=\kappa_\text{ex}/\kappa$, leading to the approximation
\begin{equation}
\mathcal{R}(\omega,\Delta)=1-\frac{i\kappa_\text{ex}}{\omega+\Delta+2g_0^2\bar{n}\frac{\Omega}{\Omega^2+\frac{1}{4}\Gamma^2}+i\frac{1}{2}\kappa},
\end{equation} 
where $|\mathcal{R}|^2$ is plotted as the dot-dashed curve in Fig. \ref{Fig16}. More accurate solutions can be found by taking the scattering matrix element $\mathcal{R}=Y_{11}$, also shown in Fig. \ref{Fig16}.

\begin{figure}[ht!]
	\centering
	\includegraphics[width=2.1in]{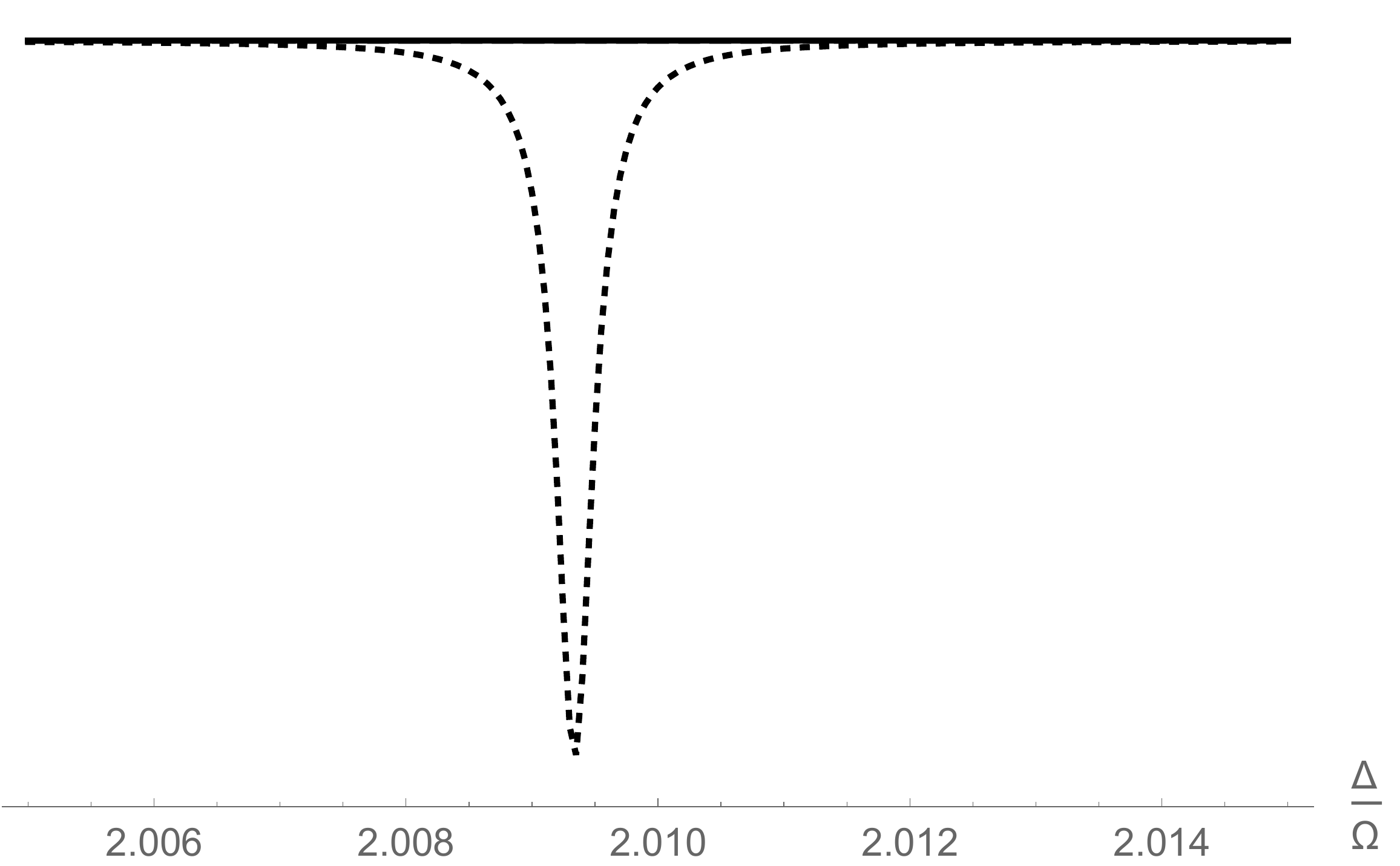}
	\includegraphics[width=2.1in]{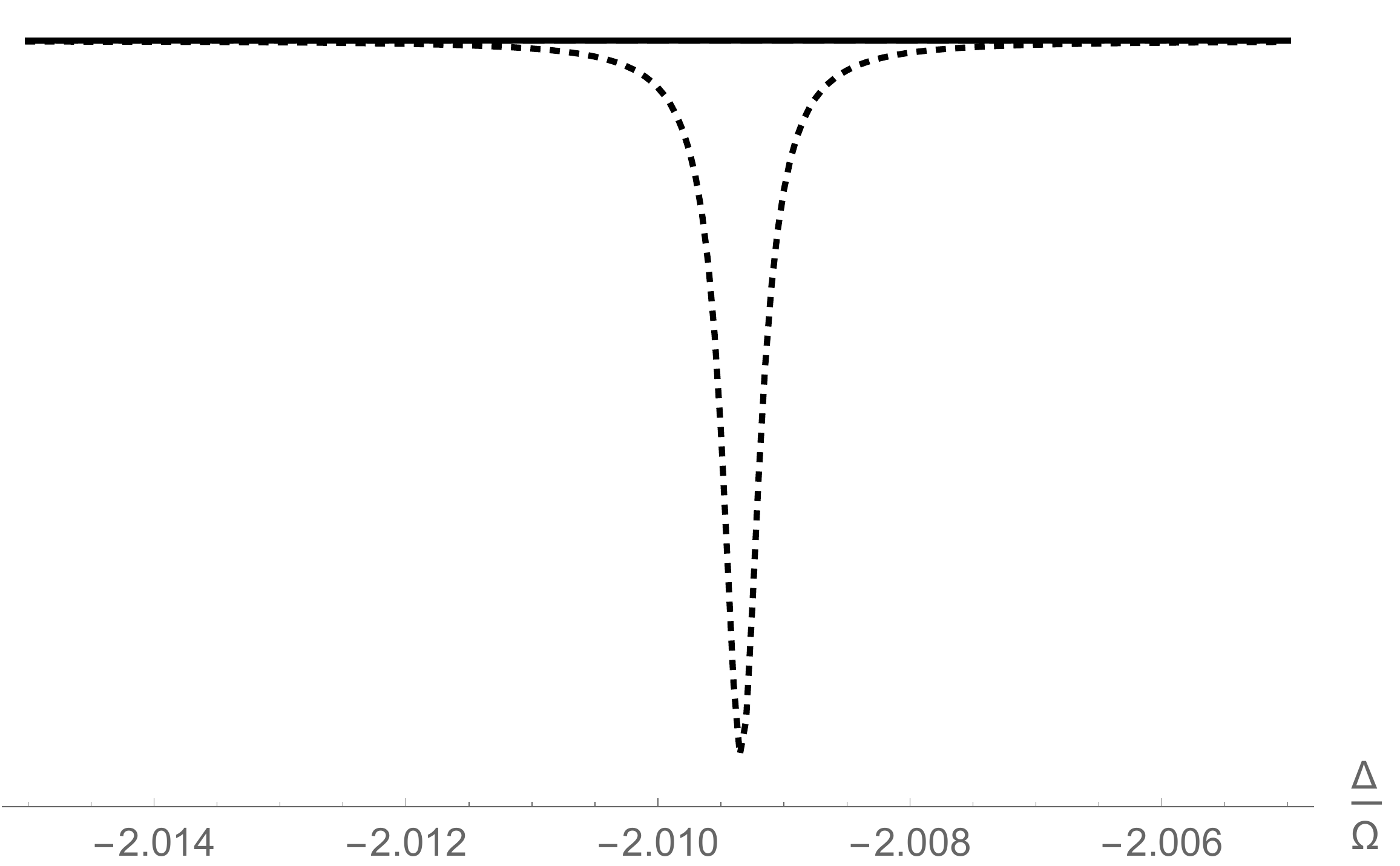}
	\caption{Illustration of tiny second-order sidebands around $\pm 2\Omega$ by calculation of reflectivity $|Y_{11}(\omega)|^2$ from a side-band resolved optomechanical cavity under single pump with varying frequency. Only the higher-order optomechanics with the bases (\ref{eq54}) shown in dotted curve may exhibit non-trivial behavior near second-order side-bands: second-order red sideband (left); second-order blue sideband (right). \label{Fig17}}
\end{figure}

The existence of tiny second-order sidebands around $\pm 2\Omega$ is illustrated by calculation of the reflectivity near the corresponding resonances. This has been shown in Fig. \ref{Fig17} for both of the second-order red and blue sidebands at $P_\text{op}=3\text{mW}$. The depths of these resonances are rather small, being only around $-0.13\text{mdB}$.

\section{Multiplicative Noise}\label{Noise}

Rewriting (\ref{eq55}) without the zeroth-order approximation for multplicative noise, gives the exact higher-order set of Langevin equations 
\begin{eqnarray}
\nonumber
\frac{d}{dt}\left\{ \begin{array}{c}
\hat{a} \\ 
\hat{a}\hat{b} \\ 
\hat{a}{\hat{b}}^{\dagger } \\ 
\hat{a}\hat{b}^{2} \\ 
\hat{a}\hat{b}^{\dagger 2} 
\end{array}
\right\}&=&\left[ 
\begin{array}{ccccc}
i\Delta-\frac{\kappa }{2} & ig_0 & ig_0 & 0 & 0 \\ 
ig_0(\bar{m}+\bar{n}+1) & -i(\Omega-\Delta)-\frac{\gamma }{2} & 0 & ig_0 & 0 \\ 
ig_0(\bar{m}-\bar{n}) & 0 & i(\Omega+\Delta)-\frac{\gamma }{2} & 0 & ig_0 \\ 
0 & ig_0(\bar{m}+2\bar{n}+2) & 0 & -i(2\Omega-\Delta)-\frac{\theta}{2} & 0 \\ 
0 & 0 & ig_0(\bar{m}-2\bar{n}-1) & 0 & i(2\Omega+\Delta)-\frac{\theta}{2}  
\end{array}
\right]
\left\{ \begin{array}{c}
\hat{a} \\ 
\hat{a}\hat{b} \\ 
\hat{a}{\hat{b}}^{\dagger } \\ 
\hat{a}\hat{b}^2 \\
\hat{a}\hat{b}^{\dagger 2}
\end{array}
\right\}\\ \label{Noise1}
&-&\left[ 
\begin{array}{ccc}
\sqrt{\kappa} & 0 & 0\\
\sqrt{\kappa}\hat{b} &  \sqrt{\Gamma}\hat{a} & 0\\
\sqrt{\kappa}\hat{b}^\dagger &  0 & \sqrt{\Gamma}\hat{a}\\ 
\sqrt{\kappa}\hat{b}^2 & \sqrt{\Gamma}\hat{a}\hat{b} & 0\\
\sqrt{\kappa}\hat{b}^{\dagger 2} & 0 & \sqrt{\Gamma}\hat{a}\hat{b}^\dagger
\end{array}
\right]
\left\{ \begin{array}{c}
\hat{a}_\text{in} \\ 
\hat{b}_\text{in} \\ 
\hat{b}_\text{in}^\dagger 
\end{array}
\right\}+
\left[ 
\begin{array}{cc}
1 & 0\\
\bar{b} & 0\\
\bar{b}^\ast & 0\\ 
\bar{b}^2 & 0\\
\bar{b}^{\ast 2} & 0\\
\end{array}
\right]
\left\{ 
\begin{array}{c}
\alpha \\ 
\alpha^\ast
\end{array}
\right\}.
\end{eqnarray}

To illustrate how the multiplicative noise terms on the second line are to be treated, let us assume that a simple equation is given as 
\begin{equation}
\label{Noise2}
\frac{d}{dt}\mathscr{A}(t)=(i\Delta-\frac{1}{2}\kappa)\mathscr{A}(t)-\sqrt{\kappa}\hat{x}(t)\hat{a}_\text{in}(t),
\end{equation}
\noindent
where $\hat{x}$ is some dimensionless and time-dependent operator and $\hat{a}_\text{in}$ corresponds to a white noise random process. The spectral density of the zero-average operator $\mathscr{A}$ by definition is
\begin{equation}
\label{Noise3}
S_{\mathscr{A}\mathscr{A}}(w)=\int_{-\infty}^{\infty}d\tau e^{iw\tau}\braket{\mathscr{A}^\dagger(t)\mathscr{A}(t+\tau)}.
\end{equation}
\noindent
Therefore, the symmetrized spectral density via symmetrization operator $\mathcal{S}$ which is the actual quantity measured in experiments is
\begin{eqnarray}
\mathcal{S}S_{\mathscr{A}\mathscr{A}}(w)&=&\int_{-\infty}^{\infty}d\tau e^{iw\tau}\braket{\mathcal{S}\{\mathscr{A}^\dagger(t)\mathscr{A}(t+\tau)\}}\\ \nonumber
&=&\int_{-\infty}^{\infty}d\tau e^{iw\tau}\braket{\mathscr{A}^\dagger(t)\mathscr{A}(t+\tau)}_\text{S}.
\end{eqnarray}
The equation (\ref{Noise2}) admits a formal solution
\begin{equation}
\label{Noise4}
\mathscr{A}(t)=-\sqrt{\kappa}\mathbb{L}(t)\hat{x}(t)\hat{a}_\text{in}(t),
\end{equation}
\noindent
where $\mathbb{L}$ is given as
\begin{equation}
\label{Noise5}
\mathbb{L}(t)=\left(\frac{d}{dt}-i\Delta+\frac{1}{2}\kappa\right)^{-1},
\end{equation}
\noindent
and is an operator which can be understood as an inverse Fourier transform such as
\begin{equation}
\label{Noise6}
\mathbb{L}(t)=\mathcal{F}^{-1}\left(\frac{1}{iw-i\Delta+\frac{1}{2}\kappa}\right)(t)=\mathcal{F}^{-1}\{L(w)\}(t).
\end{equation}
Here, we are not interested in an explicit form of $\mathbb{L}$ although it is easy to be evaluated or looked up from table of Fourier transforms.

The formal solution (\ref{Noise4}) gives rise to the symmetrized spectral density
\begin{equation}
\label{Noise7}
\mathcal{S}S_{\mathscr{A}\mathscr{A}}(w)=\kappa\int_{-\infty}^{\infty}d\tau e^{iw\tau}\braket{\hat{a}^\dagger_\text{in}(t)\hat{x}^\dagger(t)\mathbb{L}^\dagger(t)\mathbb{L}(t+\tau)\hat{x}(t+\tau)\hat{a}_\text{in}(t+\tau)}_\text{S}.
\end{equation}
\noindent
Now, we can employ the Isserlis-Wick theorem to decompose the expectation value as \cite{SPaper2,SWick1,SWick2,SWick3}
\begin{eqnarray}
\label{Noise8}
\braket{\hat{a}^\dagger_\text{in}(t)\hat{y}^\dagger(t)\hat{y}(t+\tau)\hat{a}_\text{in}(t+\tau)}_\text{S}&=&\braket{\hat{a}^\dagger_\text{in}(t)\hat{a}_\text{in}(t+\tau)}_\text{S}\braket{\hat{y}^\dagger(t)\hat{y}(t+\tau)}_\text{S}\\ \nonumber
&+&\braket{\hat{a}^\dagger_\text{in}(t)\hat{y}^\dagger(t)}_\text{S}\braket{\hat{y}(t+\tau)\hat{a}_\text{in}(t+\tau)}_\text{S}\\ \nonumber
&+&\braket{\hat{a}^\dagger_\text{in}(t)\hat{y}(t+\tau)}_\text{S}\braket{\hat{y}^\dagger(t)\hat{a}_\text{in}(t+\tau)}_\text{S}.
\end{eqnarray}
\noindent
where $\hat{y}(t)=\mathbb{L}(t)\hat{x}(t)$ is adopted for shorthand notation. Since $\hat{a}_\text{in}$ is a white noise Wiener random process, we may expect that to a very good approximation the second and third terms vanish and thus
\begin{equation}
\label{Noise9}
\mathcal{S}S_{\mathscr{A}\mathscr{A}}(w)=\kappa\int_{-\infty}^{\infty}d\tau e^{iw\tau}\braket{\hat{a}^\dagger_\text{in}(t)\hat{a}_\text{in}(t+\tau)}_\text{S}\braket{\hat{y}^\dagger(t)\hat{y}(t+\tau)}_\text{S}.
\end{equation}
The random nature of a Weiner process requires that \cite{SBowen} 
\begin{eqnarray}
\label{Noise10}
\mathcal{S}S_{\mathscr{A}\mathscr{A}}(w)&=&\left|\kappa\int_{-\infty}^{\infty}d\tau e^{iw\tau}\braket{\hat{y}^\dagger(t)\hat{y}(t+\tau)}_\text{S}\right|^2\mathcal{S}S_{AA}(w)\\ \nonumber
&=&\left|L(w)\ast\hat{x}(w)\right|^2\mathcal{S}S_{AA}(w).
\end{eqnarray}
\noindent
Here, $\ast$ represents convolution in frequency and $L(w)$ is defined in (\ref{Noise6}) and actually represents the equivalent to the scattering matrix element. The expression (\ref{Noise10}) presents a mathematically exact solution to the spectral density problem of multiplicative noise (\ref{Noise2}).

In the context of higher-order quantum optomechanics and referring to (\ref{Noise1}) the operator $\hat{x}$ may represent either of the operators within the set $\{\hat{a},\hat{b},\hat{b}^\dagger,\hat{a}\hat{b},\hat{a}\hat{b}^\dagger,\hat{b}^2,\hat{b}^{\dagger 2}\}$. However, not only these are not yet known, but also, they are influenced by random processes from the correspondingly lower-order interactions with the optical field and mechanical bath. The only approximation needed here is to replace these with corresponding non-operator functions which can be already obtained from the solution to lower-order equations. Doing this results in a set of equations for $\{\bar{a},\bar{b},\bar{b}^\ast,\overline{ab},\overline{ab^\ast},\bar{b}^2,\bar{b}^{\ast 2}\}$, where solutions for $\{\bar{b},\bar{b}^\ast,\bar{b}^2,\bar{b}^{\ast 2}\}$ can be obtained by having $\bar{b}$. This is here calculated from the $3\times 3$ linearized optomechanics, giving rise to the expression 
\begin{equation}
\label{Noise11}
\bar{b}(w)=\frac{\alpha }{i (w+\Omega)+\frac{1}{2}\Gamma}.
\end{equation}
In a similar manner to \S\ref{BMWm8}, the next required expressions can be explicitly obtained  by ${\sf Mathematica}$ as
\begin{eqnarray}
\label{Noise12}
\bar{a}(w)&=&\frac{-\alpha ^2 g_0 (w -\Delta -\Omega-i \frac{1}{2}\gamma)}{(w+\Omega-\frac{1}{2}i \Gamma) \left\{2 g_0^2\left[ (\Delta-w+\frac{1}{2} i \gamma)  (\bar{m}+\frac{1}{2})+\Omega(\bar{n}+\frac{1}{2})\right]+(w-\Delta -\frac{1}{2}i \kappa) \left[\left(w-\Delta -\frac{1}{2}i \gamma \right)^2-\Omega ^2\right]\right\}}, \\ \nonumber
\overline{ab}(w)&=&\frac{\alpha ^2 \left[g_0^2 (\bar{m}-\bar{n})-(w-\Delta -\frac{1}{2}i \kappa) (w -\Delta -\Omega -\frac{1}{2}i \gamma)\right]}{(w+\Omega-\frac{1}{2}i \Gamma) \left\{2 g_0^2 \left[(\Delta-w+\frac{1}{2}i \gamma)(\bar{m}+\frac{1}{2})+\Omega(\bar{n}+\frac{1}{2})\right]+(w-\Delta -\frac{1}{2}i \kappa) \left[\left(w-\Delta -\frac{1}{2}i \gamma \right)^2-\Omega ^2\right]\right\}}, \\ \nonumber
\overline{ab^\ast}(w)&=&\frac{-\left| \alpha \right| ^2 \left[g_0^2 (\bar{m}+\bar{n}+1)-(w-\Delta -\frac{1}{2}i \kappa) (w-\Delta+\Omega-\frac{1}{2}i \gamma)\right]}{(w+\Omega+\frac{1}{2}i \Gamma) \left\{2 g_0^2 \left[(\Delta-w+\frac{1}{2}i \gamma)(\bar{m}+\frac{1}{2})+\Omega(\bar{n}+\frac{1}{2})\right]+(w-\Delta -\frac{1}{2}i \kappa) \left[\left(w-\Delta -\frac{1}{2}i \gamma \right)^2-\Omega ^2\right]\right\}}.
\end{eqnarray}
\noindent
In the above equations, it has to be noticed that $\alpha$ is a complex quantity which satisfies $|\alpha|=\sqrt{\kappa_\text{ex}P_\text{op}/\hbar\omega}$, and also by (\ref{eq11}) we have
\begin{equation}
\label{Noise13}
\alpha=\sqrt{\bar{n}}\left[-\frac{\kappa}{2}+i\left(\Delta+\frac{2g_0^2\Omega}{\Omega^2+\frac{1}{4}\Gamma^2}\bar{n}\right)\right].
\end{equation}

Now, we can rewrite the Langevin equations (\ref{Noise1}) as
\begin{eqnarray}
\nonumber
\frac{d}{dt}\{\delta A\}&=&[\textbf{M}]\{\delta A\}-[\hat{\textbf{G}}]\{ A_\text{in}\},\\ \label{Noise14}
[\textbf{M}]&=&\left[ 
\begin{array}{ccccc}
i\Delta-\frac{\kappa }{2} & ig_0 & ig_0 & 0 & 0 \\ 
ig_0(\bar{m}+\bar{n}+1) & -i(\Omega-\Delta)-\frac{\gamma }{2} & 0 & ig_0 & 0 \\ 
ig_0(\bar{m}-\bar{n}) & 0 & i(\Omega+\Delta)-\frac{\gamma }{2} & 0 & ig_0 \\ 
0 & ig_0(\bar{m}+2\bar{n}+2) & 0 & -i(2\Omega-\Delta)-\frac{\theta}{2} & 0 \\ 
0 & 0 & ig_0(\bar{m}-2\bar{n}-1) & 0 & i(2\Omega+\Delta)-\frac{\theta}{2}  
\end{array}
\right], \\ \nonumber
\{\delta A\}^\text{T}&=&\left\{\delta\hat{a},\delta(\hat{a}\hat{b}),\delta(\hat{a}{\hat{b}}^{\dagger }),\delta(\hat{a}\hat{b}^{2}),\delta(\hat{a}\hat{b}^{\dagger 2})\right\},\\ \nonumber
\{ A_\text{in}\}^\text{T}&=&\left\{\hat{a}_\text{in},\hat{b}_\text{in},\hat{b}_\text{in}^\dagger\right\}, \\ \nonumber
[\hat{\textbf{G}}]&=&\left[ 
\begin{array}{ccc}
\sqrt{\kappa} & 0 & 0\\
\sqrt{\kappa}\hat{b} &  \sqrt{\Gamma}\hat{a} & 0\\
\sqrt{\kappa}\hat{b}^\dagger &  0 & \sqrt{\Gamma}\hat{a}\\ 
\sqrt{\kappa}\hat{b}^2 & \sqrt{\Gamma}\hat{a}\hat{b} & 0\\
\sqrt{\kappa}\hat{b}^{\dagger 2} & 0 & \sqrt{\Gamma}\hat{a}\hat{b}^\dagger
\end{array}
\right].
\end{eqnarray}
After defining the decay matrix
\begin{equation}
\label{Noise15}
[\sqrt{\Gamma}]=\left[\begin{array}{ccccc}
\sqrt{\kappa} & 0 & 0 & 0 & 0\\
0 & \sqrt{\gamma} & 0 & 0 & 0\\
0 & 0 & \sqrt{\gamma} & 0 & 0\\
0 & 0 & 0 & \sqrt{\theta} & 0\\
0 & 0 & 0 & 0 & \sqrt{\theta}
\end{array}\right],
\end{equation}
\noindent
taking the Fourier transform, and using the input-output relation 
\begin{equation}
\label{Noise16}
\{A_\text{out}(\omega)\}=\{A_\text{in}(\omega)\}+[\sqrt{\Gamma}]^\text{T}\{\delta A(\omega)\},
\end{equation}
\noindent
we arrive at the definition of the scattering matrix
\begin{equation}
\label{Noise17}
[\textbf{Y}(\omega)]=\textbf{I}-[\sqrt{\Gamma}]^\text{T}\left([\textbf{M}]-i\omega[\textbf{I}]\right)^{-1}[\sqrt{\Gamma}],
\end{equation}
by which and (\ref{Noise10}) we can evaluate the desired symmetrized spectral density of output optical field as
\begin{eqnarray}
\label{Noise18}
S(\omega)&=&|Y_{11}(\omega)|^2 S_{AA}(\omega)\\ \nonumber
&+&\frac{1}{\gamma^2}\left|\left[Y_{12}(\omega)+Y_{13}(\omega)\right]\ast\bar{a}(\omega)\right|^2 S_{BB}(\omega) \\ \nonumber
&+&\frac{1}{\theta^2}\left|\left[Y_{14}(\omega)\ast\overline{ab}(\omega)+Y_{15}(\omega)\ast\overline{ab^\ast}(\omega)\right]\right|^2 S_{BB}(\omega),
\end{eqnarray}
where spectral power densities $S_{AA}$ and $S_{BB}$ are already introduced in (\ref{eq13}) and convolutions $\ast$ take place over the entire frequency axis. 

\section{Elements of Higher-order Scattering Matrices}
This section reports the explicit elements of the first row of scattering matrix $[\textbf{Y}]$ in (\ref{Noise17}), as needed for calculation of the spectral density according to (\ref{Noise18}). These might be useful only when the method of residues are to be used for exact evaluation of complex convolution integrals, otherwise full numerical simulation of (\ref{Noise18}) is much preferable.  

\subsection{Second-order $3\times 3$ Formalism}\label{3x3}

The elements of the scattering matrix are explicitly found using the supplementary ${\sf Mathematica}$ packages, and after some simplification they take the form
\begin{eqnarray}
Y_{11}(\omega)&=&1-\frac{2i\kappa_\text{ex}\left[\left(\omega-\Delta -\frac{1}{2}i \gamma \right)^2-\Omega ^2\right]}{2 g_0^2\left[ (\Delta-\omega+\frac{1}{2} i \gamma)  (\bar{m}+\frac{1}{2})+\Omega(\bar{n}+\frac{1}{2})\right]+(\omega-\Delta -\frac{1}{2}i \kappa) \left[\left(\omega-\Delta -\frac{1}{2}i \gamma \right)^2-\Omega ^2\right]}, \\ \nonumber
Y_{12}(\omega)&=&\frac{-ig_0\sqrt{\gamma\kappa_\text{ex}}\left(\omega-\Delta-\Omega -\frac{1}{2}i \gamma \right)}{2 g_0^2\left[ (\Delta-\omega+\frac{1}{2} i \gamma)  (\bar{m}+\frac{1}{2})+\Omega(\bar{n}+\frac{1}{2})\right]+(\omega-\Delta -\frac{1}{2}i \kappa) \left[\left(\omega-\Delta -\frac{1}{2}i \gamma \right)^2-\Omega ^2\right]}, \\ \nonumber
Y_{13}(\omega)&=&\frac{-ig_0\sqrt{\gamma\kappa_\text{ex}}\left(\omega-\Delta+\Omega -\frac{1}{2}i \gamma \right)}{2 g_0^2\left[ (\Delta-\omega+\frac{1}{2} i \gamma)  (\bar{m}+\frac{1}{2})+\Omega(\bar{n}+\frac{1}{2})\right]+(\omega-\Delta -\frac{1}{2}i \kappa) \left[\left(\omega-\Delta -\frac{1}{2}i \gamma \right)^2-\Omega ^2\right]}.
\end{eqnarray}
These expressions are useful in speed up of the code, as well as wherever the method of residues is to be used.

\subsection{Third-order $5\times 5$ Formalism}\label{5x5}

The convergence of $3\times 3$ is sufficiently good for most practical purposes, and also the explicit expressions for $5\times 5$ matrices decompose into products of fourth-order polynomials in terms of $\omega$ in their denominators, which severely limits the usefulness of applicability of the method of residues. For this reason, their explicit expressions are not included here. The interested reader may find them in the supplementary ${\sf Mathematica}$ packages instead.

\section*{Mathematica Packages}
These are brief descriptions of supplied ${\sf Mathematica}$ packages along with this article, written by the author:
\begin{itemize}[noitemsep,nolistsep]
	\item ${\sf SuppleMath1.nb}$: Derivation of coherent phonon population $\bar{m}(\Delta)$ in \ref{m8}.
	\item ${\sf SuppleMath2.nb}$: Derivation of expressions corresponding to \S\ref{3x3} and \S\ref{5x5}.
	\item ${\sf SuppleMath3.nb}$: Code for generation of noise spectra in the main article including animated graphs.
	\item ${\sf SuppleMath4.nb}$: Code for generation of stability diagrams in the main article.
	\item ${\sf SuppleMath5.nb}$: Code for generation of higher-order optical spring effect in Fig. \ref{FigA1}.
	\item ${\sf SuppleMath6.nb}$: Code for generation of higher-order optical spring effect in  Fig. \ref{FigA2}.
	\item ${\sf SuppleMath7.nb}$: Code for generation of higher-order optical spring effect in  Fig. \ref{FigA3}.
	\item ${\sf SuppleMath8.nb}$: Numerical calculation of side-band inequivalence.
	\item ${\sf SuppleMath9.nb}$: Code for generation of normalized side-band inequivalence in of the main article.
	\item ${\sf SuppleMath10.nb}$: Code for generation of reflection spectra in Fig. \ref{Fig16}.
	\item ${\sf SuppleMath11.nb}$: Code for generation of second-order side-bands in Fig. \ref{Fig17}.
\end{itemize}

\end{document}